\documentclass[10pt]{article}
\usepackage{amsmath}
\usepackage{graphicx}
\usepackage{color}
\usepackage{orcidlink}
\usepackage{hyperref}
\hypersetup{colorlinks=true, linkcolor=gray, citecolor=gray, urlcolor=blue}
\usepackage{subcaption}
\RequirePackage[numbers,sort&compress]{natbib}
\begin{document}
\baselineskip=14pt

\begin{center}
\LARGE{Observable signatures of Black Holes with Hernquist Dark Matter Halo having a cloud of strings: From Geodesics to Shadow}    
\end{center}

\vspace{0.3cm}

\begin{center}

{\bf Faizuddin Ahmed}\orcidlink{0000-0003-2196-9622}\\
Department of Physics, Royal Global University, Guwahati, 781035, Assam, India\\
e-mail: faizuddinahmed15@gmail.com\\

\vspace{0.15cm}
{\bf Ahmad Al-Badawi}\orcidlink{0000-0002-3127-3453}\\
Department of Physics, Al-Hussein Bin Talal University, 71111,
Ma'an, Jordan. \\
e-mail: ahmadbadawi@ahu.edu.jo\\
\vspace{0.15cm}
{\bf \.{I}zzet Sakall{\i}}\orcidlink{0000-0001-7827-9476}\\
Physics Department, Eastern Mediterranean University, Famagusta 99628, North Cyprus via Mersin 10, Turkey\\
e-mail: izzet.sakalli@emu.edu.tr (Corresponding author)
\end{center}

\vspace{0.15cm}
\begin{abstract}
We present a comprehensive theoretical investigation of a novel black hole (BH) spacetime: a Schwarzschild BH embedded in a Hernquist-type dark matter halo (HDMH) and surrounded by a cloud of cosmic strings (CSs)---collectively termed the Schwarzschild-HDMH with CS (SHDMHCS) configuration. By analyzing the spacetime geometry, we explore how key parameters---the core radius and halo density of the dark matter, along with the string tension---affect the geodesic motion of both massless and massive particles. Our results reveal that the combined influence of HDMH and CSs modifies the effective potentials for null and timelike geodesics, leading to distinct dynamical behavior compared to standard Schwarzschild geometry. We perform a perturbative analysis for scalar (spin-0), electromagnetic (spin-1), and Dirac (spin-1/2) fields, deriving the associated effective potentials and showing how both the dark halo and CSs alter field propagation and potential barriers. The shadow of the BH is studied in detail: we derive analytical expressions for photon sphere and shadow radii, finding that CSs tend to enlarge the shadow, while HDMH properties tend to shrink it. The combined effects of these parameters significantly influence the shadow's shape and size, producing potentially observable signatures. Our results establish that the SHDMHCS configuration yields distinct observational imprints detectable by present and forthcoming astrophysical instruments. This framework provides new tools for probing exotic matter distributions via gravitational wave observations, orbital dynamics, and high-resolution black hole imaging, offering a pathway to distinguish such configurations from simpler BH models in realistic environments.
\end{abstract}

\vspace{0.15cm}

{\bf keywords}: Modified gravity; Black holes; Hernquist dark matter halo; Cloud of strings; Geodesics analysis; Black hole perturbations theory

\vspace{0.3cm}

\section{Introduction} \label{sec1}

The quest to understand the fundamental nature of dark matter (DM) and its gravitational interactions with BHs represents one of the most compelling frontiers in modern astrophysics and gravitational physics. Observational evidence from galaxy rotation curves, gravitational lensing, and cosmic microwave background radiation (CMBR) consistently points to the existence of a mysterious form of matter that comprises approximately 85\% of all matter in the universe, yet remains largely invisible to electromagnetic radiation \cite{isz01,isz02,isz03}. Among the various theoretical frameworks proposed to model DM distributions, the Hernquist profile has emerged as a particularly successful parameterization for describing the density distributions in galactic halos, especially in elliptical galaxies and the bulges of spiral galaxies \cite{isz04,isz05}. This profile, characterized by a finite total mass and a cusped central density distribution, provides an excellent fit to numerical simulations of structure formation and observational data from galaxy clusters \cite{isz06,isz07}.

The intersection of BH physics with DM halos has garnered intense theoretical and observational interest, particularly following the groundbreaking observations of supermassive BHs (SMBHs) by the Event Horizon Telescope (EHT) collaboration \cite{isz08,isz09}. These observations have not only confirmed the existence of BH shadows but have also opened new avenues for probing the environmental effects surrounding these compact objects. When BHs are embedded within the HDMH, the resulting spacetime geometry exhibits rich phenomenology that deviates significantly from the classical Schwarzschild solution \cite{isz10,isz11,isz12,bdm1}. The presence of HDMH modifies the metric components, leading to observable consequences in gravitational lensing, orbital dynamics, and the propagation of electromagnetic radiation in the strong-field regime \cite{bdm2,bdm3}. Moreover, the presence of a cloud of CSs also alters the geometric and physical properties of the BH solutions in addition to the HDMH.

Geodesic analysis constitutes the cornerstone of understanding particle motion in curved spacetime, providing fundamental insights into the behavior of both massive and massless test particles in gravitational fields. The study of geodesics in BH spacetimes has a rich history dating back to the early investigations of Schwarzschild and Reissner-Nordström solutions, where researchers discovered the existence of stable and unstable circular orbits, precession effects, and the intricate dynamics of particle trajectories \cite{isz13,isz14,isz15}. In the context of modified BH spacetimes, geodesic motion becomes significantly more complex, with additional parameters governing the effective potential landscape and the stability properties of particle orbits. Null geodesics, describing the motion of photons and other massless particles, play a particularly crucial role in understanding the observational signatures of BHs embedded in exotic matter configurations. The trajectories of light rays in the vicinity of BHs are sensitive probes of the underlying spacetime geometry, encoding information about both the central compact object and its surrounding matter distribution \cite{isz16,isz17}. The study of null geodesics has showed the existence of photon spheres-unstable circular orbits where photons can orbit the BH indefinitely-whose properties are intimately connected to the observable shadow cast by the BH \cite{isz18,isz19}. Furthermore, the analysis of null geodesics provides insights into gravitational lensing phenomena, deflection angles, and the formation of multiple images in strong gravitational fields. On the other hand, timelike geodesics, governing the motion of massive particles (like planets, comets etcs.), offer complementary insights into the dynamics of matter in BH spacetimes. The study of timelike geodesics has showed the existence of innermost stable circular orbits (ISCOs), beyond which stable circular motion becomes impossible, leading to inevitable in-fall toward the BH horizon \cite{isz20,isz21}. In modified BH spacetimes with additional matter fields, the properties of timelike geodesics can exhibit significant deviations from their counterparts in vacuum solutions, with implications for accretion disk physics, orbital period ratios, and the emission of gravitational waves from inspiraling compact objects.

Another interesting aspect of BH solutions is their perturbations theory, which represents a fundamental pillar in our understanding of compact object physics, providing a systematic framework for analyzing the stability and response characteristics of BHs under small perturbations. The mathematical formalism developed by Regge, Wheeler, Zerilli, and others has enabled researchers to study how BHs respond to external disturbances, leading to the discovery of quasinormal modes (QNMs)-characteristic oscillation frequencies that encode information about the BH's intrinsic properties \cite{isz22,isz23,isz24,isz24x1}. In modified BH spacetimes, perturbation theory becomes considerably more involved, as the presence of additional matter fields introduces new degrees of freedom and coupling terms that can significantly alter the perturbation spectrum.

The study of scalar, electromagnetic, and gravitational perturbations in BH spacetimes has provided deep insights into the stability properties of these solutions and their response to external disturbances. Scalar field perturbations, while being the simplest to analyze mathematically, often capture the essential physics of more complex field configurations and serve as valuable testing grounds for numerical methods and analytical techniques \cite{isz25,isz26,isz26x1,isz26x2,isz26x3}. Electromagnetic perturbations are particularly relevant for understanding the behavior of charged particles and electromagnetic fields in strong gravitational environments, with applications to pulsar timing, plasma physics, and high-energy astrophysical phenomena \cite{isz27,isz28}. Gravitational perturbations, while being the most complex to handle analytically, provide direct connections to gravitational wave observations and offer prospects for testing general relativity in the strong-field regime. The concept of BH shadows has emerged as one of the most powerful observational tools for probing strong-field gravity and testing alternative theories of gravitation. A BH shadow represents the dark silhouette formed against the bright background of accreting matter, arising from the existence of an event horizon and the capture of photons within a critical impact parameter \cite{isz29,isz30}. The size and shape of BH shadows are intimately connected to the underlying spacetime geometry, making them sensitive probes of both the central BH properties and any additional matter or field configurations in its vicinity. The successful imaging of shadows cast by the SMBHs in M87* and Sagittarius A* by the EHT collaboration has opened unprecedented opportunities for testing gravitational theories and constraining the properties of matter in extreme gravitational environments \cite{isz31,isz31x1,isz31x2,isz32}.

Recent theoretical investigations have demonstrated that the presence of DM halos can significantly modify BH shadow properties, leading to observable deviations from predictions based on vacuum solutions \cite{isz33,isz34,isz35}. These modifications manifest as changes in shadow size, shape asymmetries, and brightness distributions that could potentially be detected with current and future observational facilities. The study of shadows in modified BH spacetimes has thus become an active area of research, with implications for constraining DM properties, testing alternative gravity theories, and understanding the co-evolution of BHs and their host galaxies.

CSs, which arise as topological defects in many field-theoretic models of the early universe, represent another fascinating component that can significantly influence BH physics. These one-dimensional objects, predicted by grand unified theories and string theory models, carry enormous energy densities and can leave distinctive imprints on spacetime geometry \cite{isz36,isz37}. When BHs are surrounded by clouds of CSs, the resulting spacetime exhibits unique properties that differ from both pure vacuum solutions and solutions with conventional matter distributions. The study of BHs in the presence of CS has shown novel phenomena such as modified horizon structures, altered thermodynamic properties, and distinctive observational signatures \cite{isz38}. Recent studies have investigated BH solutions surrounded by a cloud of CSs, both with and without the presence of a cosmological constant, as reported in \cite{AHEP3, AHEP7, PRD2, PLB2, CPC2, PDU2} and the references therein.

The motivation for this comprehensive investigation stems from the recognition that realistic astrophysical BHs are unlikely to exist in isolation, but instead are embedded in complex environments containing various forms of matter and energy. Although previous studies have examined BHs with DM halos or CS separately, the combined system of a BH embedded in an HDMH and surrounded by a cloud of CSs, which we term the SHDMHCS configuration, represents a more realistic and physically motivated scenario that captures the complexity of actual astrophysical environments. The primary objectives of this work are multifold. First, we seek to derive and analyze the complete spacetime geometry of the SHDMHCS configuration, investigating its mathematical properties, singularity structure, and asymptotic behavior. Second, we aim to conduct a comprehensive geodesic analysis, examining both null and timelike particle trajectories and their dependence on the various physical parameters characterizing the system. Third, we intend to perform a detailed study of field perturbations, analyzing scalar, electromagnetic, and fermionic field propagation in this modified spacetime background. Finally, we aim to investigate the observational signatures of this system, particularly focusing on BH shadow properties and their potential for discriminating between different theoretical models.

This paper is organized as follows: Section \ref{sec2} presents the derivation of the SHDMHCS metric and analyzes its fundamental properties, including curvature invariants and singularity structure. Section \ref{sec3} provides a comprehensive geodesic analysis, examining both null and timelike particle motion, effective potentials, orbital stability, and the properties of circular orbits. Section \ref{sec4} investigates BH perturbations, deriving effective potentials for scalar, electromagnetic, and fermionic field perturbations and analyzing their behavior in the SHDMHCS background. Section \ref{sec5} focuses on BH shadow analysis, deriving the shadow radius and examining how the presence of HDMH and CS affects the observable shadow properties. Finally, Sec. \ref{sec6} summarizes our main findings and discusses their implications for observational astronomy and fundamental physics.

{\color{black}

\section{SHDMHCS spacetime} \label{sec2}

We start this section by considering the following DM profile \cite{isz05}
\begin{equation}
\rho(r) = \rho_s \left(\frac{r}{r_s}\right)^{-\gamma } \left[\left(\frac{r}{r_s}\right)^{\alpha }+1\right]^{\frac{\gamma -\beta }{\alpha }},\label{aa1}
\end{equation}
which, depending on the values of $(\alpha, \beta, \gamma)$, models different DM distributions. $\gamma$ and $\beta$ govern the dependence of the distribution at small and large $r$, whereas $\alpha$ determines the sharpness of the profile's transition. We consider the Hernquist distribution which corresponds to $(\alpha, \beta, \gamma) = (1,4,1)$. The density for the Hernquist profile takes the following form:
\begin{equation}
\rho(r)=\rho_s\,\left(\frac{r}{r_s}\right)^{-1}\,\left[1+\frac{r}{r_s}\right]^{-3}.\label{aa2}    
\end{equation}
Here, $\rho_s$ and $r_s$ are the central density and the core radius of the DM halo, respectively. The mass profile for the DM distribution is
\begin{equation}
M_{H} = \int_0^r 4\pi \rho (r')r'^2 dr' = \frac{2 \pi  r^2 r_s^3 \rho_s}{ \left(r_s+r\right)^2}.\label{aa3}
\end{equation}
We then assume the spherically symmetric metric for a pure DM halo to be
\begin{equation}
ds^2 = -A_1(r) dt^2 +B_1(r)^{-1} dr^2 + r^2 (d\theta^2 + \sin^2 \theta d\phi^2),\label{aa4}
\end{equation}
where $A_1(r)$ is the redshift function and $B_1(r)$ is the shape function. Utilizing the relation between the tangential velocity of a particle $v_t$ and the redshift function $A_1(r)$ on one hand and the relation between $v_t$ and $M_H$ on the other, we can now obtain $A_1(r)$ from the following equation \cite{mo,vt}:
\begin{equation}
v_t^2 = \frac{M_H}{r}=  r \frac{d}{dr} \ln \sqrt{A_1}\quad \Rightarrow A_1=e^{\int \frac{2\,M_H}{r^2}\,dr}.\label{aa5}    
\end{equation}

Combining the above equation with Eq. (\ref{aa3}) and assuming $A_1=B_1$, we get, after retaining leading order terms, the following expression
\begin{equation}
A_1=B_1=e^{-\frac{4\pi \rho_s r_s^3}{r+r_s}} \approx 1-\frac{4\pi \rho_s r_s^3}{r+r_s},\label{aa6}    
\end{equation}

From the Einstein field equations
\begin{equation}
R_{\mu\nu}- \frac{1}{2}\, R\, g_{\mu\nu}= \kappa^2 T_{\mu\nu}^H,\label{aa7}
\end{equation}
for the line element (\ref{aa4}), we obtain the energy-momentum tensor $T_{\mu\nu}^H$ for the spacetime where $T_{\mu\nu}= g^{\nu \lambda} T_{\mu \lambda} = \text{diag} [-\rho, p_r, p, p]$. In Eq. (\ref{aa7}), $R_{\mu\nu}$ is the Ricci tensor, $R$ is the Ricci scalar, and $g_{\mu\nu}$ is the metric tensor for the line element (\ref{aa4}). Utilizing the field equations (\ref{aa7}) provides the following equations:
\begin{equation}
\begin{aligned}
T_t^{t (H)}&=B_1\,\left(\frac{1}{r} \frac{B'_1}{B_1}+\frac{1}{r^2}\right)-\frac{1}{r^2}, \\
T_r^{r (H)} &=B_1\,\left(\frac{1}{r^2} + \frac{1}{r} \frac{A'_{1}}{A_1}\right) - \frac{1}{r^2}, \\
T_\theta^{\theta (H)}&=T_\phi^{\phi (H)} = \frac{1}{2}\,B_1\, \left[\frac{A''_1\,A_1-(A'_{1})^2}{A^2_{1}}+ \frac{1}{2}\, \left(\frac{A'_1}{A_1}\right)^2+\frac{1}{r}\,\left(\frac{A'_1}{A_1}+\frac{B'_1}{B_1}\right) + \frac{A'_{1}\,B'_{1}}{2\,A_1\,B_1} \right].
\end{aligned}
\label{aa8}
\end{equation}

Next, we consider the following metric for the combined system of BH and DM halo:
\begin{equation}
   ds^2 = -f(r)\,dt^2 + g(r)^{-1}\, dr^2 + r^2\, (d\theta^2 + \sin^2 \theta\, d\phi^2),\label{aa9}
\end{equation}
where we write the functions as
\begin{equation}
   f(r)=A_1+A_2\quad \text{and} \quad g(r)=B_1+B_2.\label{aa10}
\end{equation}

We will obtain functions $A_2$ and $B_2$ from field equations (\ref{aa7}) combined with the following field equations for the combined system:
\begin{equation}
R_{\mu\nu}-\frac{1}{2}\,R\,g_{\mu\nu}=T_{\mu\nu}^H+T_{\mu\nu}^{BH},\label{aa11}
\end{equation}
where $T_{\mu\nu}^{BH}$ is the energy-momentum tensor for the Schwarzschild BH. Since there exists no non-zero component of $T_{\mu\nu}^{BH}$, each component of the energy-momentum tensor for the metric (\ref{aa4}) must be equal to the corresponding component for the line element (\ref{aa9}). This yields the following equations whose solutions produce $A_2$ and $B_2$:
\begin{equation}
\begin{aligned}
& \left(B_1+B_2\right)\left(\frac{1}{r^2}+\frac{1}{r} \frac{B'_1+B'_2}{B_1+B_2}\right)=B_1\,\left(\frac{1}{r^2}+\frac{1}{r} \frac{B'_1}{B_1}\right), \\
&\left(B_1+B_2\right)\left(\frac{1}{r^2}+\frac{1}{r} \frac{A'_1+A'_2}{A_1+A_2}\right)=B_1\,\left(\frac{1}{r^2}+\frac{1}{r} \frac{A'_1}{A_1}\right) .
\end{aligned}\label{aa12}
\end{equation}
The above equations with $A_1=B_1$ produce $A_2=B_2$. Since the metric (\ref{aa9}) must revert back to that for a Schwarzschild BH in the absence of DM, we have $A_2=B_2=-\frac{2\,M}{r}$.

Recently, a static and spherically symmetric Schwarzschild BH immersed in HDMH was introduced in Ref. \cite{AK1}. The density for the Hernquist profile takes the form given in Eq. (\ref{aa2}). With this, the metric for the combined system of BH and DM halo becomes \cite{AK1}
\begin{equation}
ds^2 =-f(r)\,dt^2 + f(r)^{-1}\, dr^2 + r^2\, (d\theta^2 + \sin^2 \theta\, d\phi^2),\label{aa13}    
\end{equation}
where the metric function is given by
\begin{equation}
f(r)=1-\frac{2\,M}{r}-\frac{b}{r+r_s},\quad\quad b=4\,\pi\,\rho_s\,r_s^3.\label{aa14}    
\end{equation}

\begin{figure*}
   \centering
   \setlength{\tabcolsep}{0pt}
   \begin{minipage}{0.2\textwidth}
       \centering
       \includegraphics[width=\textwidth]{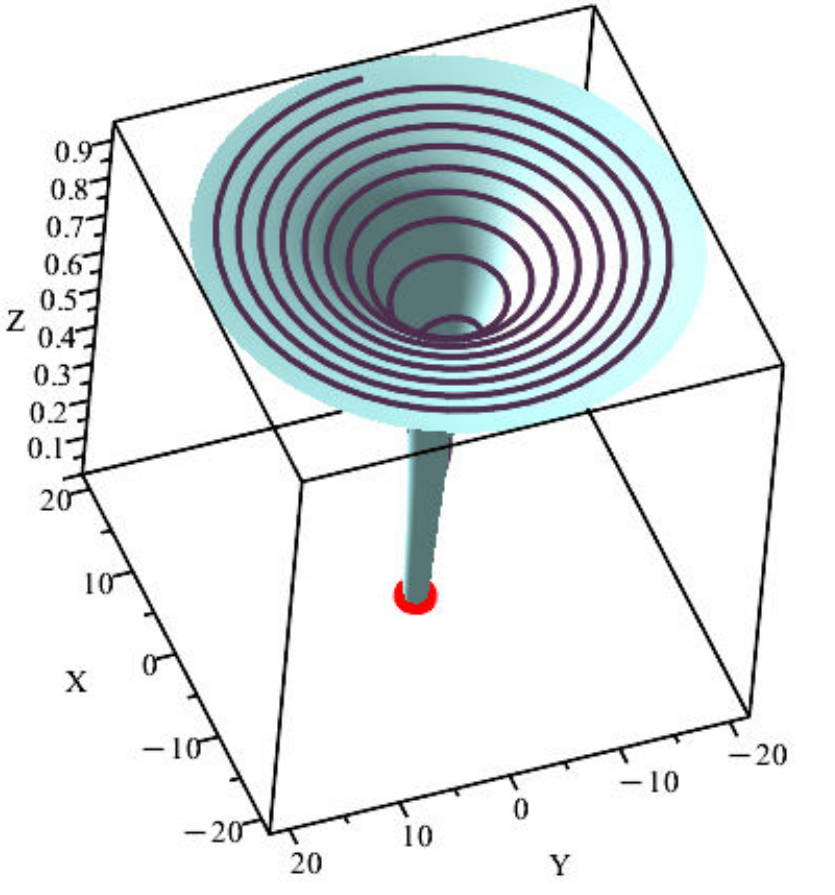}
       \subcaption{[$\rho_s=0$, $r_s=1$] \newline SBH without HDMH.}
       \label{fig:is1}
   \end{minipage}
   \begin{minipage}{0.2\textwidth}
       \centering
       \includegraphics[width=\textwidth]{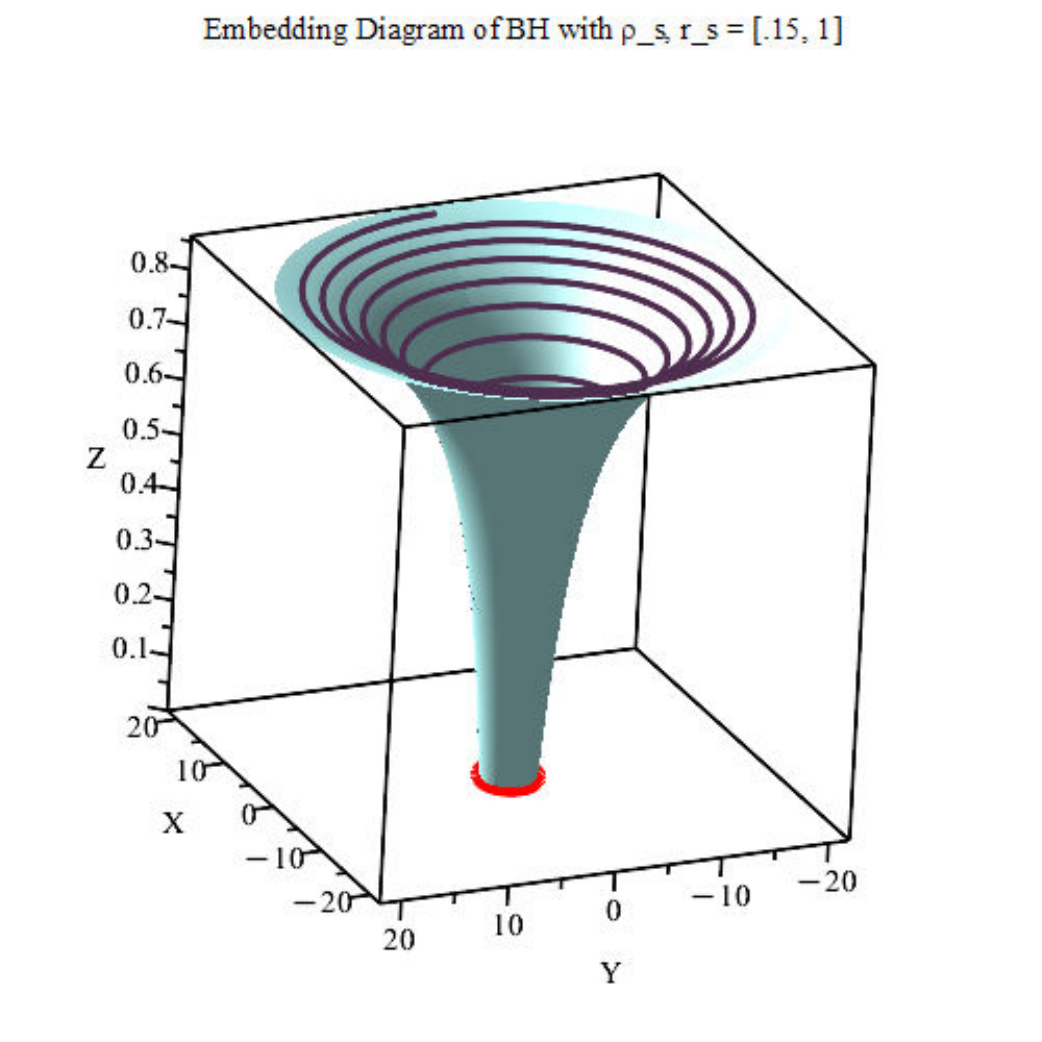}
       \subcaption{[$\rho_s=0.15$, $r_s=1$] \newline SBH with HDMH.}
       \label{fig:is2}
   \end{minipage}
   \begin{minipage}{0.2\textwidth}
       \centering
       \includegraphics[width=\textwidth]{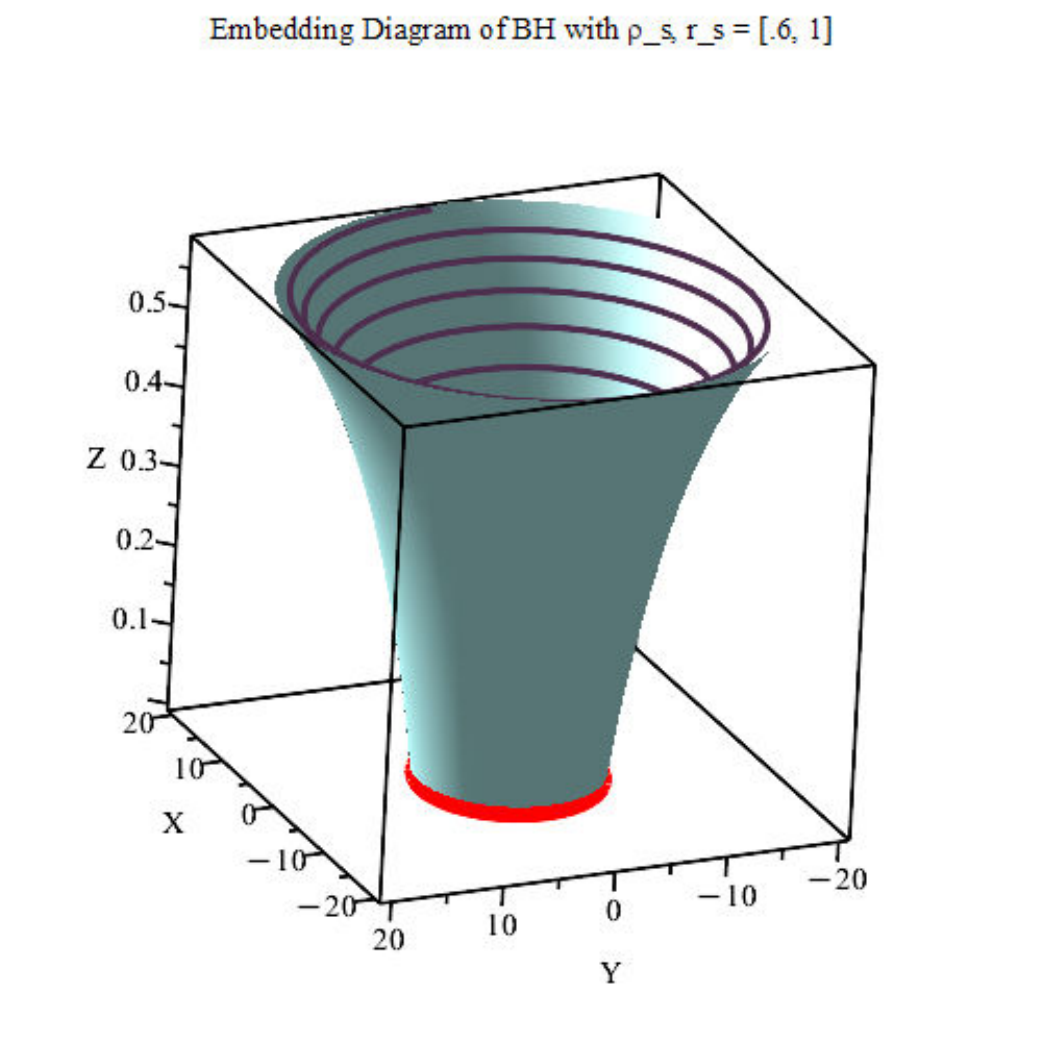}
       \subcaption{[$\rho_s=0.6$, $r_s=1$] \newline SBH with HDMH.}
       \label{fig:is3}
   \end{minipage}
       \begin{minipage}{0.2\textwidth}
       \centering
       \includegraphics[width=\textwidth]{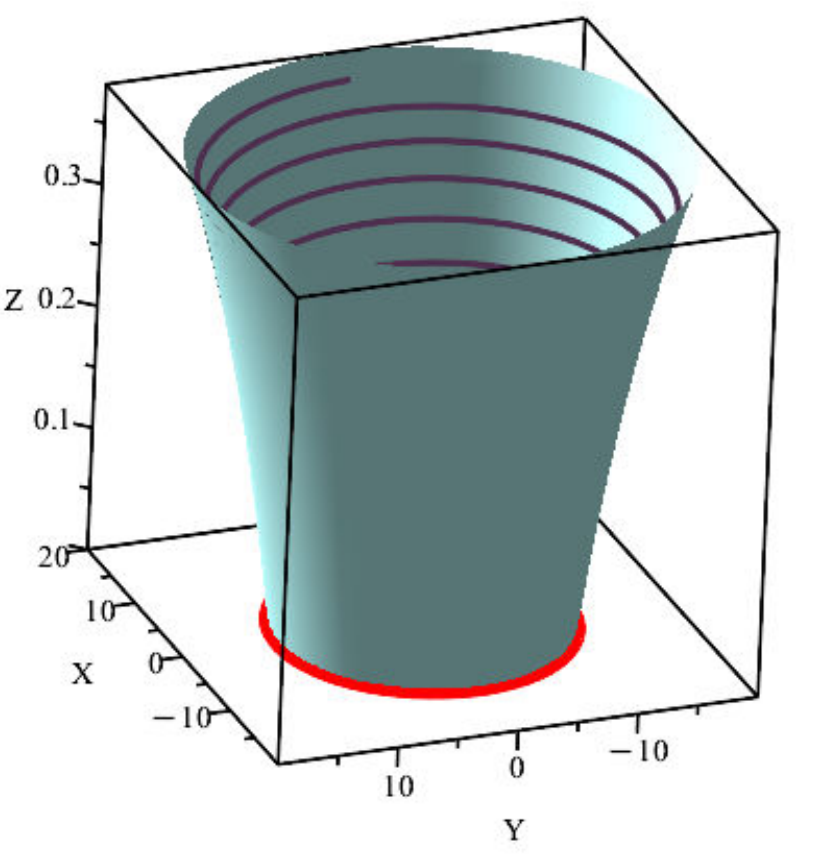}
       \subcaption{[$\rho_s=0.95$, $r_s=1$] \newline SBH with HDMH.}
       \label{fig:is4}
   \end{minipage}

\vspace{0.5em}
   \begin{minipage}{0.2\textwidth}
       \centering
       \includegraphics[width=\textwidth]{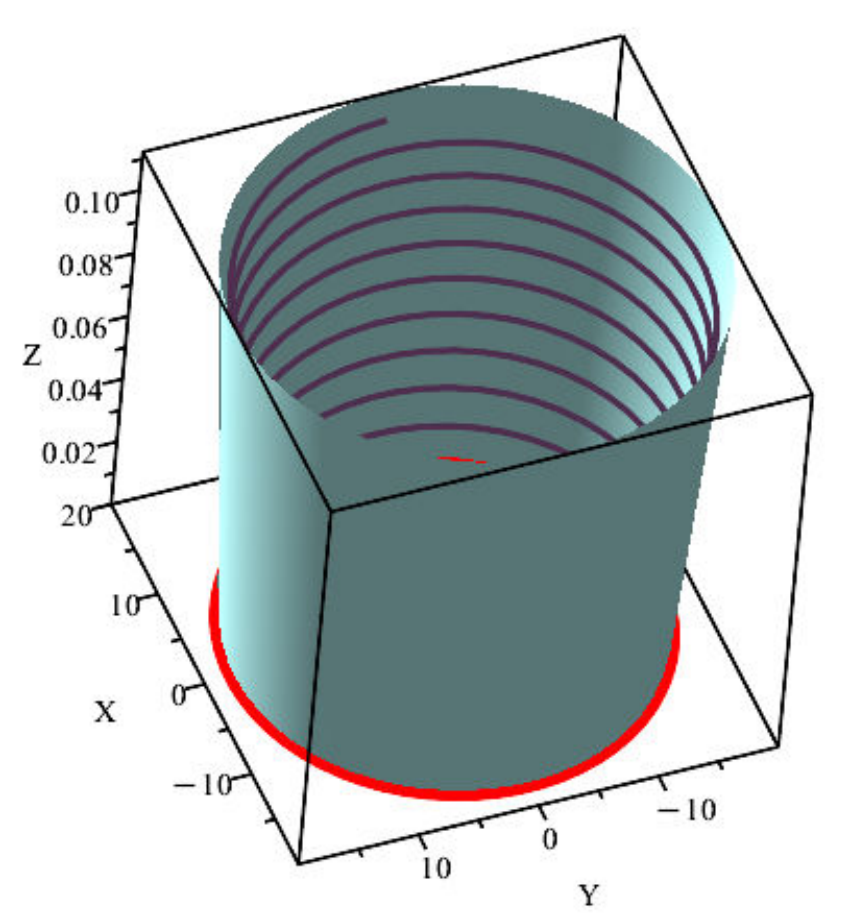}
       \subcaption{[$\rho_s=1.4$, $r_s=1$] \newline SBH with HDMH.}
       \label{fig:is5}
   \end{minipage}
   \begin{minipage}{0.2\textwidth}
       \centering
       \includegraphics[width=\textwidth]{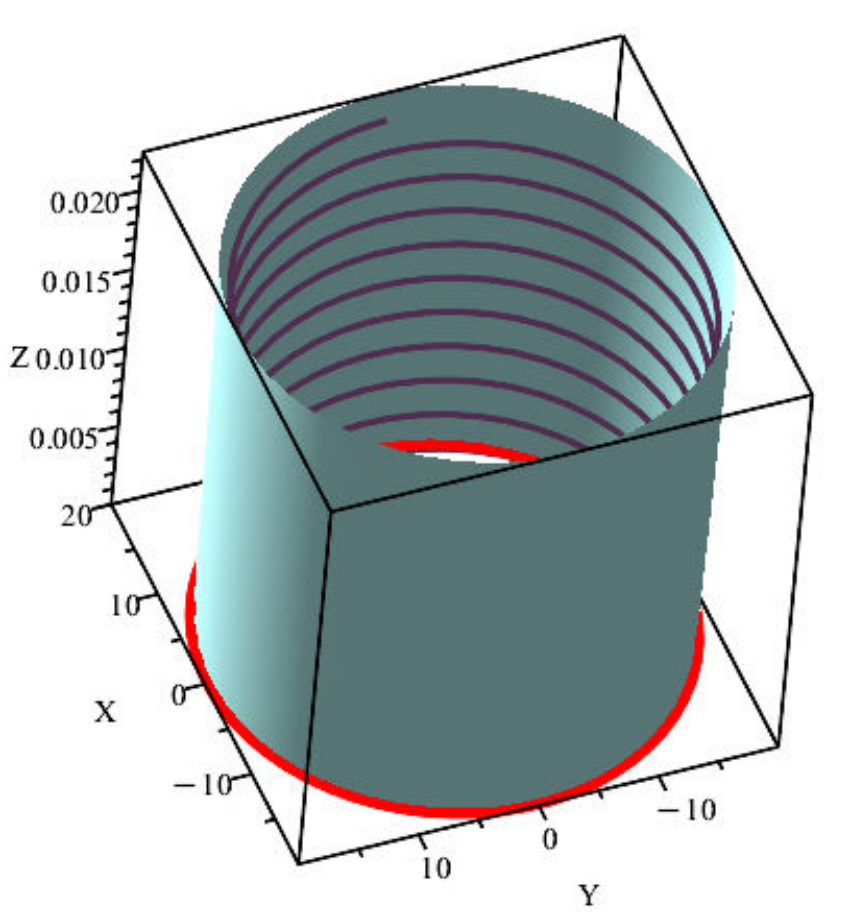}
       \subcaption{[$\rho_s=1.55$, $r_s=1$] \newline SBH with HDMH.}
       \label{fig:is6}
   \end{minipage}
      \begin{minipage}{0.2\textwidth}
      \centering
       \includegraphics[width=\textwidth]{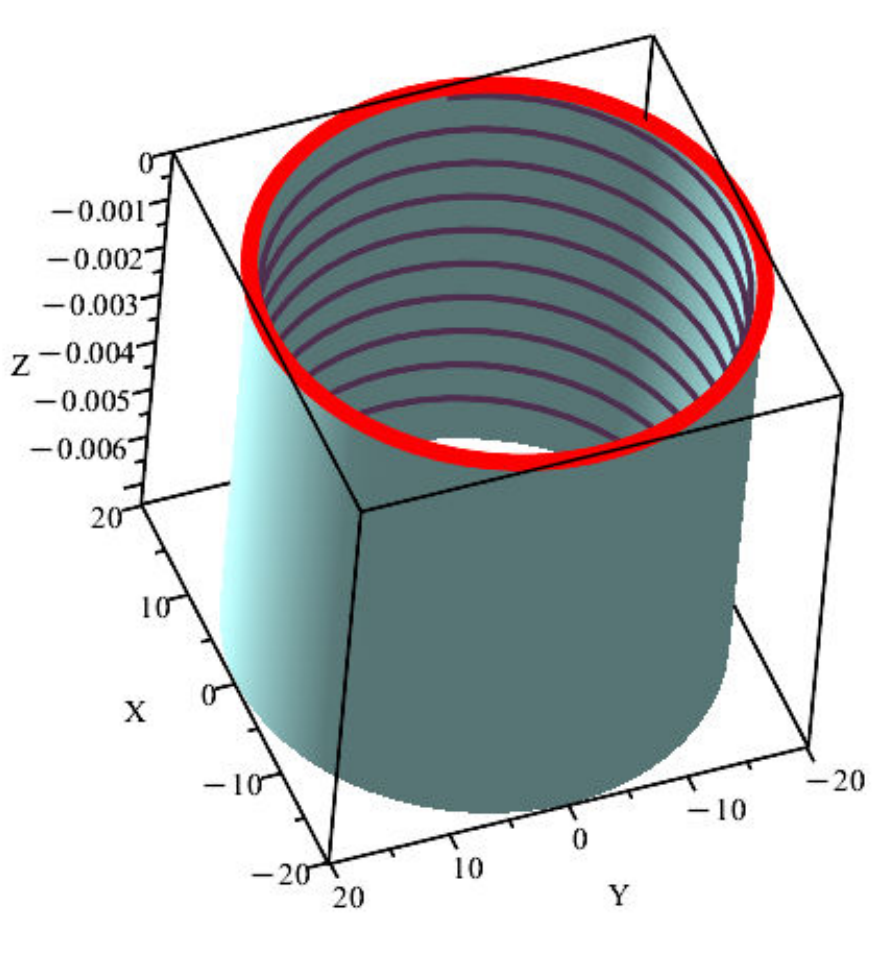}
       \subcaption{[$\rho_s=1.6$, $r_s=1$] \newline Negative metric function case (inner BH geometry).}
       \label{fig:is7}
   \end{minipage}
   \begin{minipage}{0.2\textwidth}
      \centering
       \includegraphics[width=\textwidth]{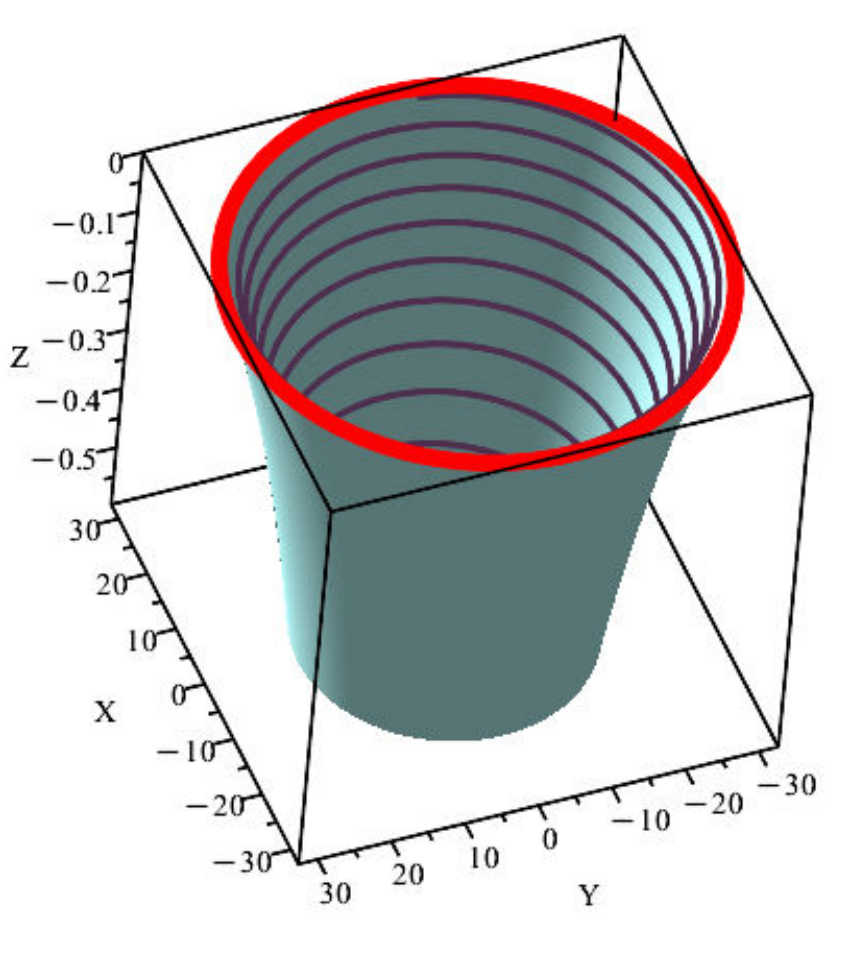}
       \subcaption{[$\rho_s=1.51$, $r_s=1.2$] \newline Negative metric function case (inner BH geometry).}
       \label{fig:is8}
   \end{minipage}
   \caption{\footnotesize Embedding diagrams of the SBH for various HDMH values. The BH mass is set to $M=1$. Plots are governed by the metric function \eqref{aa14}.}
   \label{figizzet}
\end{figure*}

Figure \ref{figizzet} presents a compelling visualization of the embedding diagrams for the SBH in an HDMH, demonstrating the profound influence of DM parameters on spacetime geometry \cite{isz101,isz102}. The sequence of eight panels reveals a systematic evolution of the BH's spatial curvature as the core density $\rho_s$ increases from 0 to 1.6, while maintaining a fixed core radius $r_s = 1$ and BH mass $M = 1$. Panel (a) shows the classical Schwarzschild geometry without HDMH ($\rho_s = 0$), exhibiting the characteristic funnel-like depression with smooth, symmetric curvature extending to spatial infinity. As the DM density progressively increases through panels (b)-(f), the embedding diagrams reveal increasingly dramatic modifications to the spacetime geometry, with the throat region becoming more constricted and the overall curvature becoming more pronounced \cite{isz103}. The transition from panels (f) to (g)-(h) marks a critical threshold where the metric function $f(r)$ becomes negative, indicating the emergence of what we term "inner BH geometry" - a regime where the standard interpretation of the embedding diagram breaks down and the spacetime exhibits exotic properties. This geometric evolution illustrates how the HDMH fundamentally alters the BH's gravitational field structure, with higher DM densities leading to more extreme spacetime curvature and eventually to configurations that challenge conventional BH physics, thereby providing a vivid geometric representation of the interplay between DM and BH spacetime.

The first study of a BH solution with a cloud of CSs as the source, within the framework of general relativity, was presented by Letelier \cite{isz38}. In that work, he derived a generalization of the Schwarzschild BH surrounded by a spherically symmetric cloud of strings, characterized by an energy-momentum tensor given by:
\begin{equation}
   T^{t\, (CS)}_{t}=T^{r\, (CS)}_{r}=\rho_c=\frac{\alpha}{r^2},\quad T^{\theta\, (CS)}_{\theta}=T^{\phi\, (CS)}_{\phi}=0.\label{pp1}
\end{equation}
where $\rho_c$ is the energy density of the cloud and $\alpha$ is an integration constant associated with the presence of the string. The Letelier BH solution is given by \cite{isz38}
\begin{equation}
   ds^2=-\left(1-\alpha-\frac{2\,M}{r}\right)\,dt^2+\left(1-\alpha-\frac{2\,M}{r}\right)^{-1}\,dr^2+r^2\,(d\theta^2+\sin^2 \theta\,d\phi^2).\label{pp2}
\end{equation}

In this study, we combine a system of BH in an anti-de Sitter (AdS) spacetime embodied with an HDMH and a CS, assuming that HDMH and CS do not interact with each other \cite{isz104,isz105}. Therefore, a static spherically symmetric line-element describing a BH with DM halo surrounded by a cloud of strings is given by the following line-element
\begin{equation}
   ds^2 =-f(r)\,dt^2 + f(r)^{-1}\, dr^2 + r^2\, (d\theta^2 + \sin^2 \theta\, d\phi^2),\label{final}
\end{equation}
where the metric function is given by
\begin{equation}
f(r)=1-\alpha-\frac{2\,M}{r}-\frac{b}{r+r_s}+\frac{r^2}{\ell^2_p},\quad\quad b=4\,\pi\,\rho_s\,r_s^3,\label{function}    
\end{equation}
where $\alpha$ is the cosmic string or CS parameter and $\ell_p$ is the radius of curvature related to the cosmological constant as $\frac{1}{\ell^2_p}=-\frac{\Lambda}{3}$ \cite{isz106}.

It is noteworthy that the above spacetime reduces to some well-known solutions in the literature. In the limit where $\rho_s=0$ (implying $b \to 0$) and $\ell_p \to\infty$, the above metric reduces to a well-known BH solution with a cloud of strings, reported in \cite{isz38}. Moreover, in the limit where $\alpha=0$ and $\rho_s=0$, we recover the standard Schwarzschild-AdS BH solution which has been widely studied in the literature \cite{isz107,isz108}.

\begin{figure*}[ht!]
   \centering
   \setlength{\tabcolsep}{0pt}
   \begin{minipage}{0.3\textwidth}
       \centering
       \includegraphics[width=\textwidth]{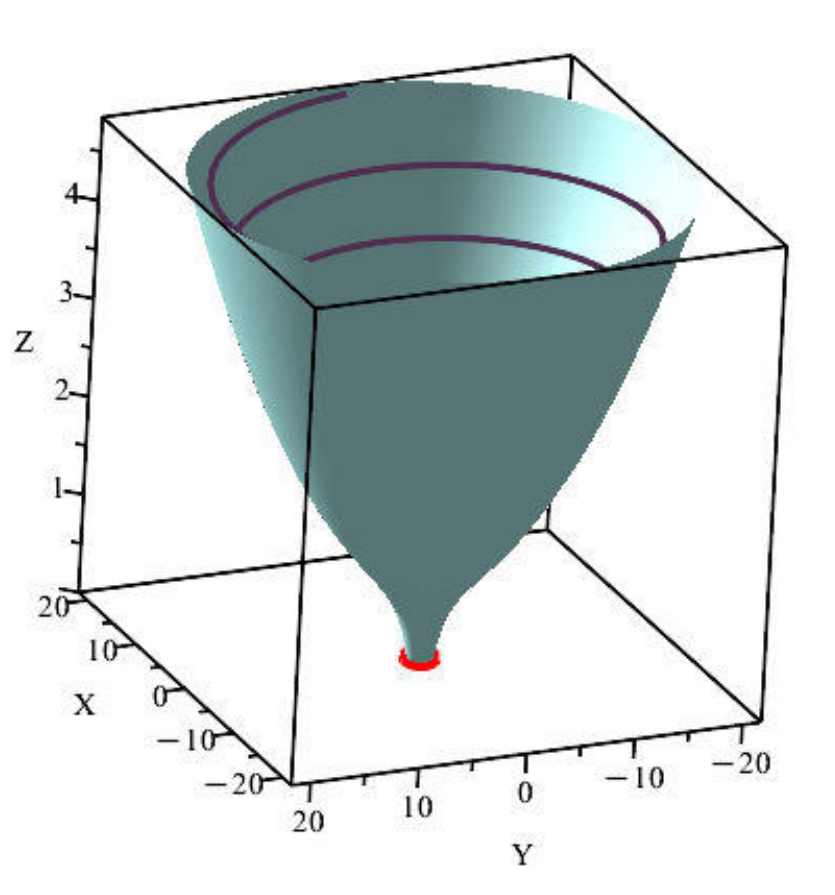}
       \subcaption{[$\rho_s=0$, $r_s=1$, $\alpha=0.1$, $\ell_p=10$] \newline \centering SBH with CS and without HDMH.}
       \label{fig:isn1}
   \end{minipage}
   \begin{minipage}{0.3\textwidth}
       \centering
       \includegraphics[width=\textwidth]{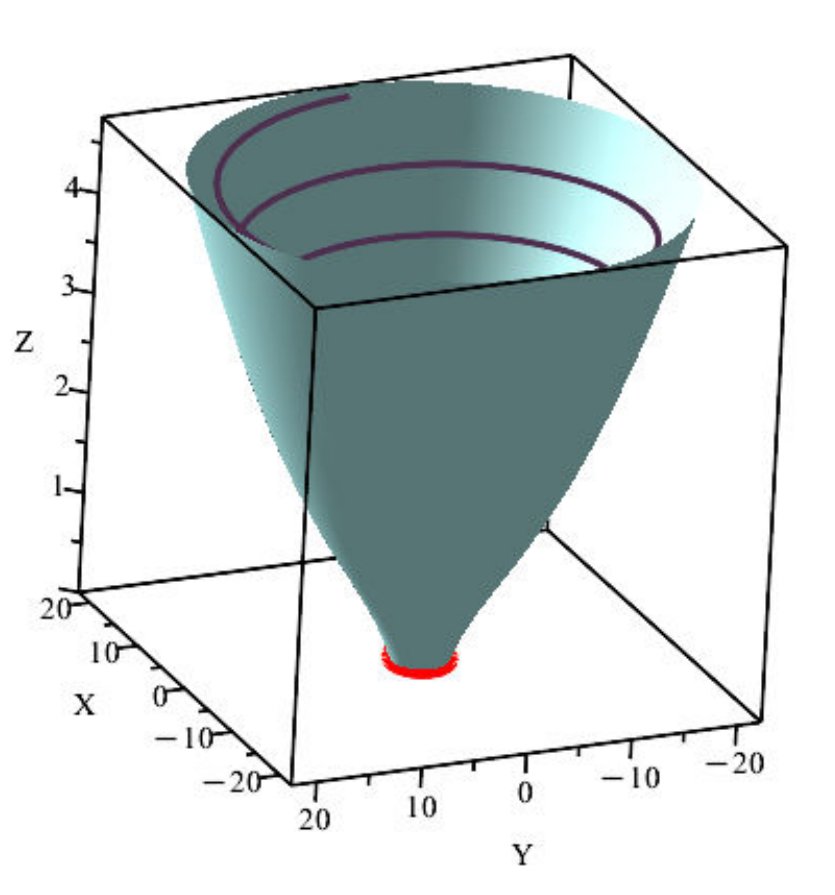}
       \subcaption{[$\rho_s=0.15$, $r_s=1$, $\alpha=0.1$, $\ell_p=10$] \newline \centering SHDMHCS}
       \label{fig:isn2}
   \end{minipage}
   \vspace{0.5em}
   \begin{minipage}{0.3\textwidth}
       \centering
       \includegraphics[width=\textwidth]{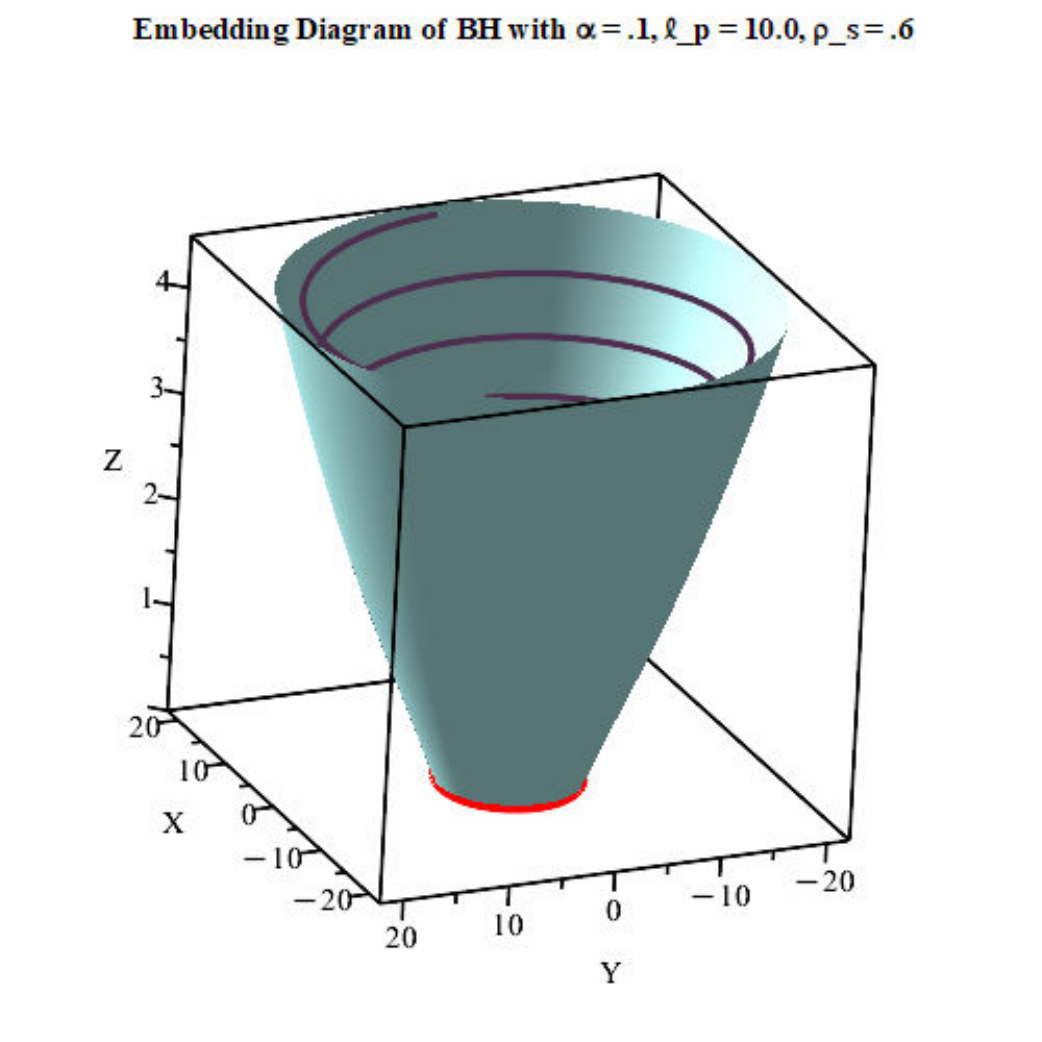}
       \subcaption{[$\rho_s=0.6$, $r_s=1$, $\alpha=0.1$, $\ell_p=10$] \newline \centering SHDMHCS}
       \label{fig:isn3}
   \end{minipage}
       \begin{minipage}{0.3\textwidth}
       \centering
       \includegraphics[width=\textwidth]{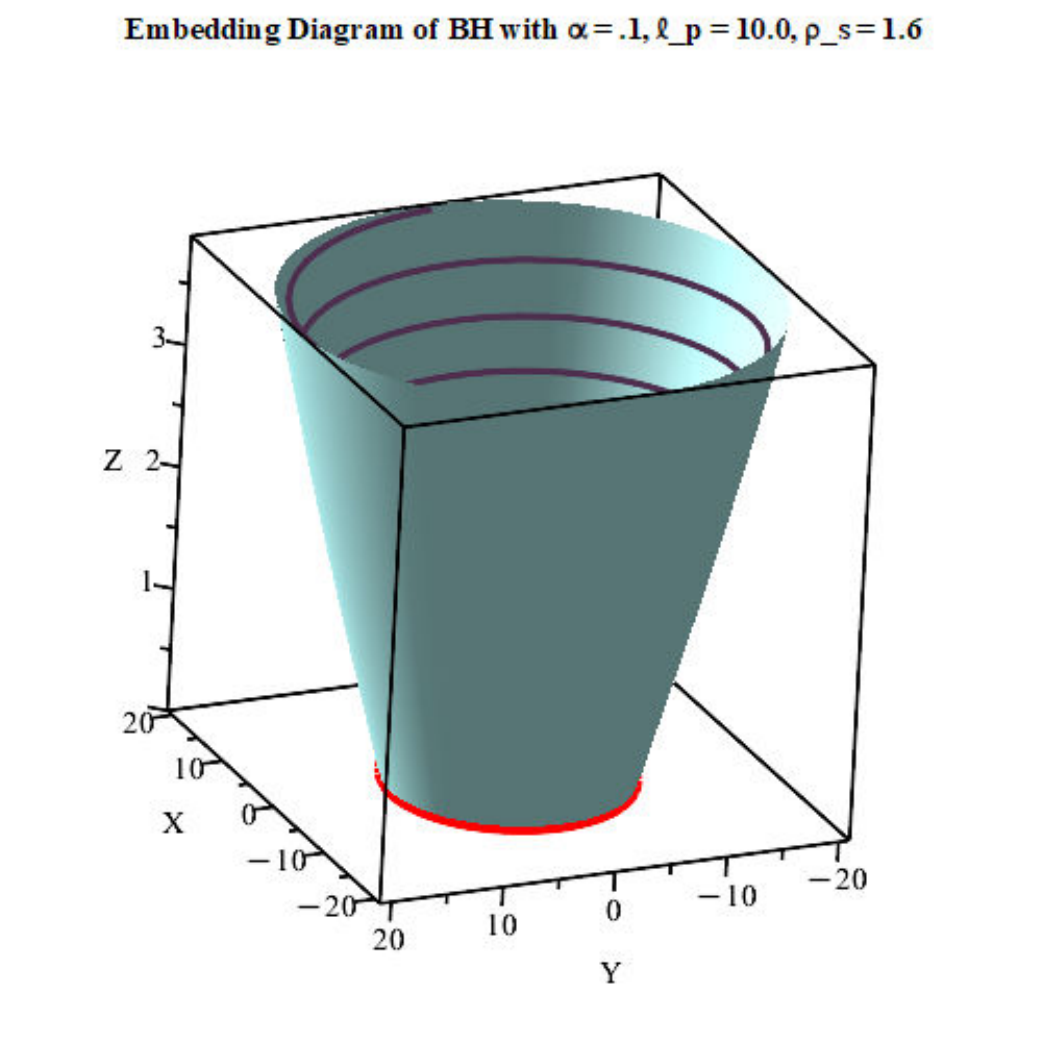}
       \subcaption{[$\rho_s=1.6$, $r_s=1$, $\alpha=0.1$, $\ell_p=10$] \newline \centering SHDMHCS}
       \label{fig:isn4}
   \end{minipage}
   \vspace{0.5em}
   \begin{minipage}{0.3\textwidth}
       \centering
       \includegraphics[width=\textwidth]{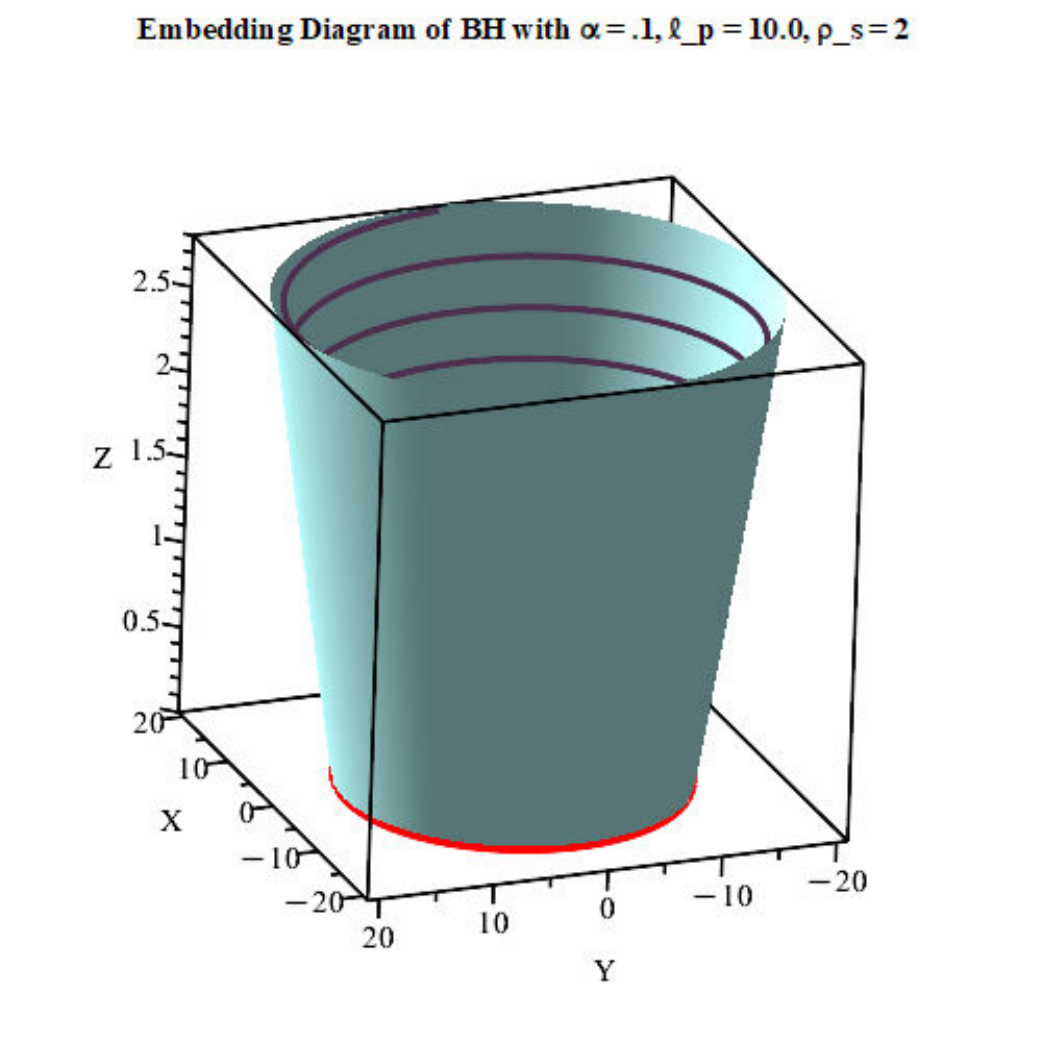}
       \subcaption{[$\rho_s=2$, $r_s=1.2$, $\alpha=0.1$, $\ell_p=10$] \newline \centering SHDMHCS}
       \label{fig:isn5}
   \end{minipage}
   \begin{minipage}{0.3\textwidth}
       \centering
       \includegraphics[width=\textwidth]{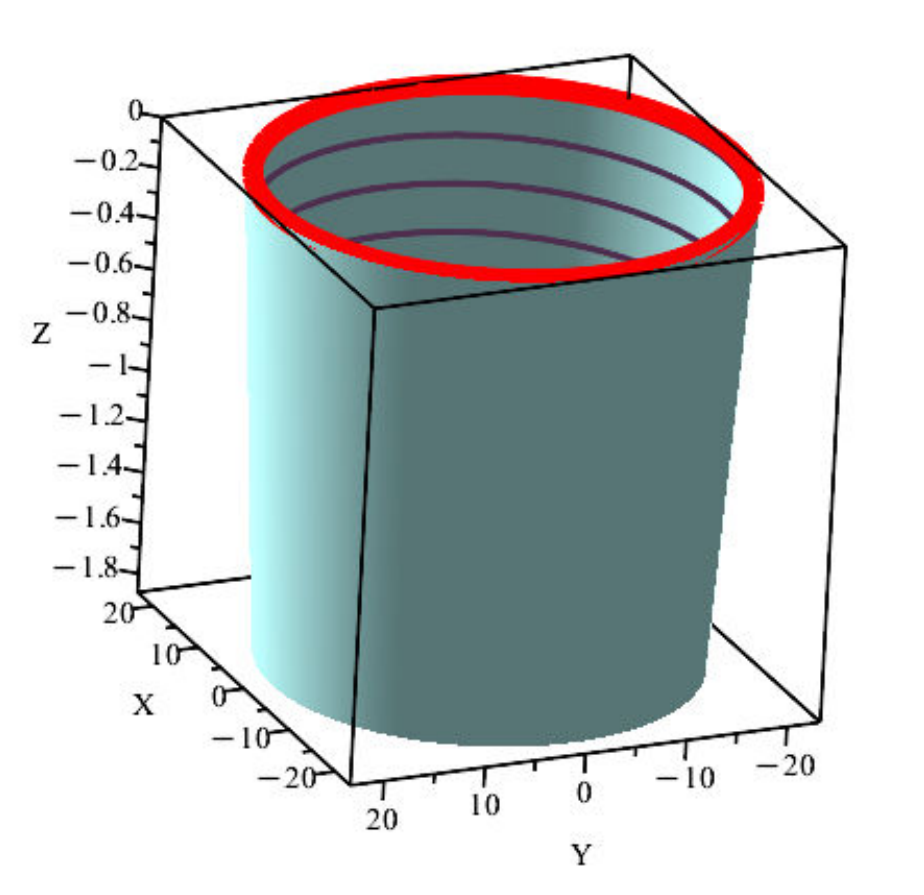}
       \subcaption{[$\rho_s=2$, $r_s=1.8$, $\alpha=0.1$, $\ell_p=10$] \newline \centering SHDMHCS: Negative metric function case}
       \label{fig:isn6}
   \end{minipage}
   \caption{\footnotesize Embedding diagrams of the SBH for various HDMH and CS values. The BH mass is set to $M=1$. Plots are governed by the metric function \eqref{function}.}
   \label{figizzet2}
\end{figure*}

Figure \ref{figizzet2} illustrates the three-dimensional embedding diagrams for the complete SHDMHCS configuration, revealing the combined gravitational effects of both the HDMH and CS on the BH spacetime geometry. The six panels systematically demonstrate how the CS parameter $\alpha = 0.1$ and cosmological parameter $\ell_p = 10$ modify the embedding surfaces as the core density $\rho_s$ increases from 0 to 2 while maintaining $r_s = 1$. Panel (a) shows the baseline configuration with CS but without HDMH ($\rho_s = 0$), displaying a conical spacetime structure that differs markedly from the pure Schwarzschild case due to the topological defect introduced by the CS \cite{isz37}. As the HDMH density progressively increases through panels (b)-(e), the embedding diagrams reveal a systematic deepening and narrowing of the gravitational well, with the CS-induced conical deficit angle becoming more pronounced in the presence of DM. The transition to panel (f) represents a critical regime where $\rho_s = 2$ and $r_s = 1.8$, resulting in the emergence of "inner BH geometry" characterized by the exotic spacetime structure highlighted by the red circular boundary. This evolution demonstrates that the combined presence of CS and HDMH creates a rich geometric landscape where the topological defects of CSs interact synergistically with the matter distribution of the DM halo, leading to increasingly complex spacetime curvature patterns that culminate in novel gravitational configurations not present in pure Schwarzschild or single-component modified BH solutions \cite{isz112}.

For spacetime (\ref{final}), we determine various scalar curvature invariants. These are given by {\small
\begin{eqnarray}
   &&R=g^{\mu\nu}\,R_{\mu\nu}=\frac{2\,(r^2_s\,b+r^3_s\,\alpha + 3\, r^2_s\, r\, \alpha + 3\, r_s\, r^2\, \alpha + 
  r^3\, \alpha + 2\, r^2\, (r_s + r)^3\, \Lambda)}{r^2\,(r_s + r)^3},\label{scalar1}\\
   &&R^{\mu\nu}\,R_{\mu\nu}=\frac{2\,[(r_s + r)^2\,(r_s\, b + r^2_s\, \alpha + 2\, r_s r\, \alpha + r^2\, \alpha + 
     r^2\, (r_s + r)^2\, \Lambda)^2 + 
  r^2\, (r^2_s b^2 - 2\, r_s\, b\, r\, (r_s + r)^3\, \Lambda + 
     r^2\, (r_s + r)^6\,\Lambda^2)]}{r^4\, (r_s + r)^6},\label{scalar2}\\
   &&R^{\mu\nu\rho\sigma}\,R_{\mu\nu\rho\sigma}=\frac{4}{3\, r^6_s\, (r_s + r)^6}\,\Big[r^6_s (36\, M^2\, + 12\, M\, r\,\alpha + 3 \,r^2\, \alpha^2 + 
  2\, r^4\, \alpha\, \Lambda+2\,r^6\,\Lambda^2)\nonumber\\
  &&+6\, r_s\, r^5\, (3\,b^2 + 24\, b\, M + 36\, M^2 + 5\, b\, r\,\alpha + 12\, M\, r\, \alpha + 
  3\, r^2\,\alpha^2 + 2\, r^4\, \alpha\, \Lambda+ 
  2\, r^6\, \Lambda^2)\nonumber\\
  &&+r^6\, \left(9\, b^2 + 36\, M^2 + 12\, M\, r\, \alpha + 3\, r^2\, \alpha^2 + 
  6\, b\, (6\, M + r\, \alpha) + 2\, r^4\, \alpha\, \Lambda + 
  2\, r^6\, \Lambda^2\right)\nonumber\\
  &&+2\,r^5_s\, r\, (6\, M\, (b + 18\, M) + 3\, (b + 12\, M)\, r\, \alpha + 9\, r^2\, \alpha^2 + 
  r^3\, (b + 6\, r\, \alpha)\,\Lambda+ 6\, r^6\,\Lambda^2)\nonumber\\
  &&+3\,r^4_s\, r^2\, (b^2 + 24\, b\, M + 180\, M^2 + 10\, b\, r\,\alpha + 60\, M\, r\,\alpha + 
  15\, r^2\, \alpha^2 + 2\, r^3\, (b + 5\, r\,\alpha)\,\Lambda+ 
  10\, r^6\,\Lambda^2)\nonumber\\
  &&+2\,r^3_s\, r^3\, (6\, (b^2 + 15\, b\, M + 60\, M^2 + 5\, (b + 4\, M)\, r\,\alpha + 
     5\, r^2\, \alpha^2) + r^3\, (3\, b + 20\, r\,\alpha)\,\Lambda + 
  20\, r^6\,\Lambda^2)\nonumber\\
  &&+r^2_s\, r^4\, (21\, b^2 + 228\, b\, M + 540\, M^2 + 60\, (b + 3\, M)\, r\,\alpha + 
  45\, r^2\, \alpha^2 + 2\, r^3\, (b + 15\, r\,\alpha)\, \Lambda + 
  30\, r^6\, \Lambda^2)\Big].\label{scalar3}
\end{eqnarray}
}
From the above expressions (\ref{scalar1})--(\ref{scalar3}), one can easily show that at $r=0$ and large distances, $r \to \infty$, 
\begin{eqnarray}
   &&\lim_{r \to 0} R=\infty,\quad\quad \lim_{r \to 0} R^{\mu\nu}\,R_{\mu\nu}=\infty,\quad\quad \lim_{r \to 0} R^{\mu\nu\rho\sigma}\,R_{\mu\nu\rho\sigma}=\infty,\nonumber\\
   &&\lim_{r \to \infty} R = 4\,\Lambda,\quad\quad \lim_{r \to \infty} R^{\mu\nu}\,R_{\mu\nu} = 4\,\Lambda^2,\quad\quad \lim_{r \to \infty} R^{\mu\nu\rho\sigma}\,R_{\mu\nu\rho\sigma} = \frac{8\,\Lambda^2}{3}.\label{scalar4}
\end{eqnarray}

From the above scalar curvature analysis, we observe that the BH solution surrounded with an HDMH and a cloud of CSs is a singular solution in general relativity and is asymptotically AdS spacetime \cite{isz113,isz114}. 

To visualize the metric function (see Eq. \ref{function}), we plot its behavior as a function of the radial coordinate $r$, as shown in Fig. \ref{fig:wave-function}. The figure reveals that the BH possesses a single horizon. The horizon radius decreases as the core density $\rho_s$ and core radius $r_s$ parameters increase (left and middle panels). In contrast, the horizon expands with an increase in the cloud string parameter $\alpha$ (right panel) \cite{isz115}.

\begin{figure}[ht!]
    \centering
    \subfloat[$r_s=0.5,\,\alpha=0.1$]{\centering{}\includegraphics[scale=0.36]{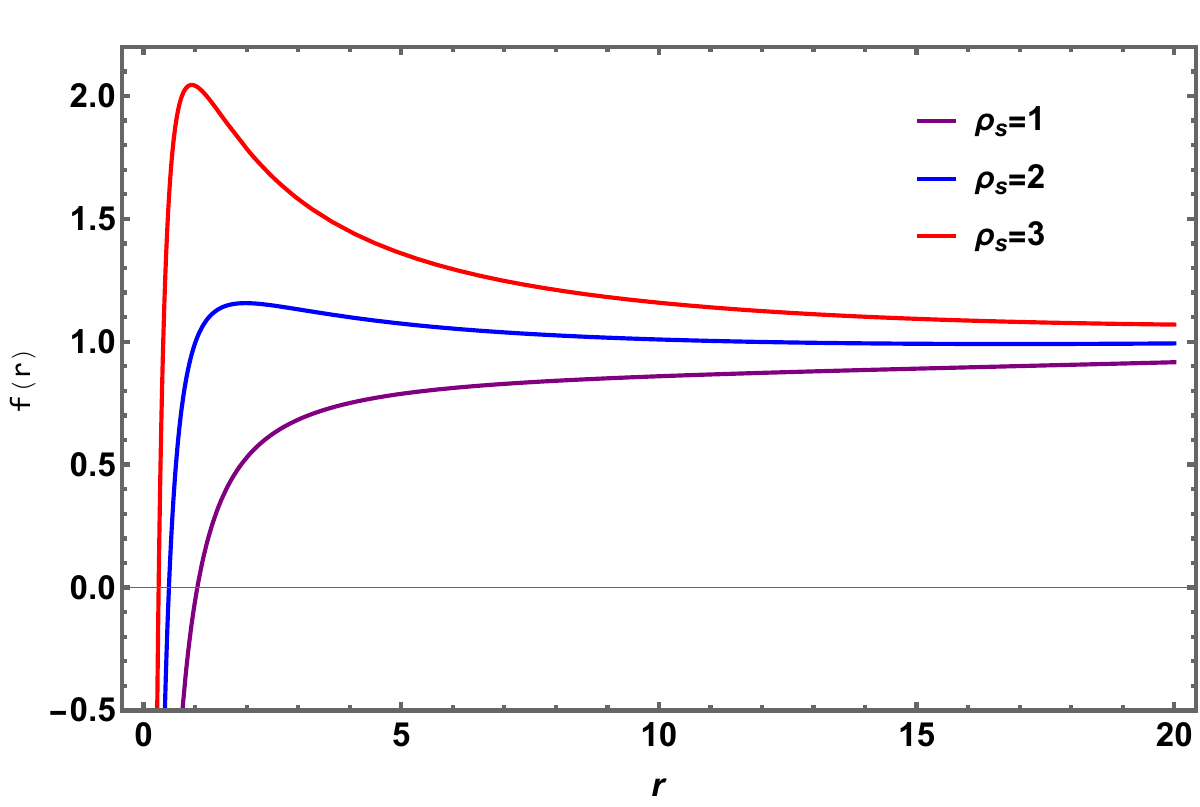}}\quad\quad
    \subfloat[$\rho_s=1,\,\alpha=0.1$]{\centering{}\includegraphics[scale=0.35]{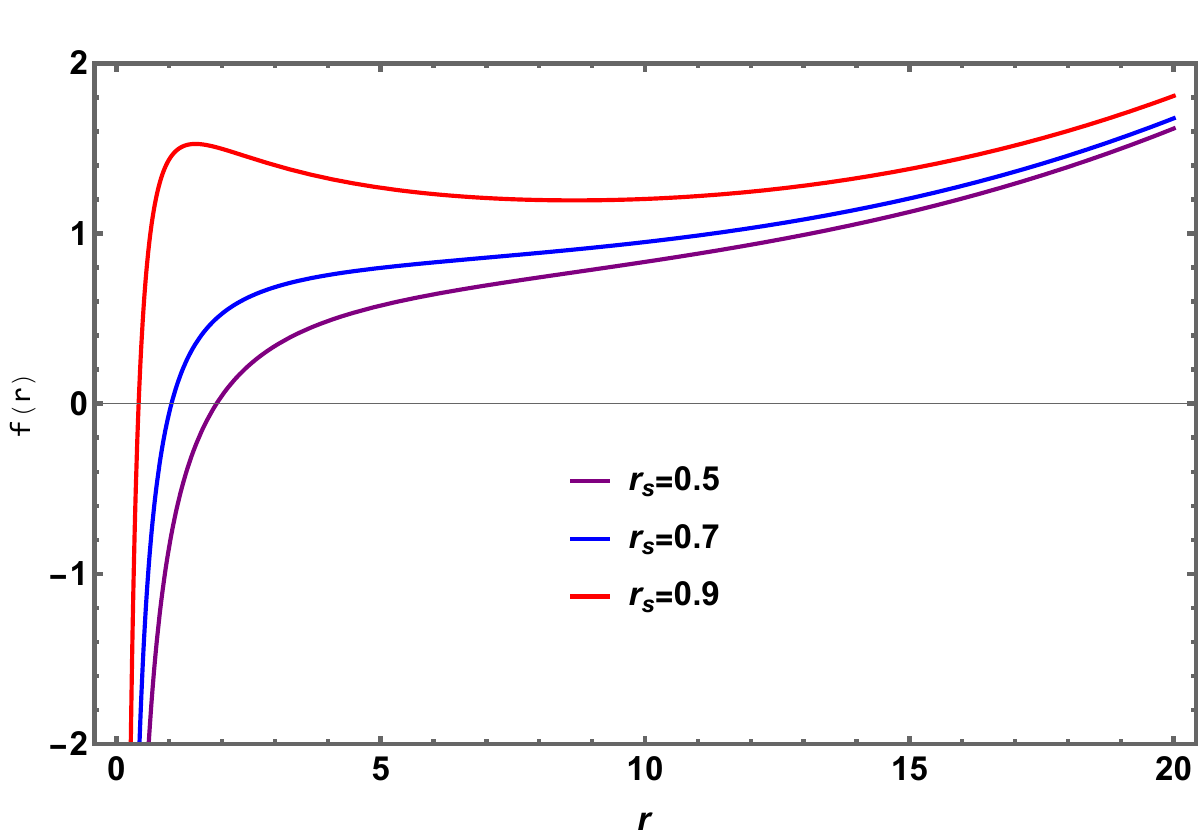}}\\
    \subfloat[$r_s=0.5=\rho_s$]{\centering{}\includegraphics[scale=0.35]{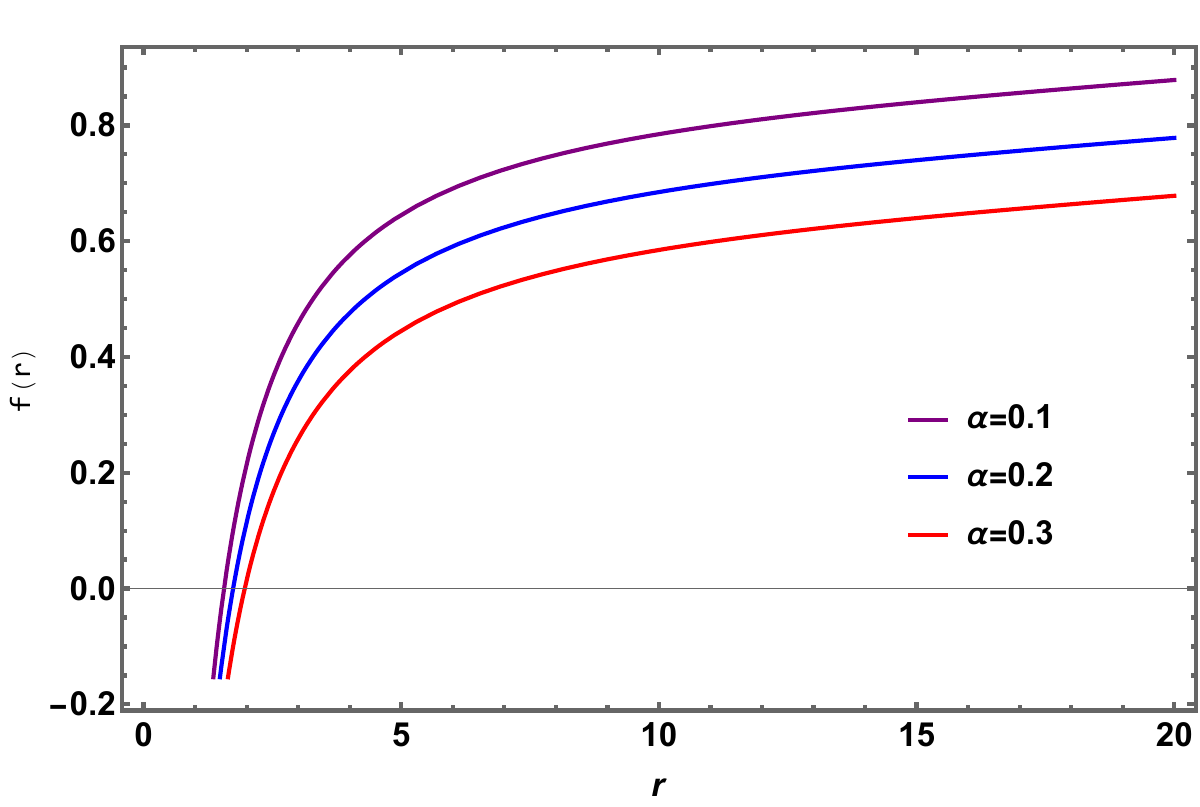}}
    \caption{\footnotesize The behavior of the metric function \( f(r) \) is illustrated as a function of the radial coordinate \( r \), considering the individual effects of varying (a) the halo density \( \rho_s \), (b) the core radius \( r_s \), and (c) the string parameter \( \alpha \). In all cases, the BH mass is fixed at \( M = 1 \), and the curvature radius is set to \( \ell_p = 100 \).}
    \label{fig:wave-function}
\end{figure}

\section{Geodesics Analysis} \label{sec3}

In spherically symmetric spacetimes, the metric functions $g_{tt} \neq 1$ and $g_{rr}$ are the redshift factor and spatial curvature that basically determines the motion of a particle (massless or massive) within the gravitational field of a BH. In such a spacetime, two Killing vectors intrinsic to the metric could be found those are associated closely with time translation and rotational symmetry-yielding two conserved quantities respectively along the geodesics. Specifically, the time-translational Killing vector $\xi_{(t)} \equiv \partial_{t}$ produces a conserved energy, $\mathrm{E}=-g_{tt}\,\dot{t}$, that remains constant along the geodesic of the particle. Similarly, the Killing vector $\xi_{(\phi)} \equiv \partial_{\phi}$ associated with rotational symmetry represents a conserved angular momentum, $\mathrm{L}=g_{\phi\phi}\,\dot{\phi}$ that makes the motion constrained to orbits (parabolic, hyperbolic or elliptical depending on their eccentricities). These intrinsic symmetries lead to conserved quantities that play a crucial role in determining the equations of motion. Further, one can derive the effective potential controlling both null and timelike particle trajectories and see whether the orbits are stable or unstable or whether precession of timelike orbits occur.

Now, in order to derive the geodesic equations we may utilize the Lagrangian density function given by \cite{FA1,FA2,FA3,NPB,AHEP1,AHEP2,CJPHY,EPJC,AHEP4,AHEP5,AHEP6,AHEP7,AHEP8}
\begin{equation}
    \mathcal{L}=\frac{1}{2}\,g_{\mu\nu}\,\left(\frac{dx^{\mu}}{d\lambda}\right)\,\left(\frac{dx^{\nu}}{d\lambda}\right),\label{bb1}
\end{equation}
where $\lambda$ is an affine parameter.

Considering the geodesics motions in the equatorial plane define by $\theta=\pi/2$ since the spacetime is spherically symmetric. Using (\ref{final}), the Lagrangian density function  becomes
\begin{equation}
    \mathcal{L}=\frac{1}{2}\,\left[-f(r)\,\left(\frac{dt}{d\lambda}\right)^2+\frac{1}{f(r)}\,\left(\frac{dr}{d\lambda}\right)^2+r^2\,\left(\frac{d\phi}{d\lambda}\right)^2\right],\label{bb2}
\end{equation}

There are two conserved quantities corresponding to cyclic coordinates ($t, \phi$) and these are given by
\begin{equation}
    \mathrm{E}=-\frac{\partial \mathcal{L}}{\partial \dot{t}}=f(r)\,\frac{dt}{d\lambda}.\label{bb3}
\end{equation}
And
\begin{equation}
    \mathrm{L}=\frac{\partial \mathcal{L}}{\partial \dot{\phi}}=r^2\,\frac{d\phi}{d\lambda},\label{bb4}
\end{equation}
where $\mathrm{E}$ is the conserved energy and $\mathrm{L}$ is the conserved angular momentum.

With these, the first order geodesics path for $r$ coordinate from Eq. (\ref{bb2}) becomes
\begin{equation}
    \left(\frac{dr}{d\lambda}\right)^2+V_\text{eff}(r)=\mathrm{E}^2\label{bb5}
\end{equation}
which is one-dimensional equation of motion of a particle of unit mass having energy $\mathrm{E}^2$ and the effective potential is given by
\begin{equation}
    V_\text{eff}(r)=\left(-\varepsilon+\frac{\mathrm{L}}{r^2}\right)\,f(r)=\left(-\varepsilon+\frac{\mathrm{L}}{r^2}\right)\,\left(1-\alpha-\frac{2\,M}{r}-\frac{b}{r+r_s}-\frac{\Lambda}{3}\,r^2\right).\label{bb6}
\end{equation}

From the expression in Eq.~(\ref{bb6}), it is evident that the effective potential for both null and timelike particles is influenced by several key parameters. These include the cosmic string parameter $\alpha$, the core radius $r_s$ and density $\rho_s$ of DM halo, the angular momentum $\mathrm{L}$, the BH mass $M$, and the cosmological constant $\Lambda$. 

\begin{figure}[ht!]
    \centering
    \subfloat[$r_s=0.5\,M,M^2\,\rho_s=1$]{\centering{}\includegraphics[width=0.4\linewidth]{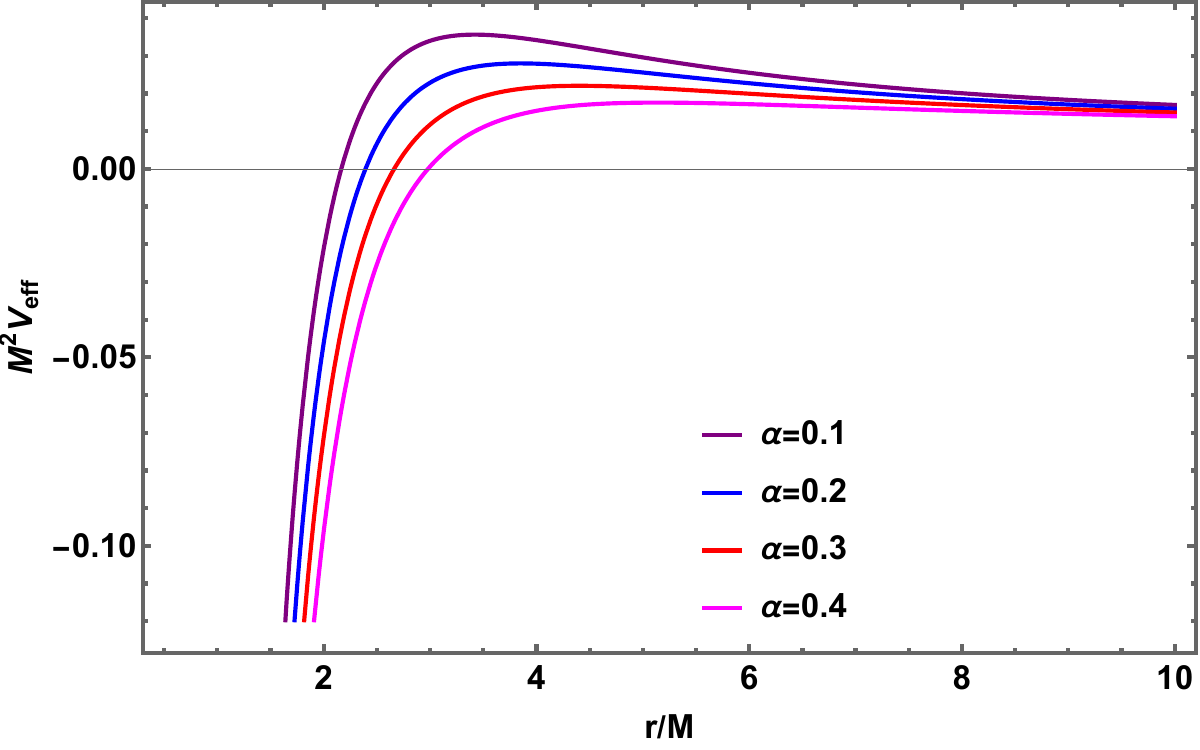}}\quad\quad
    \subfloat[$\alpha=0.1,M^2\,\rho_s=1$]{\centering{}\includegraphics[width=0.4\linewidth]{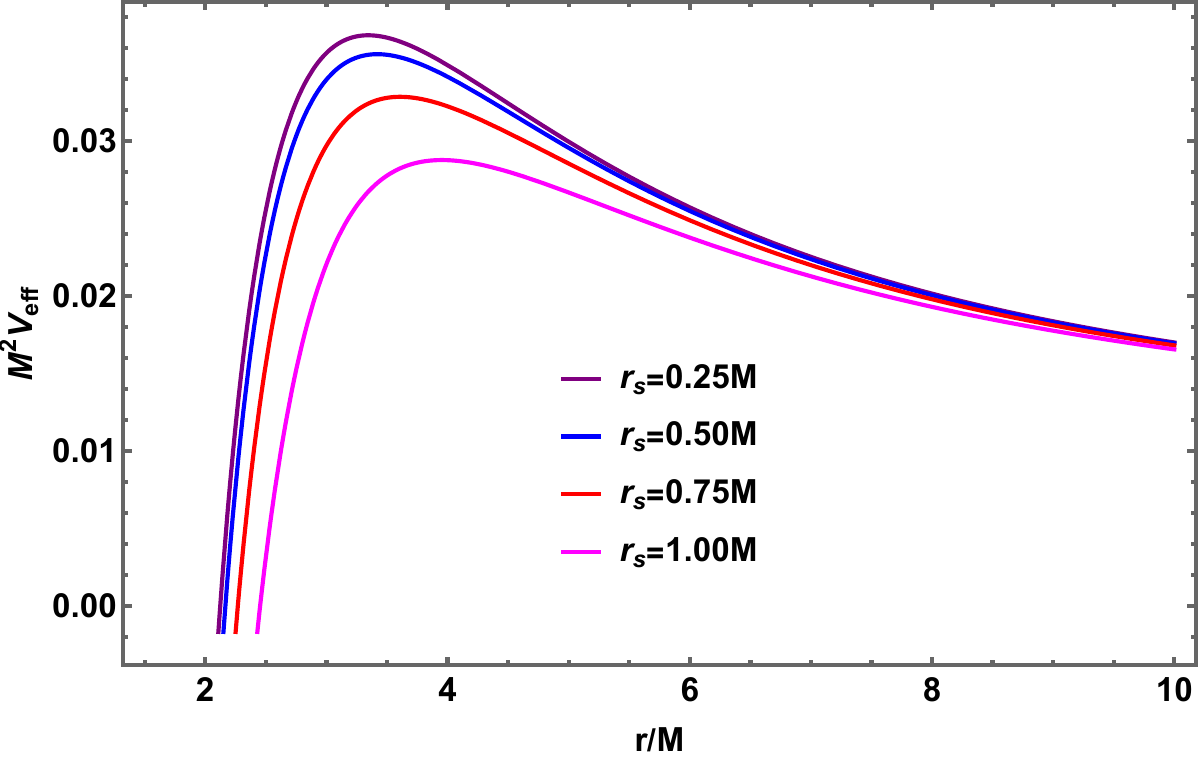}}\\
    \subfloat[$\alpha=0.1,r_s=1\,M$]{\centering{}\includegraphics[width=0.4\linewidth]{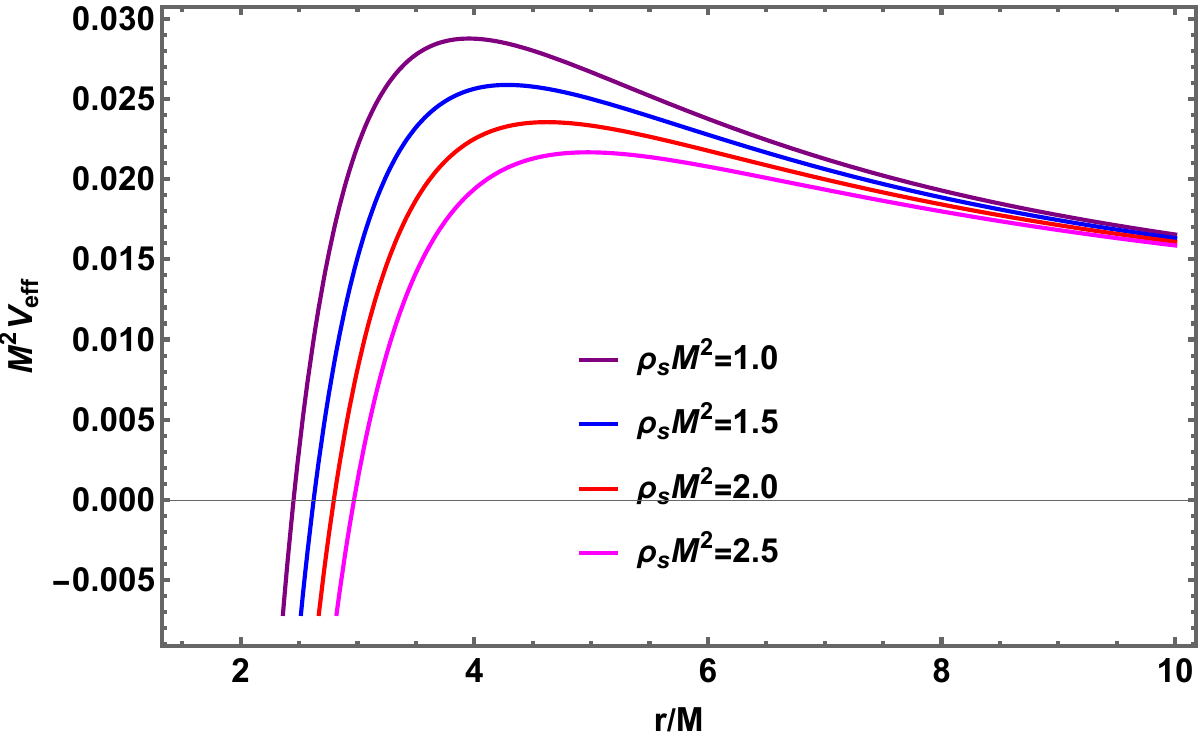}}\quad\quad
    \subfloat[$M^2\,\rho_s=1$]{\centering{}\includegraphics[width=0.4\linewidth]{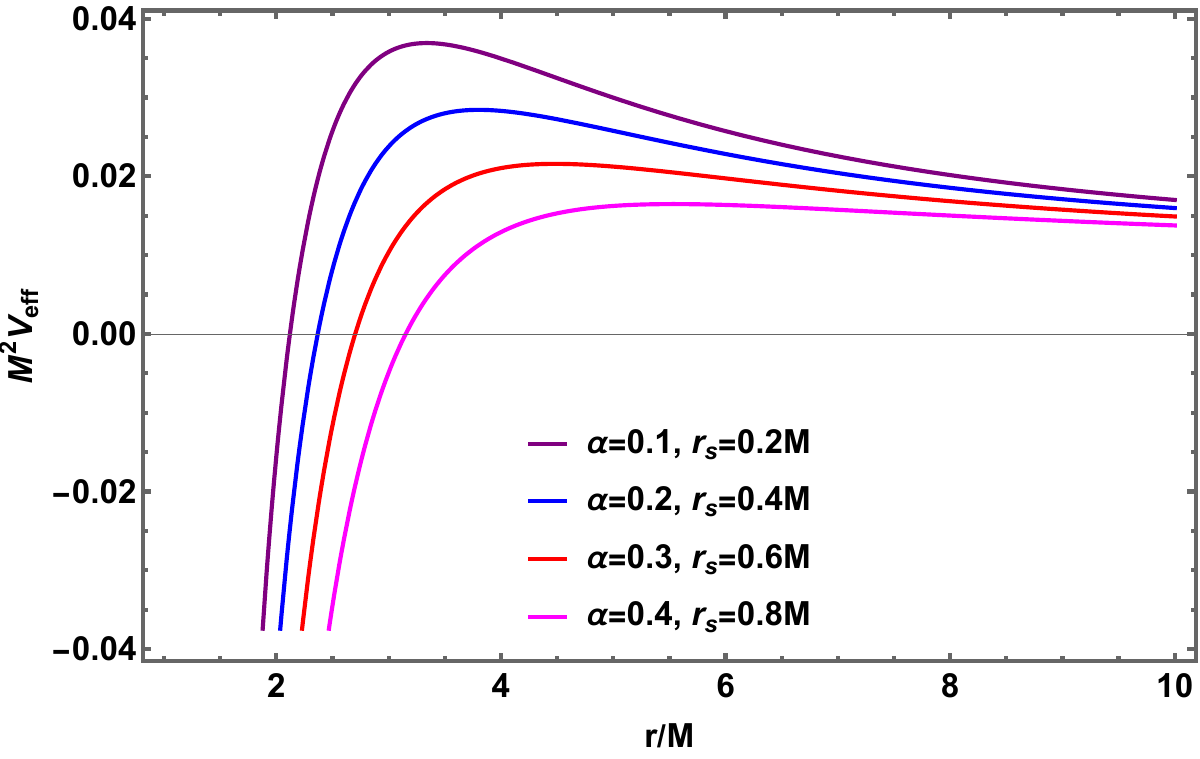}}\\
    \subfloat[$r_s=0.5\,M$]{\centering{}\includegraphics[width=0.4\linewidth]{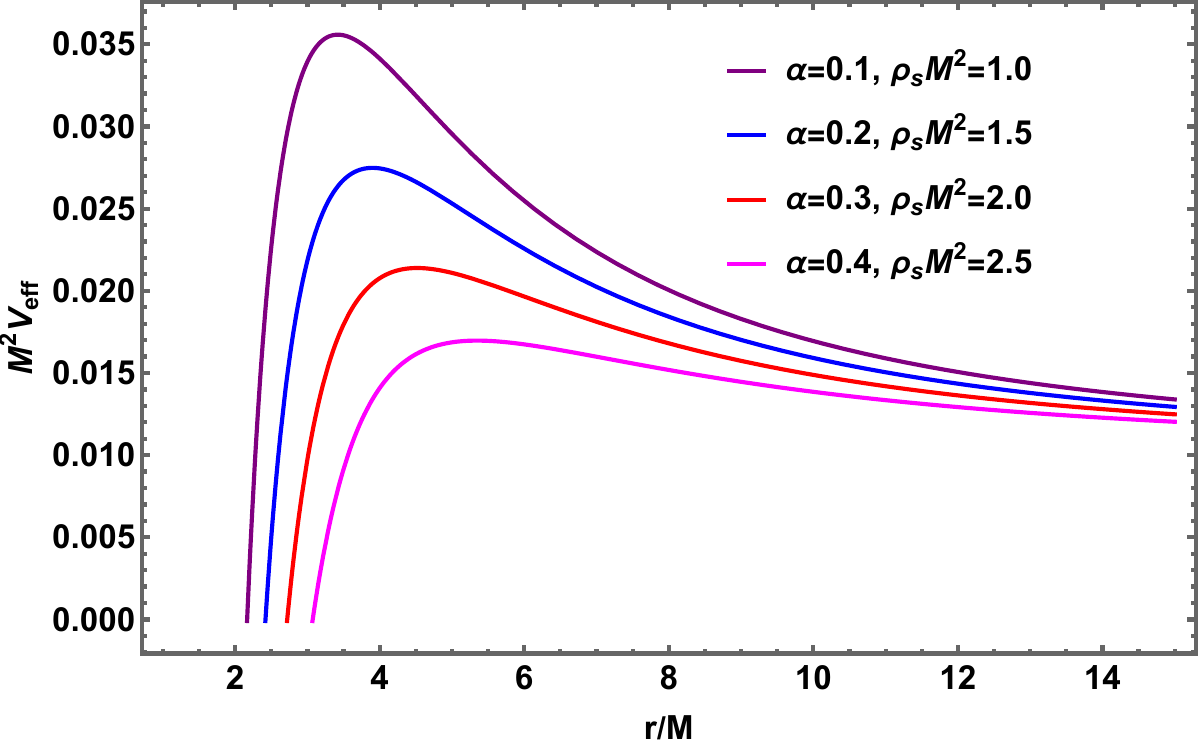}}\quad\quad
    \subfloat[]{\centering{}\includegraphics[width=0.4\linewidth]{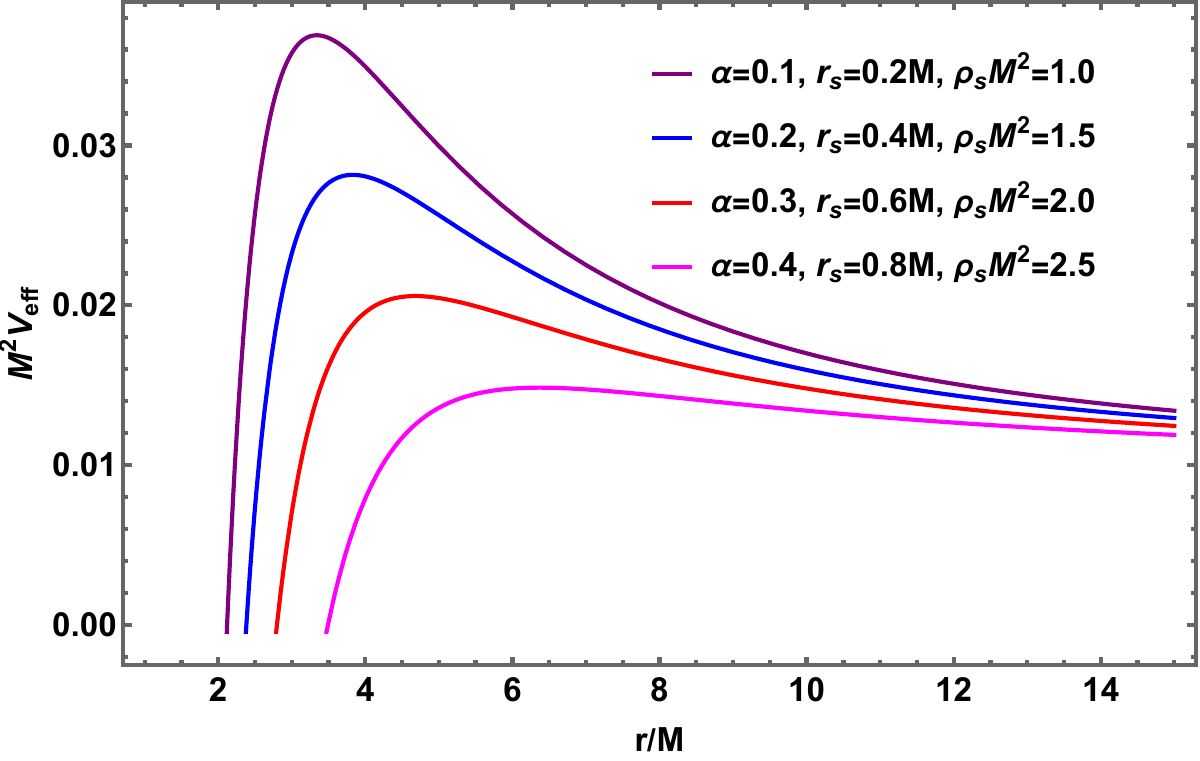}}
    \caption{\footnotesize Behavior of the effective potential for null geodesics is shown for varying values of the string parameter \(\alpha\), the core radius \(r_s\), and the DM halo density \(\rho_s\). Subfigures illustrate the individual effects of (a) \(\alpha\), (b) \(r_s\), and (c) \(\rho_s\,M^2\), as well as the combined effects of (d) \(\alpha\) and \(r_s\), (e) \(\alpha\) and \(\rho_s M^2\), and (f) \(\alpha\), \(r_s\), and \(\rho_s M^2\) together. Here, we set the angular momentum $\mathrm{L}=1$ and the dimensionless parameter $k=M\,\sqrt{-\frac{\Lambda}{3}}=0.1$, all are in natural units.}
    \label{fig:null-potential}
\end{figure}

\subsection{Null Geodesics}

Null geodesics play a crucial role in understanding the geometric properties of BHs and offer deep insights into their behavior, especially in relation to light and other massless particles. Analyzing null geodesics through the effective potential is a powerful and essential method for investigating both the geometric and physical characteristics of BHs. This approach aids in comprehending the deflection of light, the formation and size of BH shadows, the conditions for the existence of photon orbits, and the dynamics of circular null orbits around BHs. In the current study, we investigate these characteristics for the selected BH spacetime given in Eq. (\ref{final}) and discuss the outcomes.

For null geodesics, $\varepsilon=0$, and hence, the effective potential from Eq. (\ref{bb6}) becomes
\begin{equation}
    V_\text{eff}(r)=\frac{\mathrm{L}^2}{r^2}\,f(r)=\frac{\mathrm{L}^2}{r^2}\,\left(1-\alpha-\frac{2\,M}{r}-\frac{b}{r+r_s}-\frac{\Lambda}{3}\,r^2\right).\label{cc1}
\end{equation}

In Figure~\ref{fig:null-potential}, we present a series of plots illustrating the behavior of the effective potential for null geodesics as a function of the radial coordinate \( r \), under variations of the cosmic string parameter \( \alpha \), the core radius \( r_s \), and the core density \( \rho_s \) of the HDM halo. Across all panels, we observe that increasing the values of \( \alpha \), \( r_s \), \( \rho_s \), or their combinations leads to a decrease in the effective potential of the system. This suggests that all the parameters collectively shape the behavior of the potential in the system when photon particles traverse the gravitational field.

\begin{figure}[ht!]
    \centering
    \subfloat[$r_s=0.5\,M,\,M^2\,\rho_s=1$]{\centering{}\includegraphics[width=0.4\linewidth]{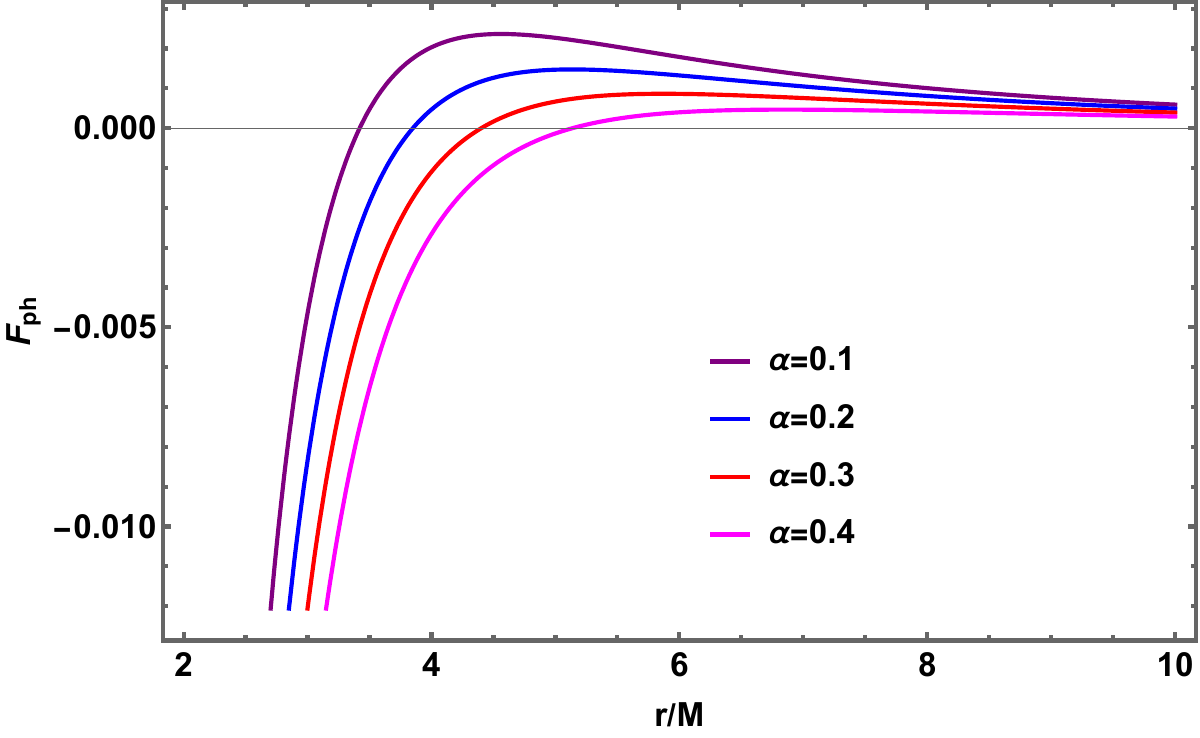}}\quad\quad
    \subfloat[$\alpha=0.1,M^2\,\rho_s=1$]{\centering{}\includegraphics[width=0.4\linewidth]{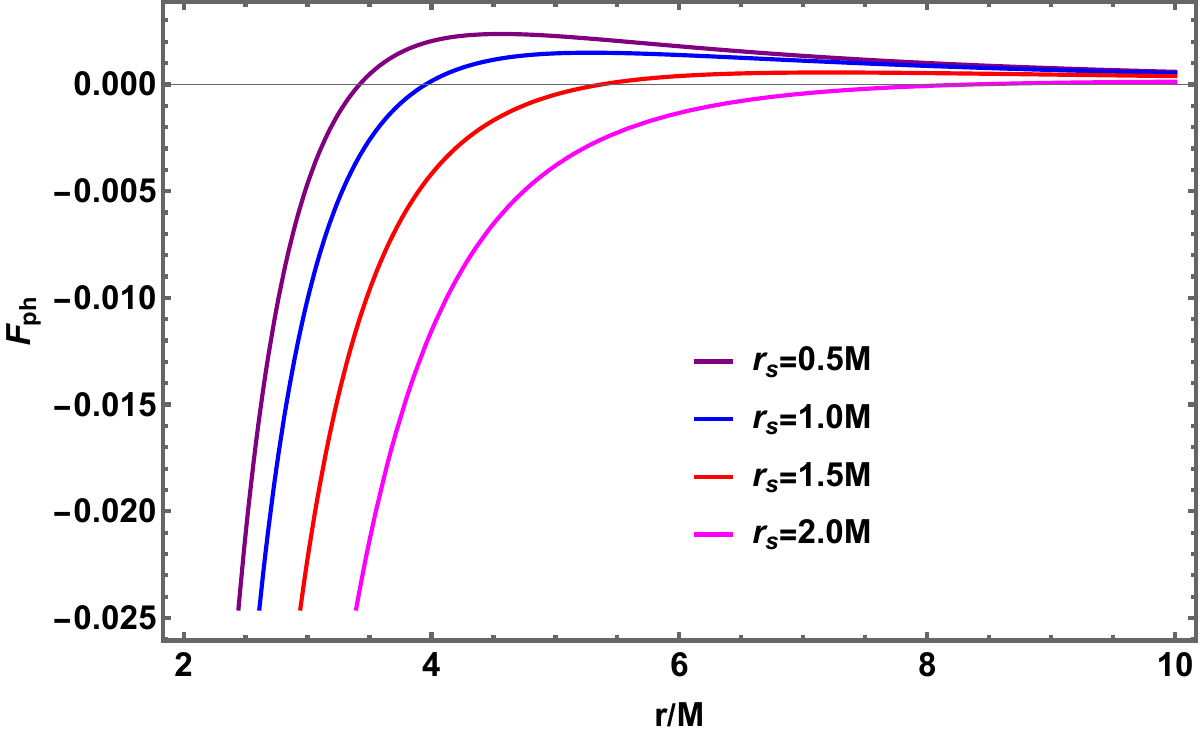}}\\
    \subfloat[$\alpha=0.1,r_s=1.5\,M$]{\centering{}\includegraphics[width=0.4\linewidth]{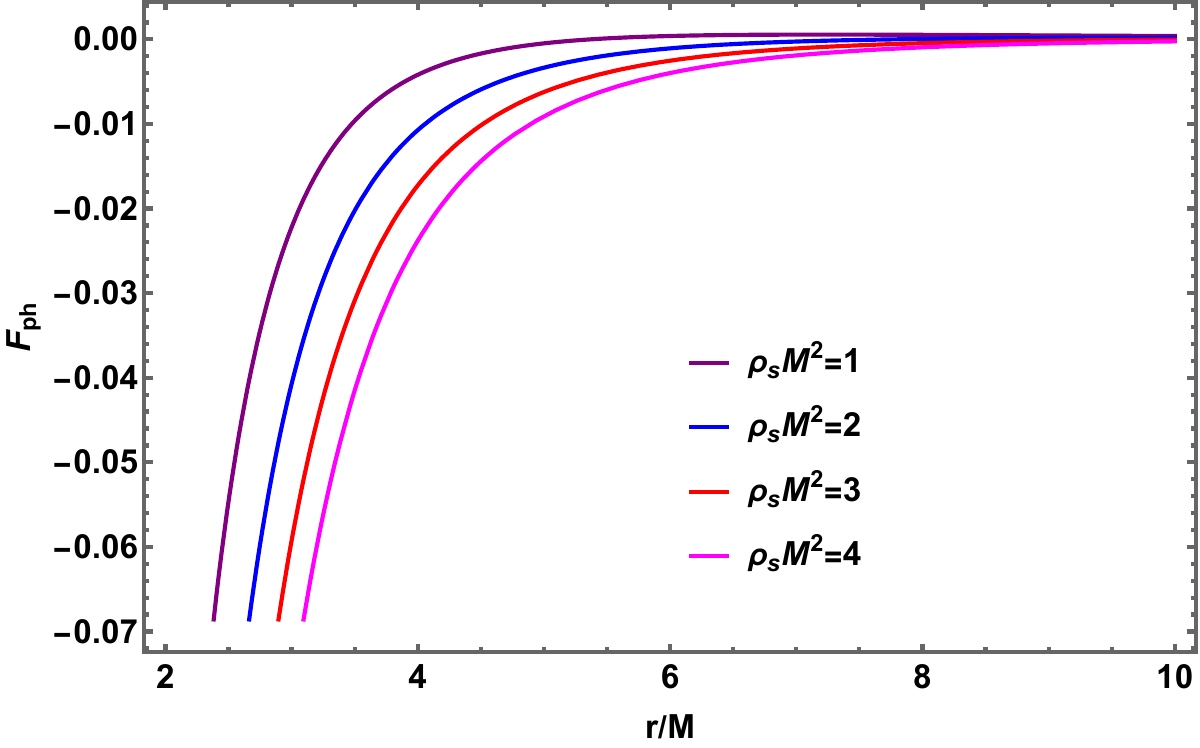}}\quad\quad
    \subfloat[$M^2\,\rho_s=2$]{\centering{}\includegraphics[width=0.4\linewidth]{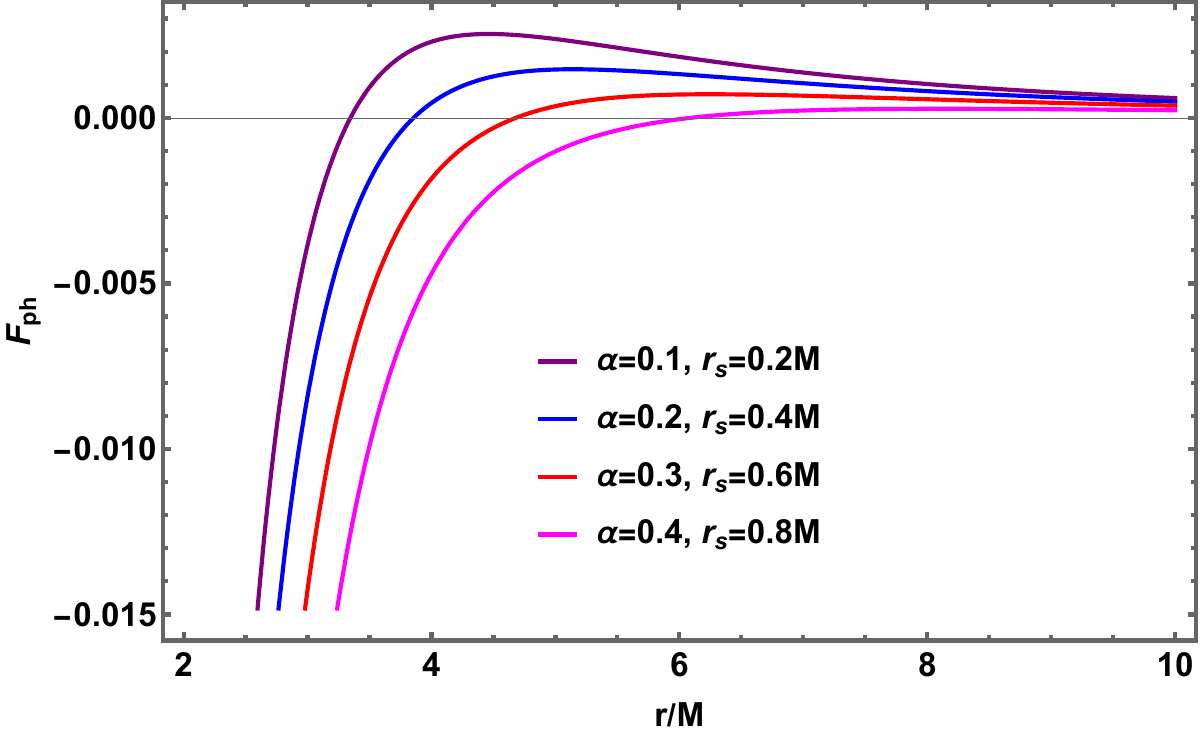}}\\
    \subfloat[$r_s=0.5\,M$]{\centering{}\includegraphics[width=0.4\linewidth]{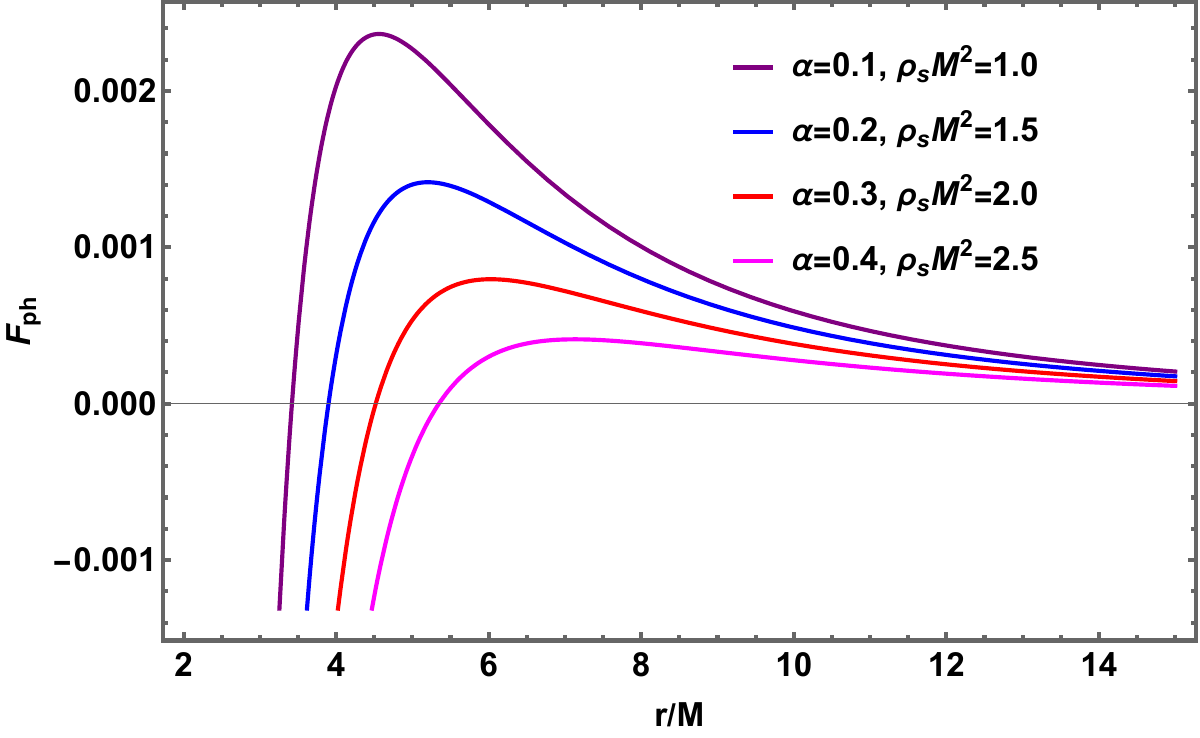}}\quad\quad
    \subfloat[]{\centering{}\includegraphics[width=0.4\linewidth]{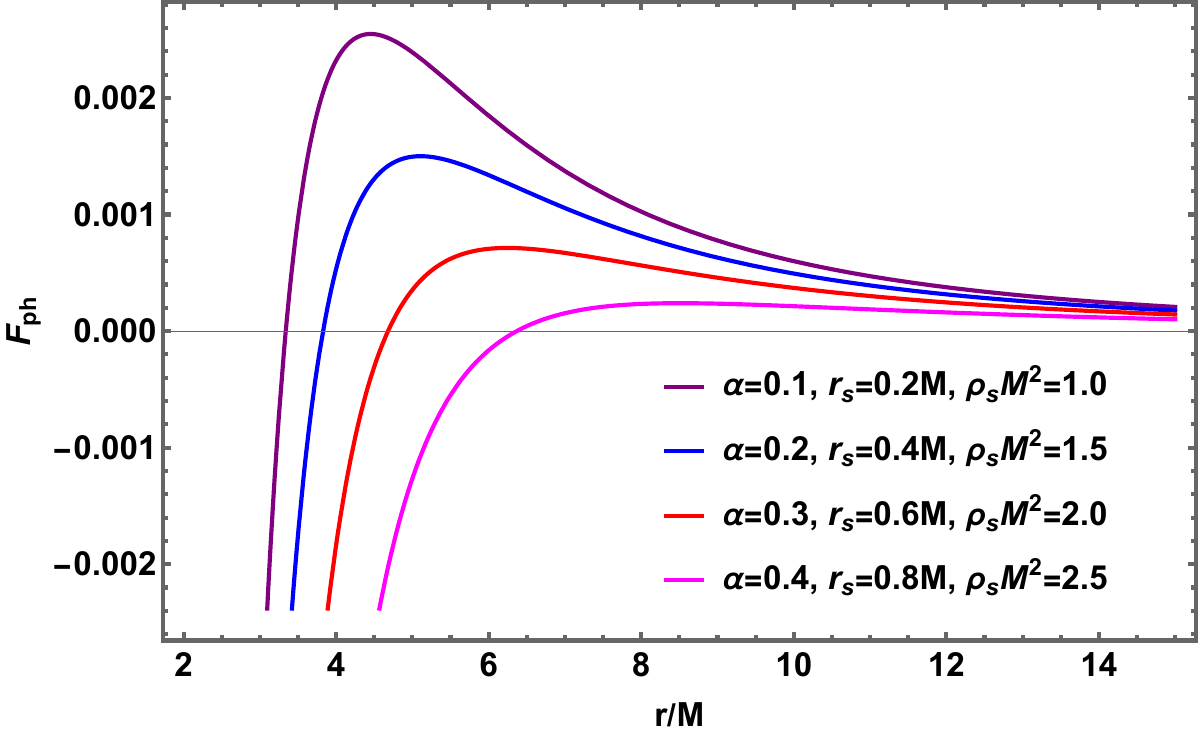}}
    \caption{\footnotesize Behavior of the force on the photon particles is shown for varying values of the string parameter \(\alpha\), the core radius \(r_s\), and the DM halo density \(\rho_s\). Sub-figures illustrate the individual effects of (a) \(\alpha\), (b) \(r_s\), and (c) \(\rho_s\,M^2\), as well as the combined effects of (d) \(\alpha\) and \(r_s\), (e) \(\alpha\) and \(\rho_s\,M^2\), and (f) \(\alpha\), \(r_s\), and \(\rho_s\, M^2\) together. Here, we set the angular momentum $\mathrm{L}=1$.}
    \label{fig:force}
\end{figure}

Using the effective potential for null geodesics given in Eq. (\ref{cc1}), we determine the force acting on the photon particles. It is defined in terms of the effective potential as, $\mathrm{F}_\text{ph}=-\frac{V'_\text{eff}}{2}$. In our case, we find 
\begin{equation}
    \mathrm{F}_\text{ph}=\frac{\mathrm{L}^2}{2\,r^3}\,(2\,f(r)-r\,f'(r))=\frac{\mathrm{L}^2}{r^3}\,\left[1-\alpha-\frac{3\,M}{r}-\frac{b}{(r+r_s)^2}\,\left(\frac{3\,r}{2}+r_s\right)\right].\label{cc2}
\end{equation}

From the above expression (\ref{cc2}), we observe that the force acting on the photon particles in the given gravitational field is affected by several parameters. These include the cosmic string parameter $\alpha$, the density $\rho_s$ of HDM, the angular momentum $\mathrm{L}$, and the BH mass $M$. 

In the limit $\rho_s=0$ (implies $b=0$), corresponding to the absence of HDM, the force from Eq. (\ref{cc2}) reduces as,
\begin{equation}
    \mathrm{F}_\text{ph}=\frac{\mathrm{L}^2}{r^3}\,\left(1-\alpha-\frac{3\,M}{r}\right).\label{cc2aa}
\end{equation}
Equation (\ref{cc2aa}) is the expression of force on the photon particles in the Letelier BH solution \cite{isz38}.

In Figure~\ref{fig:force}, we present a series of plots illustrating the behavior of the force acting on photon particles as a function of the radial coordinate \( r \), considering variations in the cosmic string parameter \( \alpha \), the core radius \( r_s \), and the core density \( \rho_s \) of the HDM halo. In all panels, it is evident that increasing the values of \( \alpha \), \( r_s \), \( \rho_s \), or their combinations leads to an enhancement of the negative force. In this context, a more negative force corresponds to a stronger attractive interaction acting on the photon. This behavior implies that, contrary to the conventional expectation in asymptotically flat Schwarzschild spacetime where gravitational attraction weakens with increasing distance, the presence of the cosmic string and the halo structure induces a confining-like effect in the modified geometry.

Such an effect suggests that the combined influence of the topological defect (cosmic string) and the matter distribution of the HDM halo significantly alters the curvature of spacetime. As a result, the dynamics of photon particles are modified, with the effective force becoming increasingly attractive at larger distances. This highlights the importance of including topological and matter-induced corrections when modeling light propagation in realistic astrophysical environments, particularly in the presence of non-trivial spacetime geometries.

We now focus on deriving the trajectory equation for photon particles in the given gravitational field and analyze the influence of the CS parameter, the density of Hernquist DM, and the cosmological constant on their motion. Using Eqs. (\ref{bb3}), (\ref{bb5}) and (\ref{cc1}), we define the equation of orbit as,
\begin{equation}
    \frac{\dot{r}^2}{\dot{\phi}^2}=\left(\frac{dr}{d\phi}\right)^2=r^4\,\left[\frac{1}{\beta^2}-\frac{1}{r^2}\,\left\{1-\alpha-\frac{2\,M}{r}-\frac{b}{r+r_s}-\frac{\Lambda}{3}\,r^2\right\}\right].\label{cc3}
\end{equation}
where $\beta=\mathrm{L}/\mathrm{E}$ is the impact parameter for photon light.

Transforming to a new variable via $u=\frac{1}{r}$ into the Eq. (\ref{cc3}), we find
\begin{equation}
    \left(\frac{du}{d\phi}\right)^2+(1-\alpha)\,u^2=\frac{1}{\beta^2}+\frac{\Lambda}{3}+2\,M\,u^3+\frac{b\,u^3}{1+r_s\,u}.\label{cc4}
\end{equation}

In order to find a second-order differential equation for photon trajectory, differentiating Eq. (\ref{cc4}) w. r. t. $\phi$ and after simplification, we find
\begin{equation}
    \frac{d^2u}{d\phi^2}+(1-\alpha)\,u=3\,M\,u^2+\frac{b\,u^2}{(1+r_s\,u)^2}\,\left(\frac{3}{2}+r_s\,u\right).\label{cc5}
\end{equation}
From the above expression (\ref{cc5}), we observe that the trajectory equation for photons in the gravitational field is influenced by the cosmic string parameter $\alpha$, the density $\rho_s$ of HDM, and the BH mass $M$.

In the limit where HDM density vanishes, $\rho_s=0$, we find the following photon trajectory equation
\begin{equation}
    \frac{d^2u}{d\phi^2}+(1-\alpha)\,u=3\,M\,u^2.\label{cc5aa}
\end{equation}
The above photon trajectory equation is similar to those result obtain in the Letelier BH solution. 

By comparing Eqs. (\ref{cc5}) and (\ref{cc5aa}), we observe that photon trajectories are further influenced by the density of Hernquist DM, in addition to the effect of the cosmic string parameter.

Now, we discuss circular null geodesics of radius $r=r_c$. For this, two important conditions $\frac{dr}{d\lambda}=0$ and $\frac{d^2r}{d\lambda^2}=0$ must hold goods. Using Eq. (\ref{bb5}), we find the following two relations
\begin{equation}
    \mathrm{E}^2=V_\text{eff}(r)=\frac{\mathrm{L}^2}{r^2}\,f(r)\quad,\quad V'_\text{eff}(r)=0,\label{cc6}
\end{equation}
where prime denotes partial derivative w. r. t. $r$. The first relation gives us the critical impact parameter for photons and second relation gives the photon sphere radius.

Substituting the effective potential for null geodesics from Eq. (\ref{cc1}) into the Eq. (\ref{cc6}) results the critical impact parameter for photon particles and the photon sphere radius $r=r_\text{ph}$, respectively. The simplification of the first relation gives us the critical impact parameter for photon particles given by
\begin{equation}
    \beta_c=\frac{r_c}{\sqrt{1-\alpha-\frac{2\,M}{r_c}-\frac{b}{r_c+r_s}-\frac{\Lambda}{3}\,r^2_c}}.\label{cc7}
\end{equation}
One can see that the critical impact parameter for photon particles is influenced by the cosmic string ($\alpha$), the density of DMH ($\rho_s$), as well as the BH mass (M) and the cosmological constant ($\Lambda$).

Let \(\beta = \frac{\mathrm{L}}{\mathrm{E}}\) denote the impact parameter for photon particles, where \(\mathrm{L}\) is the angular momentum and \(\mathrm{E}\) is the energy of the photon. Depending on the relationship between \(\beta\) and the critical impact parameter \(\beta_c\), the motion of the photon near a BH can be described as follows: \begin{itemize}
    \item[(i)] If \(\beta < \beta_c\), the photon is captured by the BH. \item[(ii)] If \(\beta = \beta_c\), the photon moves in a circular orbit at radius \(r = r_c\). \item[(iii)] If \(\beta > \beta_c\), the photon escapes and moves away from the BH region.
\end{itemize}

And the simplification of the second condition in Eq. (\ref{cc6}) yields
\begin{equation}
    r\,f'(r)=2\,f(r)\Rightarrow 1-\alpha-\frac{3\,M}{r_\text{ph}}-\frac{b}{(r_\text{ph}+r_s)^2}\,\left(\frac{3\,r_\text{ph}}{2}+r_s\right)=0.\label{cc8}
\end{equation}
An exact analytical expression for photon sphere radius $r_\text{ph}$ from Eq. (\ref{cc8}) is a challenging task. However, one can find the numerical results of this by setting suitable parameters of $\alpha$, $r_s$, $\rho_s$ and $M$ which we will discuss in the subsequent section.

Now, we investigate the stability of circular null geodesics at radius $r=r_c$ and show how various factors-such as the cosmic string, the core radius of halo and the core density of HDM influences the stability. For that, one needs to determine a physical quantity called the Lyapunov exponent which is defined by \cite{VC}
\begin{equation}
    \lambda^\text{null}_{L}=\sqrt{-\frac{V''_\text{eff}(r)}{2\,(dt/d\lambda)^2}}\Big{|}_{r=r_c}\quad\mbox{where}\quad \frac{dt}{d\lambda}=\mathrm{E}/f(r).\label{cc9}
\end{equation}

Using the effective potential given in Eq. (\ref{cc1}), we find the Lyapunov exponent in terms of the metric function as follows:
\begin{equation}
    V''_\text{eff}(r)=\frac{\mathrm{L}^2}{r^4}\,\left[r^2\,f''(r)-2\,f(r)\right],\label{cc10}
\end{equation}
where we have used the relation $r\,f'(r)=2\,f(r)$.

Thereby, substituting $\frac{dt}{d\lambda}=\mathrm{E}/f(r)$ and $V''_\text{eff}(r)$ from Eq. (\ref{cc10}) into the Eq. (\ref{cc9}) and with the help of the metric function $f(r)$ given in Eq. (\ref{function}) results
\begin{eqnarray}
    \lambda^\text{null}_{L}&=&\sqrt{f(r)\,\left(\frac{f(r)}{r^2}-\frac{f''(r)}{2}\right)}\Bigg{|}_{r=r_c}=\frac{1}{\beta_c}\,\sqrt{1-\alpha-\frac{b}{r_c+r_s}+\frac{b\,r^2_c}{2\,(r_c+r_s)^3}},\label{cc11}
\end{eqnarray}
where $\beta_c$ is given in Eq. (\ref{cc7}).

From the above expression (\ref{cc11}), we observe that the Lyapunov exponent for circular null geodesics is influenced by several key factors present in the BH geometry. These include the cosmic string parameter $\alpha$, the density of HDM $\rho_s$, the core radius $r_s$, as well as the BH mass $M$ and the cosmological constant $\Lambda$. 

In the limit $\rho_s=0$, corresponding to the absence of HDM halo density, the Lyapunov exponent from Eq. (\ref{cc11}) reduces as (here $r_c=r_\text{ph}=3M/(1-\alpha)$),
\begin{eqnarray}
    \lambda^\text{null}_{L}=\frac{\sqrt{1-\alpha}}{\beta_c}=\sqrt{1-\alpha}\,\sqrt{\frac{(1-\alpha)^3}{27\,M^2}-\frac{\Lambda}{3}}.\label{cc12}
\end{eqnarray}
Equation (\ref{cc12}) represents the Lyapunov exponent for circular null geodesics in the Letelier-AdS BH, which further reduces to the Schwarzschild-AdS BH result when $\alpha = 0$, that is $\lambda^\text{null}_{L}=\sqrt{\frac{1}{27\,M^2}-\frac{\Lambda}{3}}$.

By comparing Eqs. (\ref{cc11}) and (\ref{cc12}), we observe that the Lyapunov exponent for circular null geodesics decreases due to the presence of Hernquist DM, along with the effects of the cosmic string parameter and the cosmological constant.

The geodesic angular velocity (coordinate angular velocity) of circular null orbits is defined by \cite{VC}
\begin{equation}
    \Omega=\frac{\dot{\phi}}{\dot{t}}=\frac{f(r)}{r^2}\,\frac{\mathrm{L}}{\mathrm{E}}=\frac{\sqrt{f(r)}}{r}\Big{|}_{r=r_c}.\label{cc13}
\end{equation}
where we have used the relation given in Eq. (\ref{cc6}).

Substituting the metric function $f(r)$ given in Eq. (\ref{function}), we find the geodesic angular velocity given by
\begin{equation}
     \Omega=\frac{1}{r_c}\,\sqrt{1-\alpha-\frac{2\,M}{r_c}-\frac{b}{r_c+r_s}-\frac{\Lambda}{3}\,r^2_c}=\frac{1}{\beta_c}.\label{cc14}
\end{equation}

From the above expression (\ref{cc14}), we observe that the angular velocity of geodesics is influenced by the cosmic string parameter $\alpha$, the density of HDM $\rho_s$, the BH mass $M$, and the cosmological constant $\Lambda$. All these factors together determine the faster/slower motion of photon particles orbiting around the BH.

In the limit $\rho_s=0$, corresponding to the absence of HDM halo density, we find the geodesic angular velocity given by
\begin{equation}
     \Omega=\frac{1}{r_c}\,\sqrt{1-\alpha-\frac{2\,M}{r_c}-\frac{\Lambda}{3}\,r^2_c}=\frac{1}{\beta_c}=\sqrt{\frac{(1-\alpha)^3}{27\,M^2}-\frac{\Lambda}{3}}.\label{cc15}
\end{equation}

By comparing Eqs. (\ref{cc14}) and (\ref{cc15}), we observe that the geodesic angular velocity decreases due to the presence of Hernquist DM, along with the effects of the cosmic string parameter and the cosmological constant.

\subsection{Timelike Geodesics}

Timelike geodesics serve as a fundamental tool for analyzing the motion of massive (i.e., timelike) particles around a BH. By employing the effective potential for timelike geodesics, one can investigate the dynamics of these particles, including their trajectories, orbital stability, and gyroscopic precession frequencies.

In this subsection, we investigate the dynamics of timelike particles around an AdS BH surrounded by a HDMH and a cloud of strings. We analyze how various components of the BH geometry-such as the cloud of strings, the core radius of the HDM halo, and the cosmological constant-influence the motion of massive particles.

For massive test particles, $\varepsilon=-1$, the effective potential from (\ref{bb6}) becomes
\begin{equation}
    V_\text{eff}=\left(1+\frac{\mathrm{L}^2}{r^2}\right)\,\left(1-\alpha-\frac{2\,M}{r}-\frac{b}{r+r_s}-\frac{\Lambda}{3}\,r^2\right).\label{dd1}
\end{equation}

\begin{figure}[ht!]
    \centering
    \subfloat[$r_s=0.5\,M,\,M^2\,\rho_s=1$]{\centering{}\includegraphics[width=0.4\linewidth]{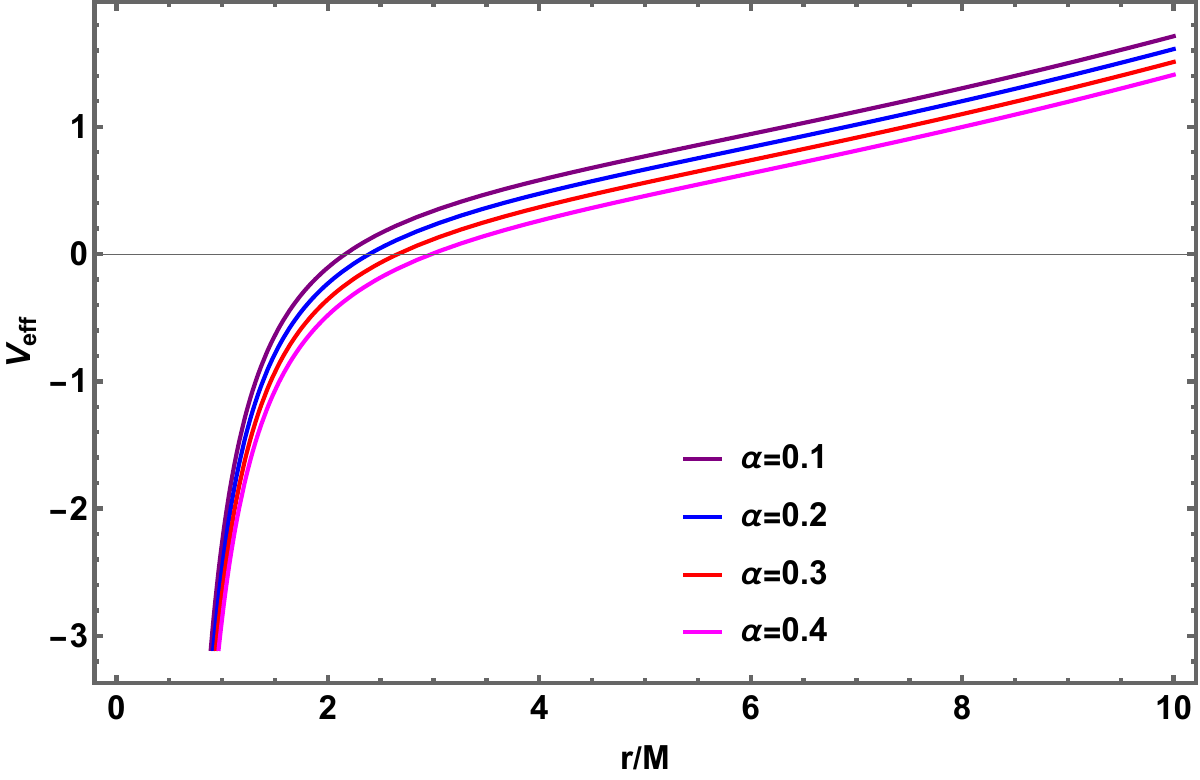}}\quad\quad
    \subfloat[$\alpha=0.1,M^2\,\rho_s=1$]{\centering{}\includegraphics[width=0.4\linewidth]{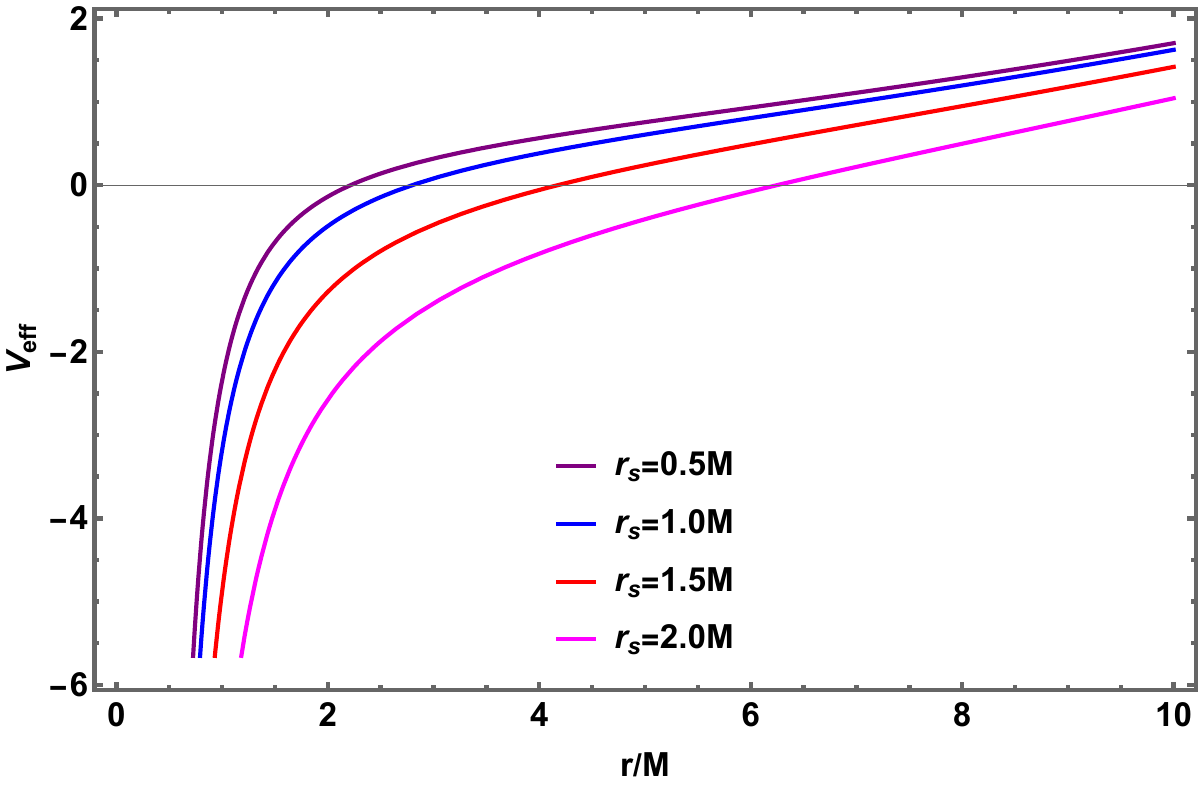}}\\
    \subfloat[$\alpha=0.1,r_s=2\,M$]{\centering{}\includegraphics[width=0.4\linewidth]{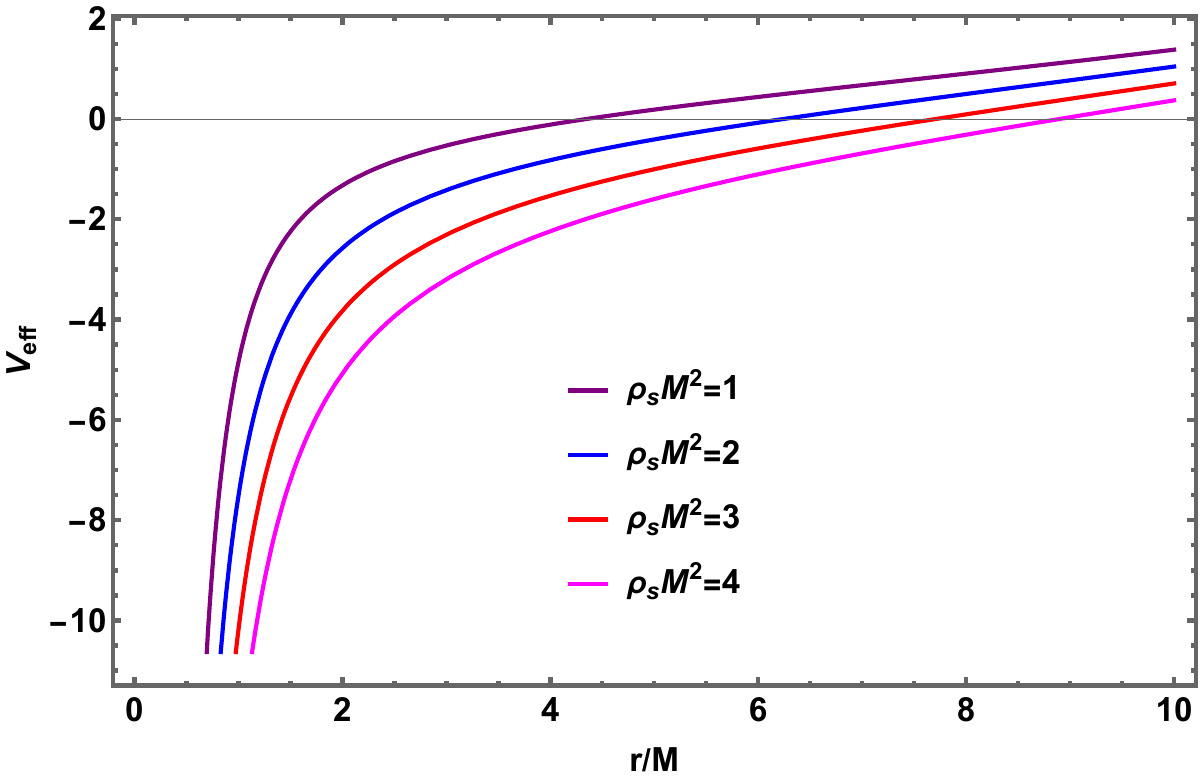}}\quad\quad
    \subfloat[$M^2\,\rho_s=1$]{\centering{}\includegraphics[width=0.4\linewidth]{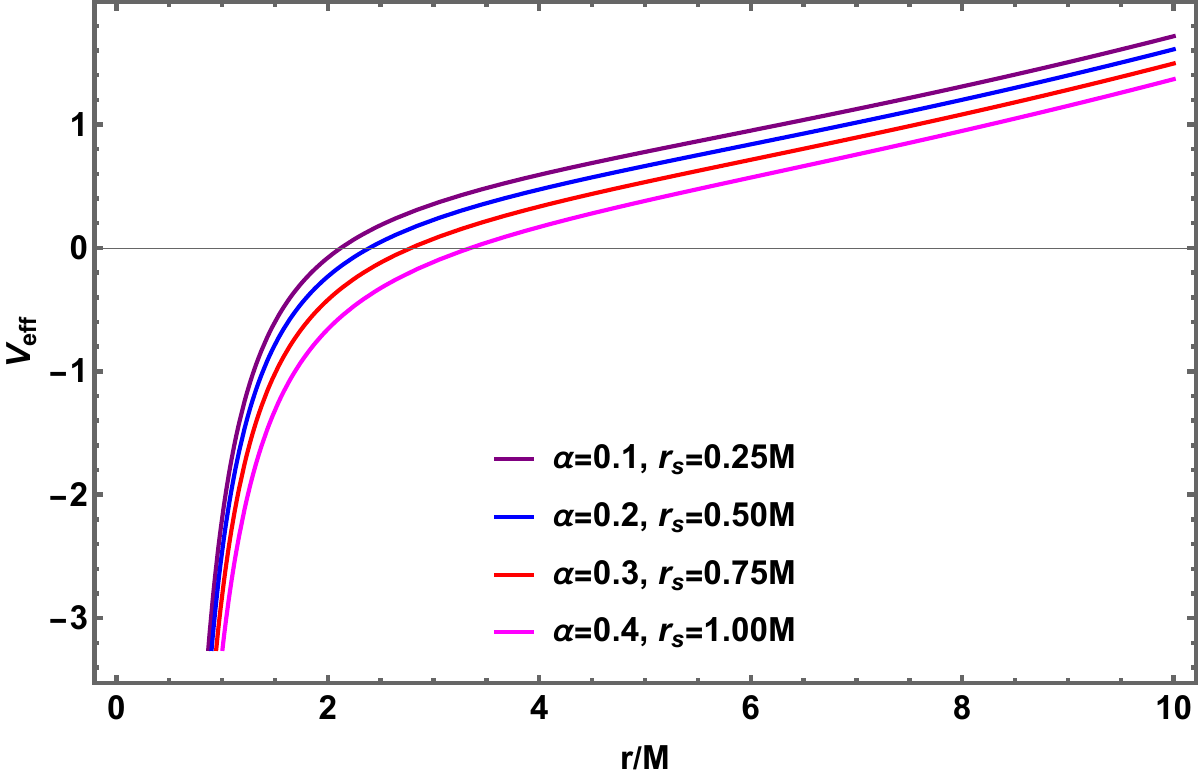}}\\
    \subfloat[$r_s=0.5\,M$]{\centering{}\includegraphics[width=0.4\linewidth]{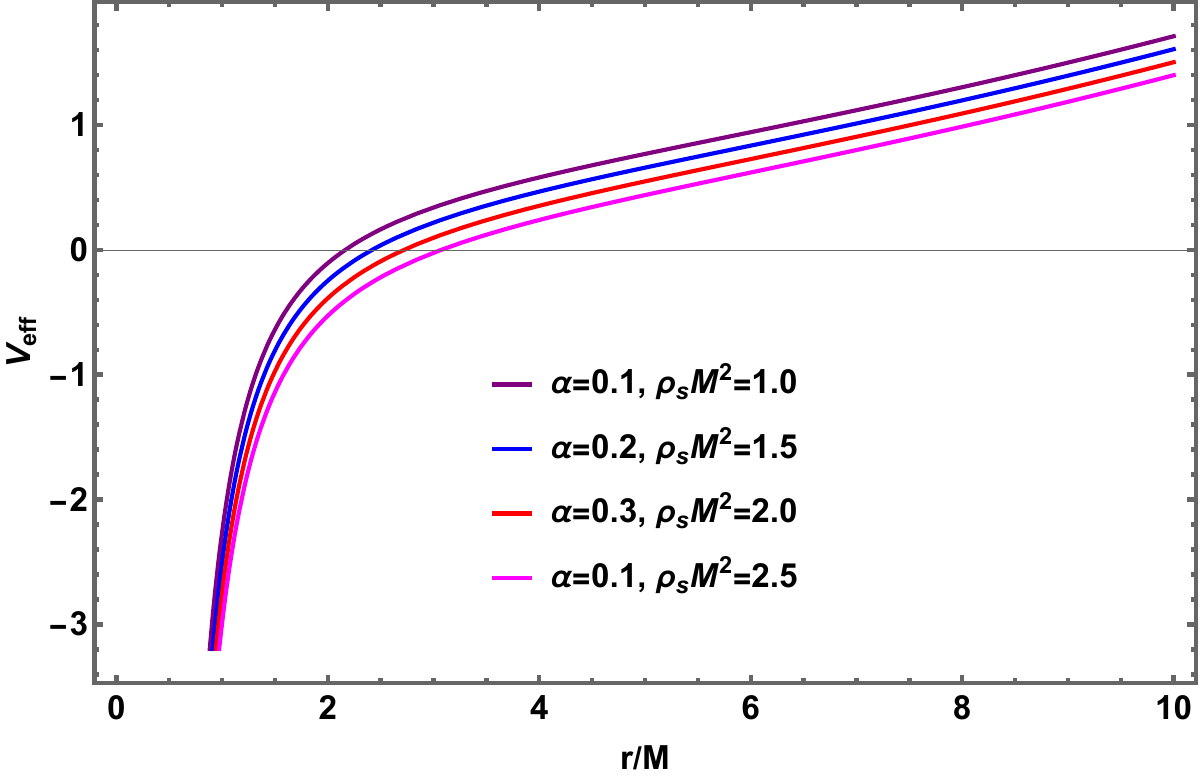}}\quad\quad
    \subfloat[]{\centering{}\includegraphics[width=0.4\linewidth]{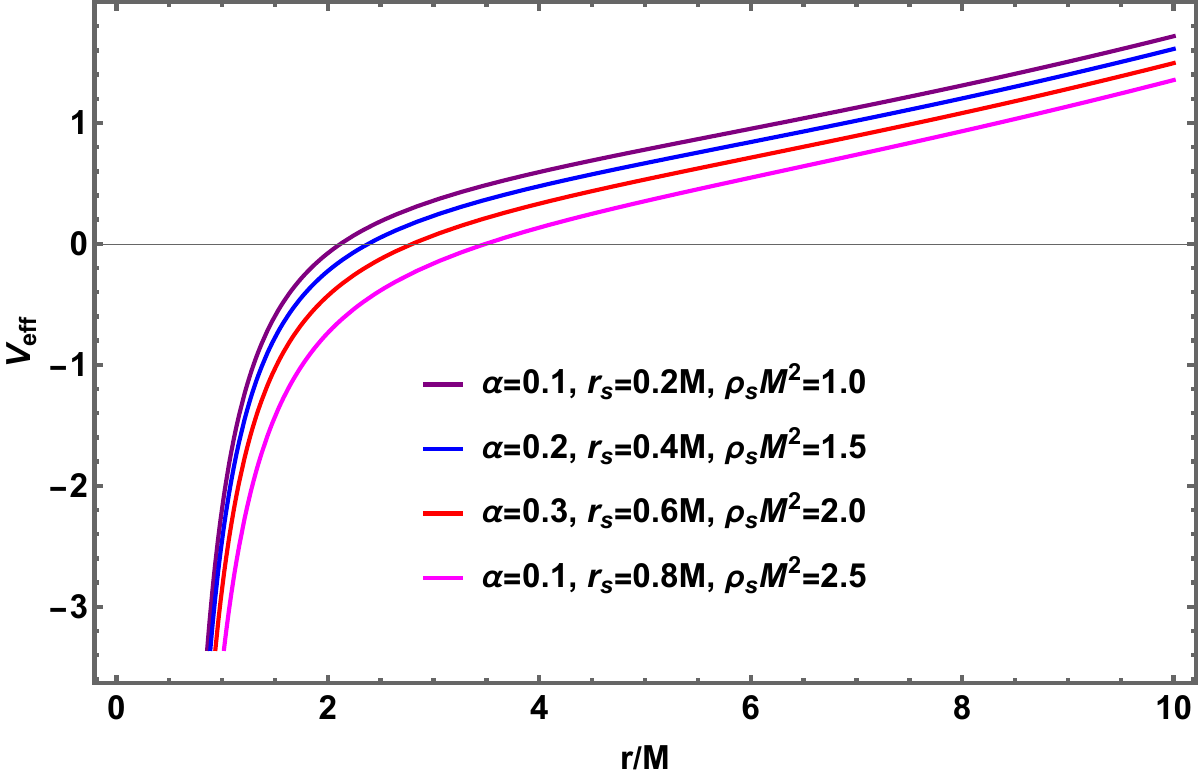}}
    \caption{\footnotesize Behavior of the effective potential for timelike is shown for varying values of the string parameter \(\alpha\), the core radius \(r_s\), and the DM halo density \(\rho_s\). Sub-figures illustrate the individual effects of (a) \(\alpha\), (b) \(r_s\), and (c) \(\rho_s\,M^2\), as well as the combined effects of (d) \(\alpha\) and \(r_s\), (e) \(\alpha\) and \(\rho_s\,M^2\), and (f) \(\alpha\), \(r_s\), and \(\rho_s\, M^2\) together.  Here, we set the angular momentum per unit mass $\mathrm{L}/M=1$ and the dimensionless parameter $k=M\,\sqrt{-\frac{\Lambda}{3}}=0.1$, all are in natural units.}
    \label{fig:timelike-potential}
\end{figure}

\begin{figure}[ht!]
    \centering
    \subfloat[$r_s=0.5\,M,\,M^2\,\rho_s=1$]{\centering{}\includegraphics[width=0.4\linewidth]{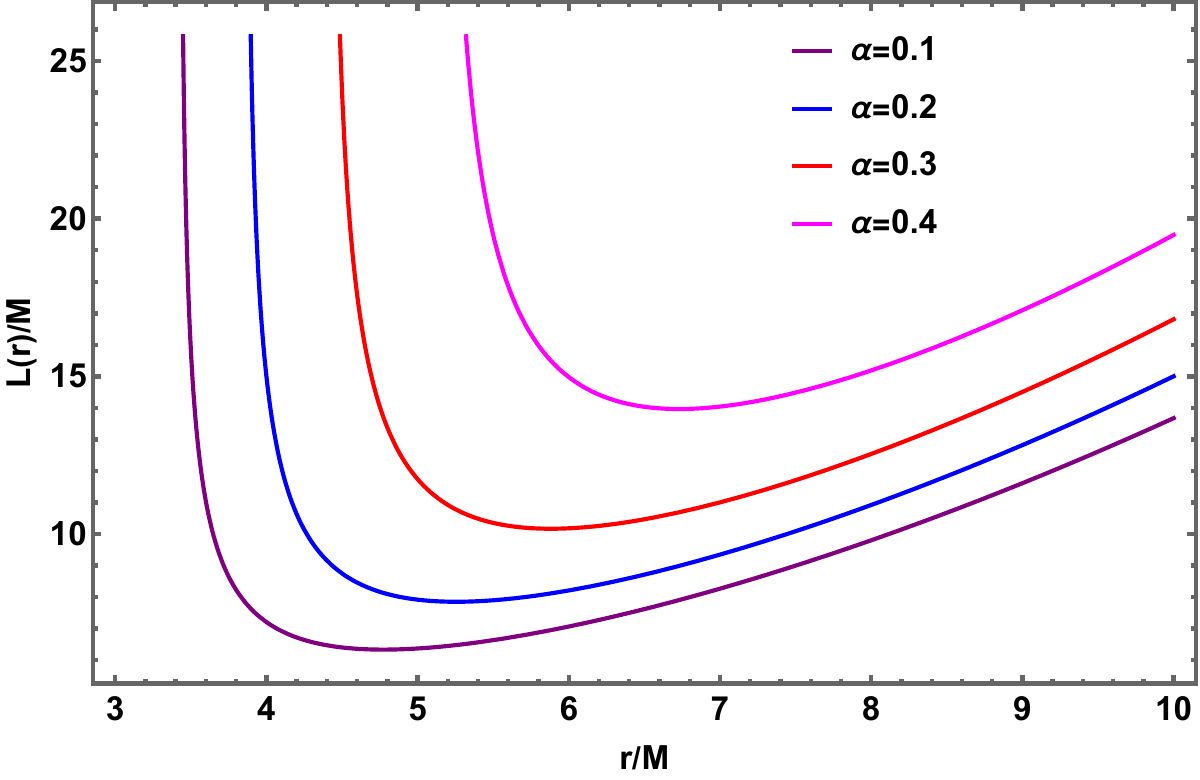}}\quad\quad
    \subfloat[$\alpha=0.1,M^2\,\rho_s=1$]{\centering{}\includegraphics[width=0.4\linewidth]{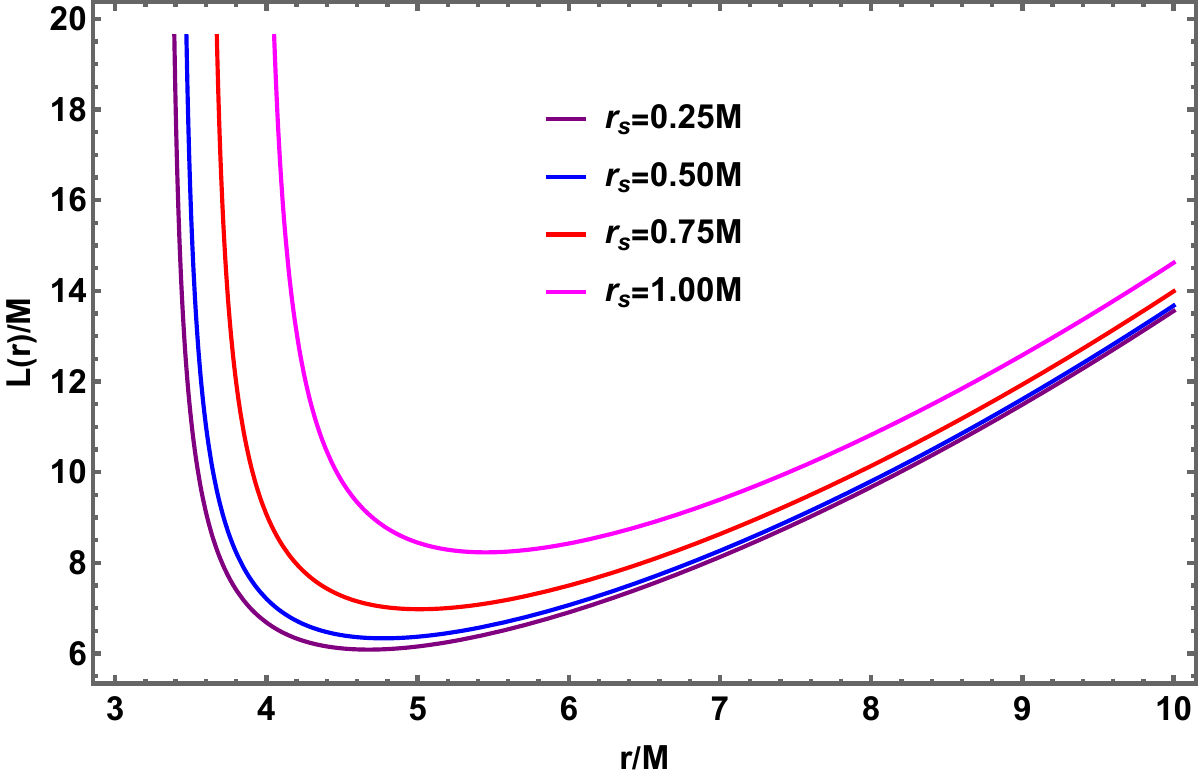}}\\
    \subfloat[$\alpha=0.1,r_s=1\,M$]{\centering{}\includegraphics[width=0.4\linewidth]{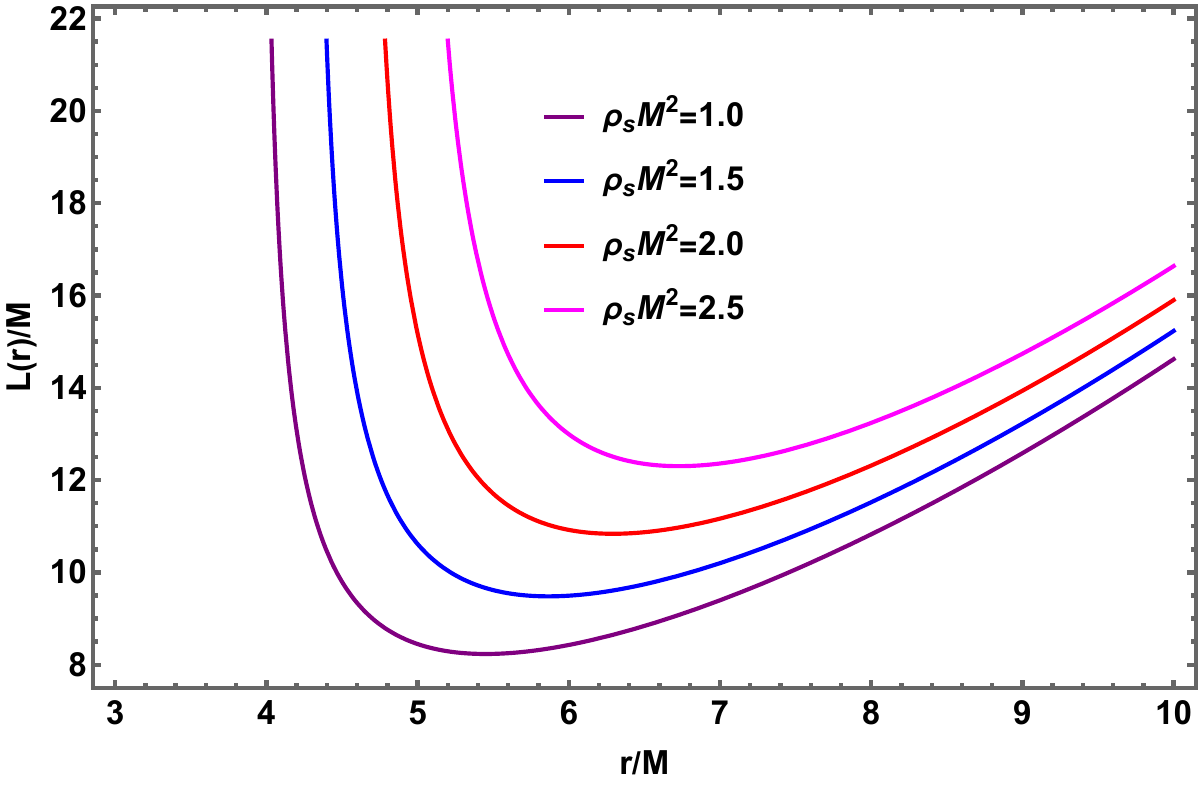}}\quad\quad
    \subfloat[$M^2\,\rho_s=1$]{\centering{}\includegraphics[width=0.4\linewidth]{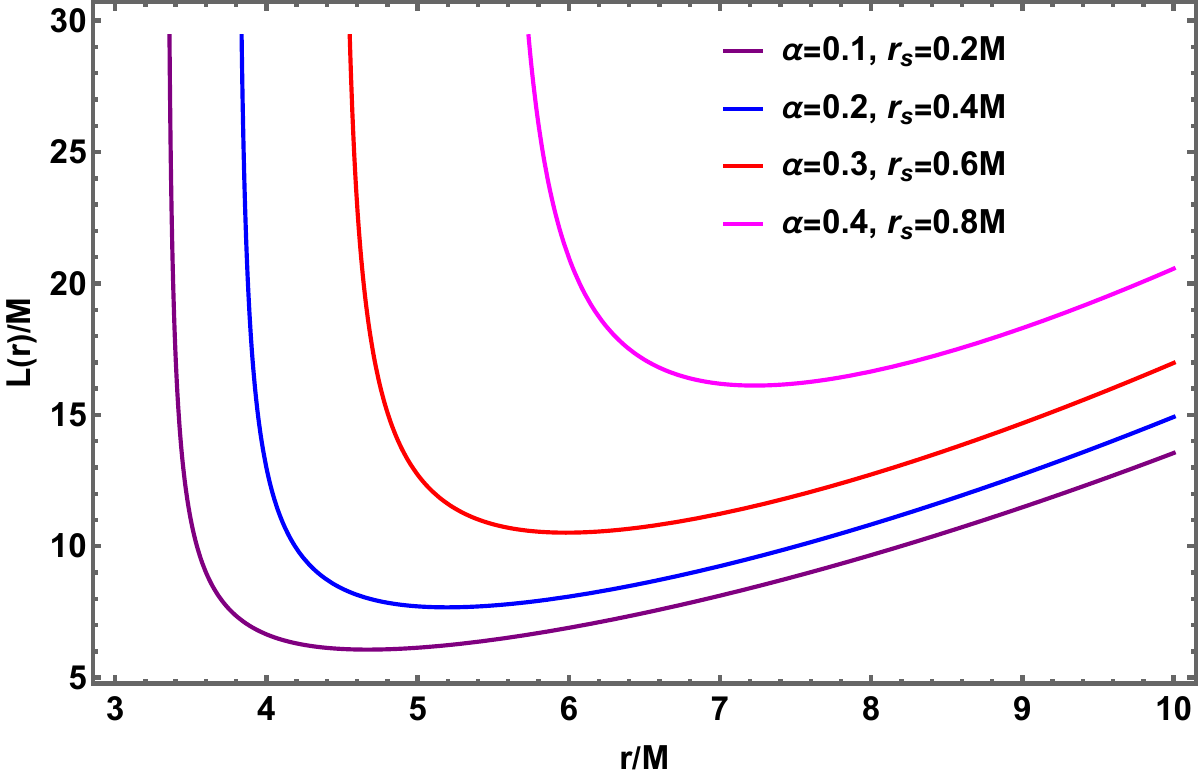}}\\
    \subfloat[$r_s=0.5\,M$]{\centering{}\includegraphics[width=0.4\linewidth]{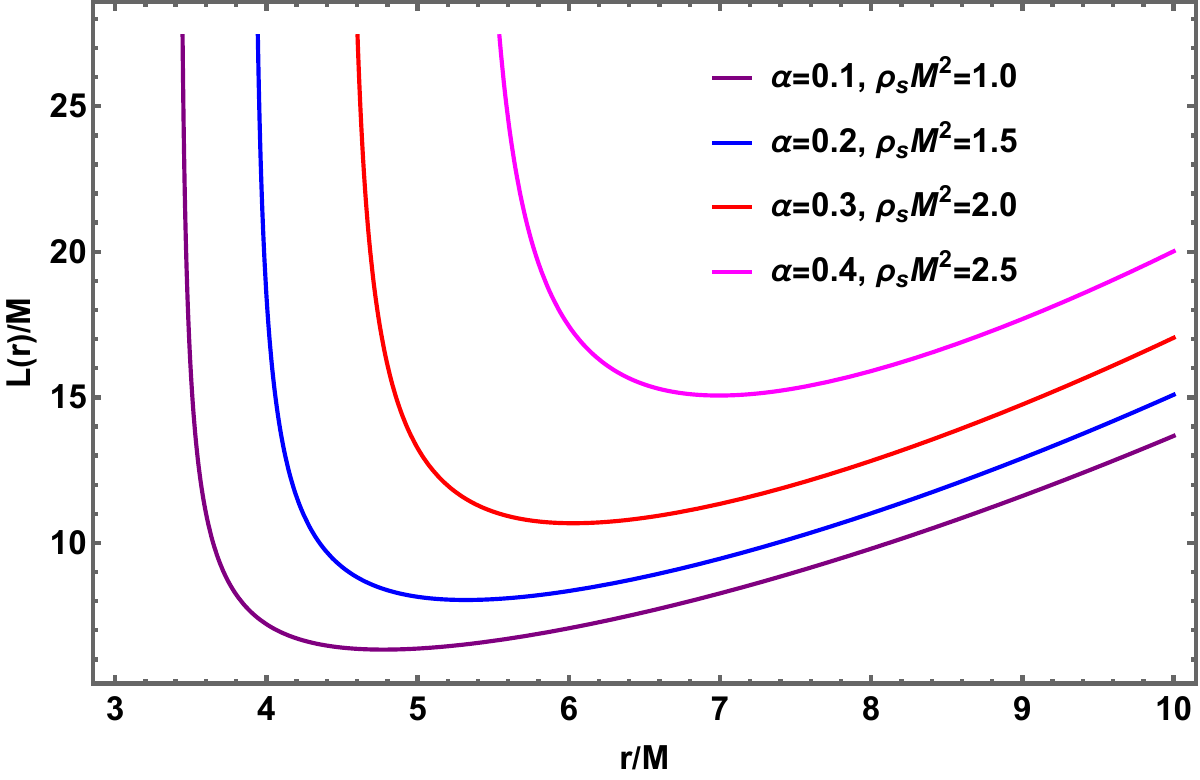}}\quad\quad
    \subfloat[]{\centering{}\includegraphics[width=0.4\linewidth]{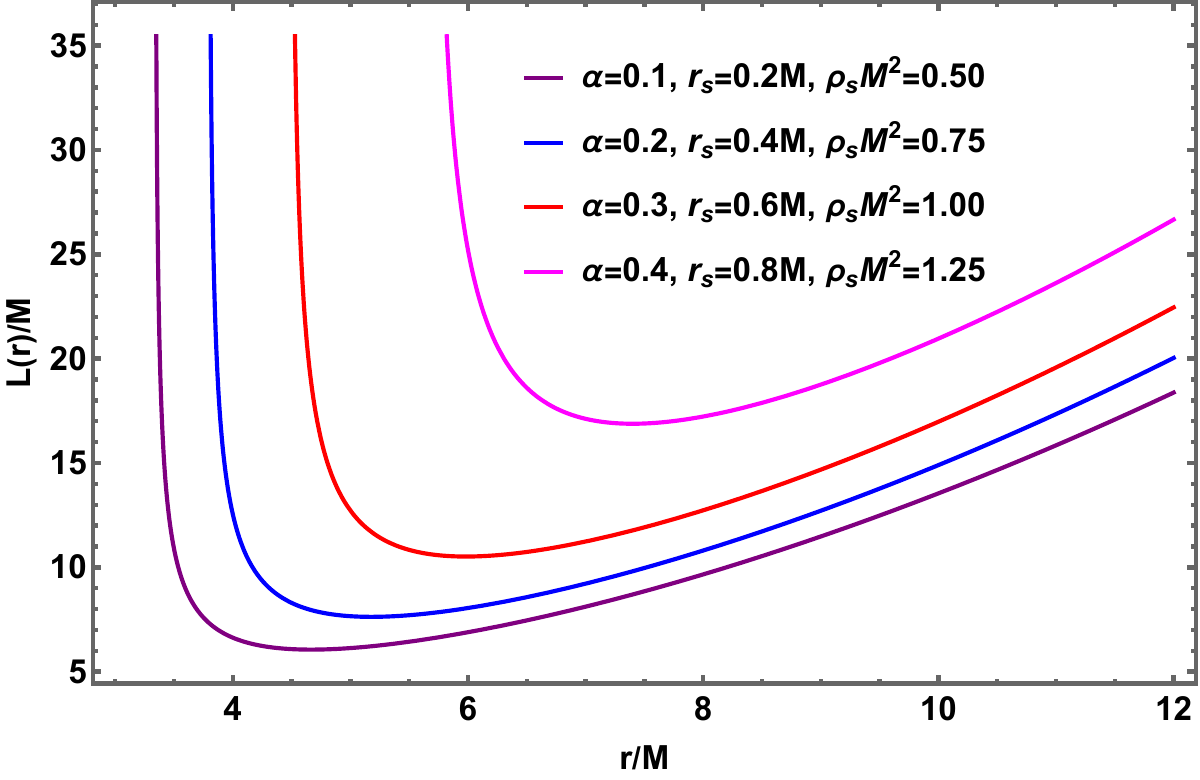}}
    \caption{\footnotesize Behavior of the specific angular momentum per unit mass ($\mathrm{L}/M$) of time-like particles orbiting in circular paths is shown for varying values of the string parameter \(\alpha\), the core radius \(r_s\), and the DM halo density \(\rho_s\). Sub-figures illustrate the individual effects of (a) \(\alpha\), (b) \(r_s\), and (c) \(\rho_s\,M^2\), as well as the combined effects of (d) \(\alpha\) and \(r_s\), (e) \(\alpha\) and \(\rho_s\,M^2\), and (f) \(\alpha\), \(r_s\), and \(\rho_s\, M^2\) together. Here, we set the dimensionless parameter $k=M\,\sqrt{-\frac{\Lambda}{3}}=0.1$.}
    \label{fig:angular-momentum}
\end{figure}

\begin{figure}[ht!]
    \centering
    \subfloat[$r_s=0.5\,M,\,M^2\,\rho_s=1$]{\centering{}\includegraphics[width=0.4\linewidth]{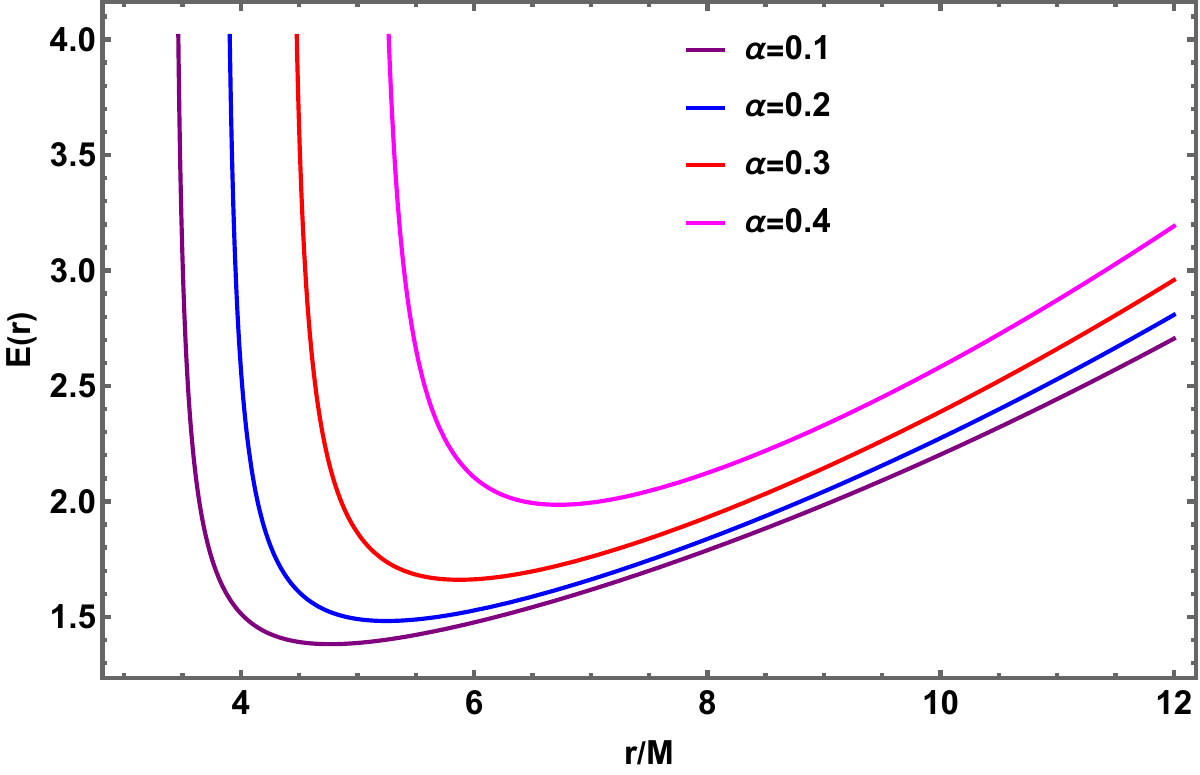}}\quad\quad
    \subfloat[$\alpha=0.1,M^2\,\rho_s=1$]{\centering{}\includegraphics[width=0.4\linewidth]{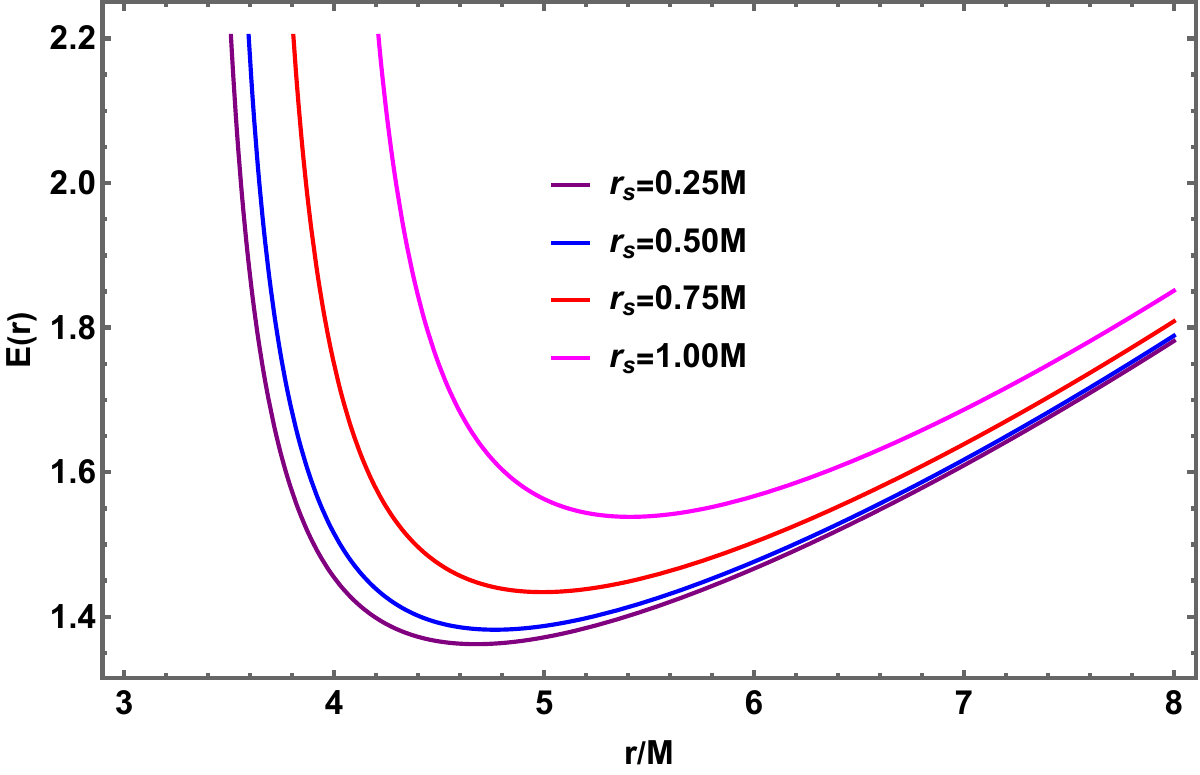}}\\
    \subfloat[$\alpha=0.1,r_s=1\,M$]{\centering{}\includegraphics[width=0.4\linewidth]{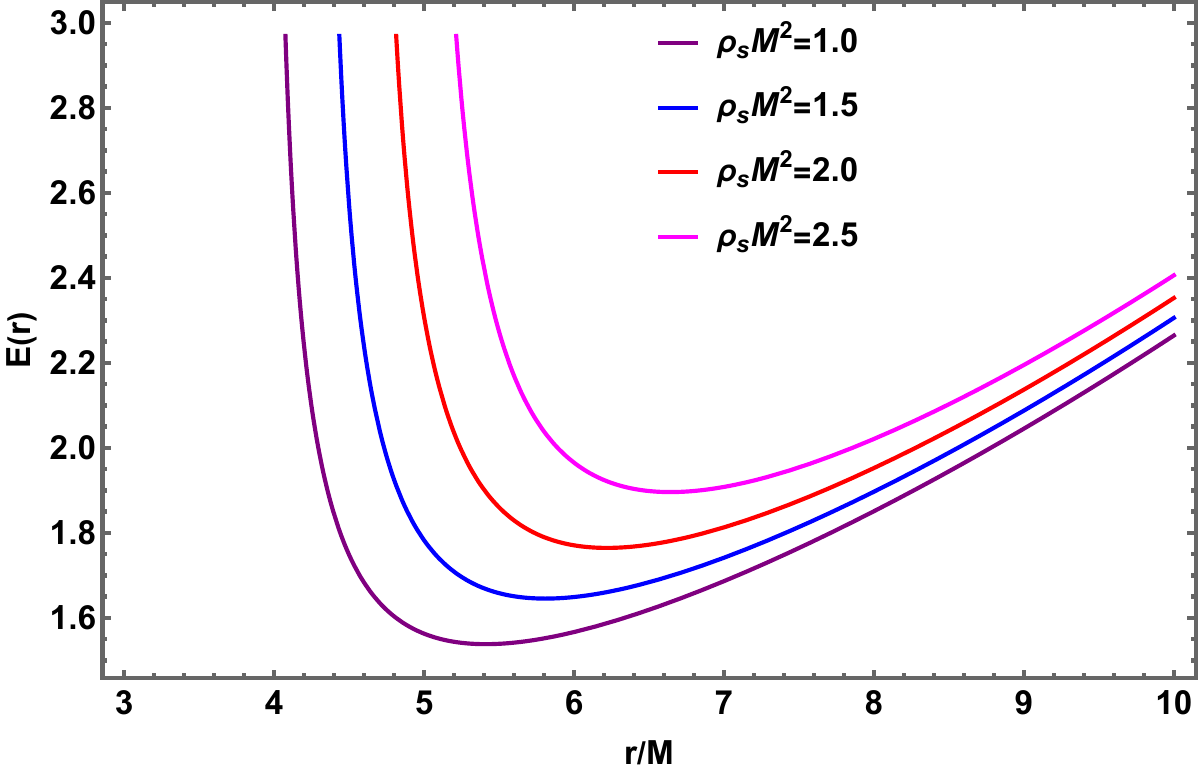}}\quad\quad
    \subfloat[$M^2\,\rho_s=1$]{\centering{}\includegraphics[width=0.4\linewidth]{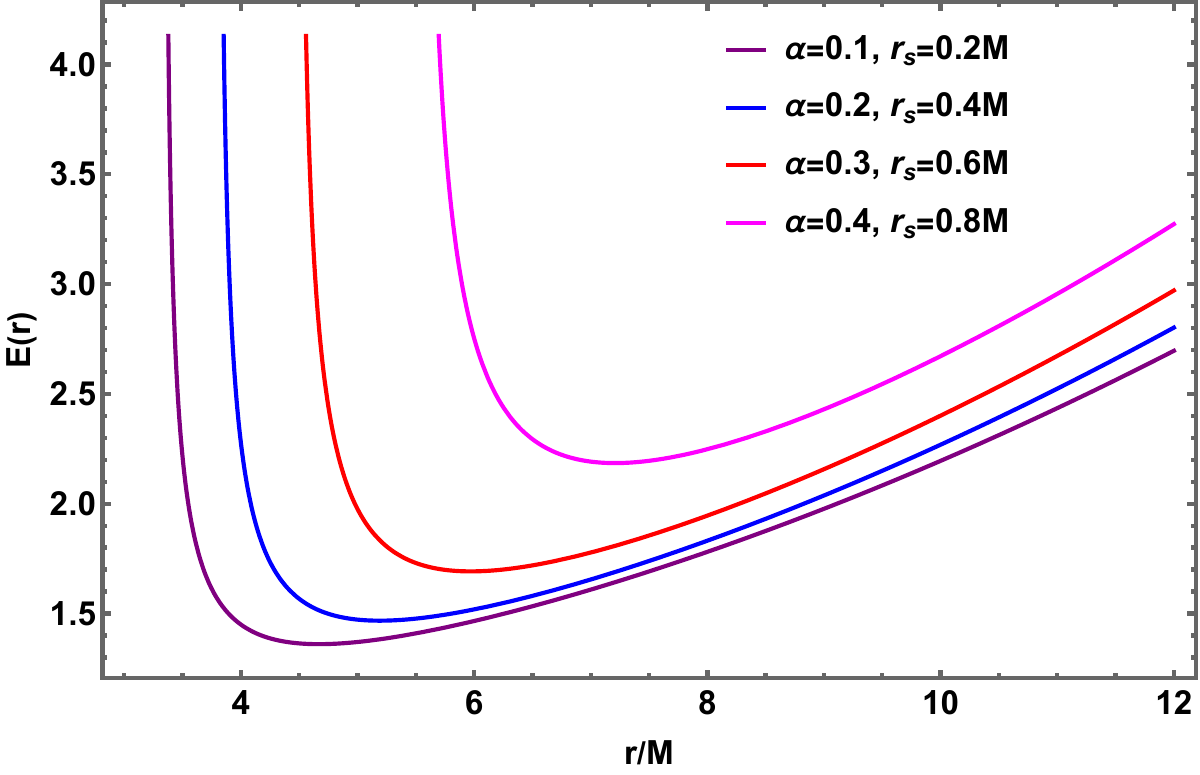}}\\
    \subfloat[$r_s=0.5\,M$]{\centering{}\includegraphics[width=0.4\linewidth]{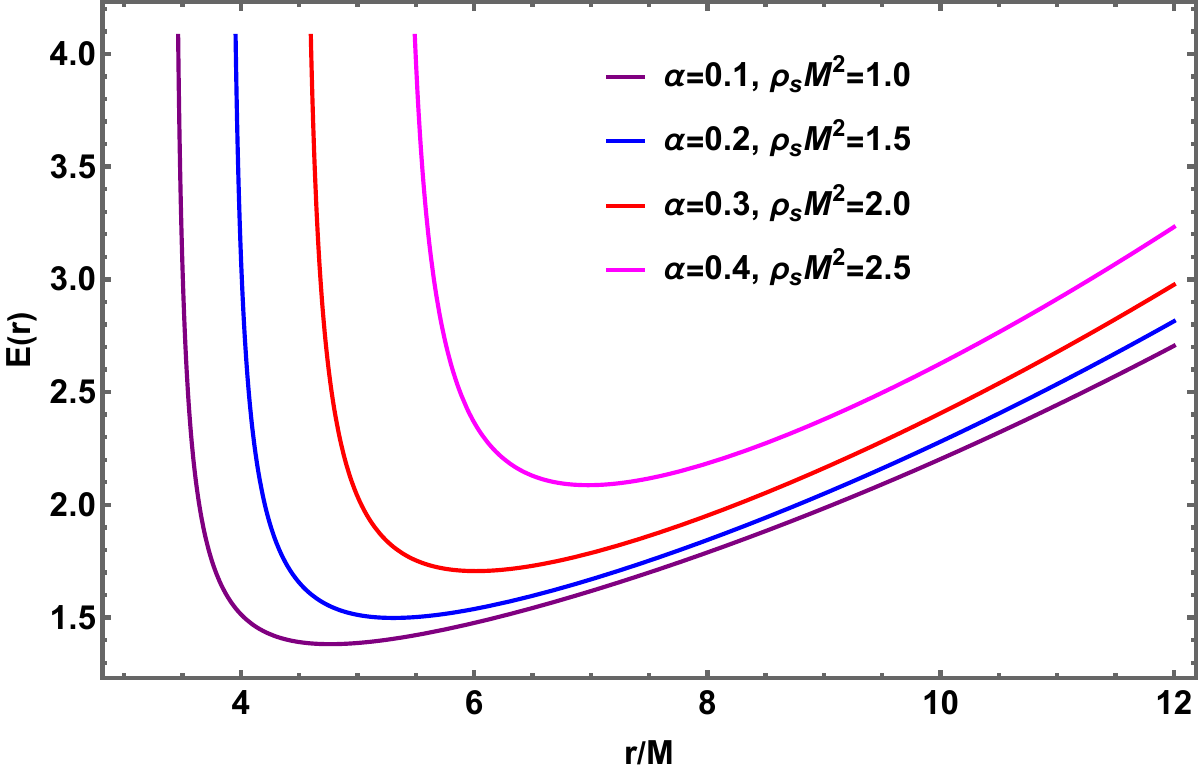}}\quad\quad
    \subfloat[]{\centering{}\includegraphics[width=0.4\linewidth]{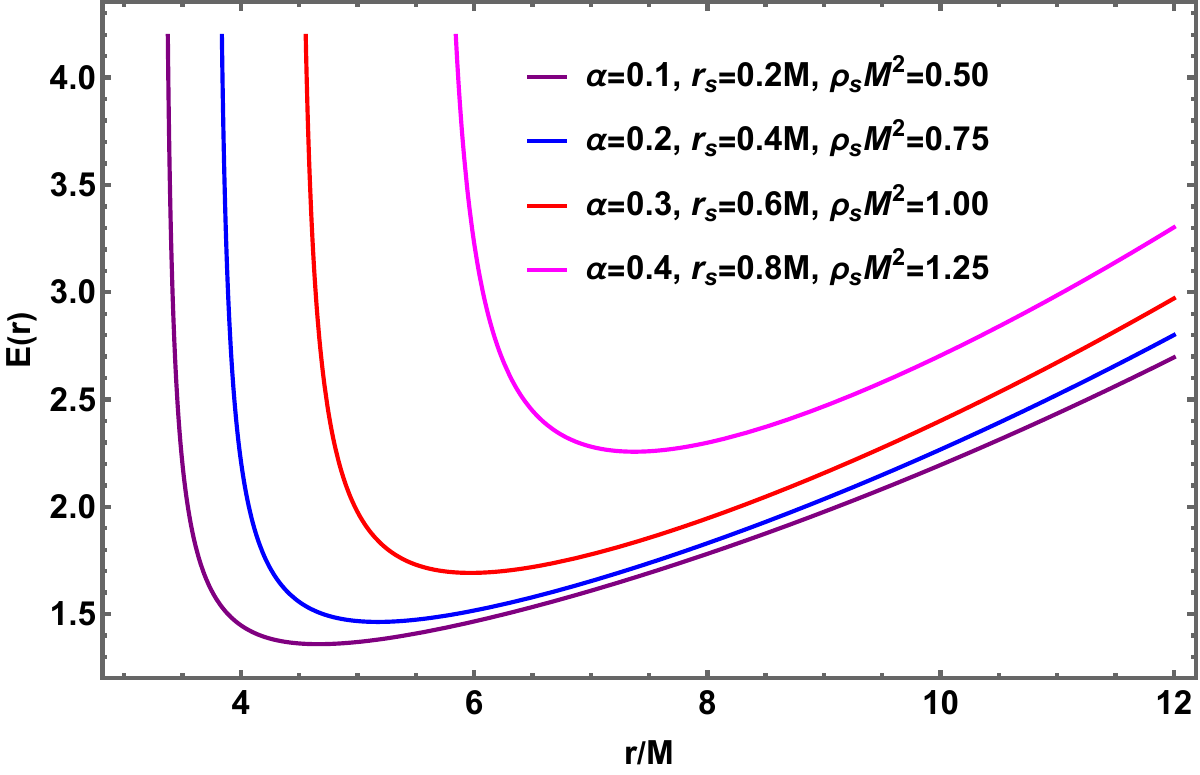}}
    \caption{\footnotesize Behavior of the specific energy ($\mathrm{E}$) of time-like particles orbiting in circular paths is shown for varying values of the string parameter \(\alpha\), the core radius \(r_s\), and the DM halo density \(\rho_s\). Sub-figures illustrate the individual effects of (a) \(\alpha\), (b) \(r_s\), and (c) \(\rho_s\,M^2\), as well as the combined effects of (d) \(\alpha\) and \(r_s\), (e) \(\alpha\) and \(\rho_s\,M^2\), and (f) \(\alpha\), \(r_s\), and \(\rho_s\, M^2\) together. Here, we set the dimensionless parameter $k=M\,\sqrt{-\frac{\Lambda}{3}}=0.1$.}
    \label{fig:energy}
\end{figure}

In Figure~\ref{fig:timelike-potential}, we present a series of plots that illustrate the behavior of the effective potential for time-like geodesics as a function of the radial coordinate \( r \), under variations of the cosmic string parameter \( \alpha \), the radius of the core \( r_s \), and the density of the core \( \rho_s \) of the HDM halo. In all panels, we observe that increasing the values of \( \alpha \), \( r_s \), \( \rho_s \), or their combinations, leads to a decrease in the effective potential of the system. This suggests that all the parameters collectively shape the behavior of the potential in the system when photon particles traverse the gravitational field.

In Figure~\ref{fig:angular-momentum}, we present a series of plots that illustrate the behavior of the specific angular momentum of time-like particles as a function of the radial coordinate \( r \), under variations of the cosmic string parameter \( \alpha \), the radius of the core \( r_s \), and the density of the core \( \rho_s \) of the HDM halo. Across all panels, we observe that increasing the values of \( \alpha \), \( r_s \), \( \rho_s \), or their combinations leads to an increases in the specific angular momentum of timelike particles.

In Figure~\ref{fig:energy}, we present a series of plots illustrating the behavior of the specific energy of timelike particles as a function of the radial coordinate \( r \), under variations of the cosmic string parameter \( \alpha \), the core radius \( r_s \), and the core density \( \rho_s \) of the HDM halo. Across all panels, we observe that increasing the values of \( \alpha \), \( r_s \), \( \rho_s \), or their combinations, leads to an increase in the specific energy of timelike particles.

For circular orbits of timelike particles around the BH, we have conditions $\dot{r}=0$ and $\ddot{r}=0$ which result in the two physical quantities called the angular momentum and the particle's energy. These are given by
\begin{eqnarray}
    \mathrm{L}(r)=r\,\sqrt{\frac{r\,f'(r)}{2\,f(r)-r\,f'(r)}}=r\,\sqrt{\frac{\frac{M}{r}+\frac{b\,r}{(r+r_s)^2}-\frac{\Lambda}{3}\,r^2}{1-\alpha-\frac{3\,M}{r}-\frac{b}{(r+r_s)^2}\,\left(\frac{3\,r}{2}+r_s\right)}}.\label{dd2}
\end{eqnarray}
And
\begin{eqnarray}
    \mathrm{E}_{\pm}(r)=\pm\,\frac{1-\alpha-\frac{2\,M}{r}-\frac{b}{r+r_s}-\frac{\Lambda}{3}\,r^2}{\sqrt{1-\alpha-\frac{3\,M}{r}-\frac{b}{(r+r_s)^2}\,\left(\frac{3\,r}{2}+r_s\right)}}.\label{dd3}
\end{eqnarray}

From the energy of the above particles, we see that this energy approaches $\mathrm{E}_{\pm} \to \pm\,\sqrt{1-\alpha}$ as $r \to \infty$. Hence the maximum particles' energy is $\sqrt{1-\alpha}$ less than unity (1). In the limit where $r_s=0$, that is, absence of the HDM halo core, these physical quantities ($\mathrm{L}$,$\mathrm{E}_{\pm}$) of time-like particles reduce to those results for the Letelier AdS BH solution.

Now, we aim to find a speed with which a timelike particle orbits around the BH at a large distance compared to the BH horizon $r>>r_\text{h}$. This is in analogy with a distant star in a galaxy moving in a circular path around the BH of the galaxy. In the zeroth approximation, the gravitational redshift function $-g_{tt}$ can be expressed in terms of the Newtonian gravitational potential $\Phi(r)$ as follows:
\begin{equation}
    f(r)=1+2\,\Phi(r),\label{dd4}
\end{equation}

Using the metric function given in Eq. (\ref{function}), we find this Newtonian gravitational potential $\Phi(r)$ given by
\begin{equation}
    \Phi(r)=\frac{1}{2}\,\left(-\alpha-\frac{2\,M}{r}-\frac{b}{r+r_s}-\frac{\Lambda}{3}\,r^2\right).\label{dd5}
\end{equation}

Now, using the Newtonian gravitational potential, one can determine an effective gravitational force which is simply given by $\mathrm{F}_c=-\frac{\partial \Phi(r)}{\partial r}$, towards the center of the BH. Thereby, using the above potential, we find the following expression of the central force as
\begin{eqnarray}
    \mathrm{F}_c=-\frac{M}{r^2}-\frac{b}{2\,(r+r_s)^2}+\frac{\Lambda}{3}\,r.\label{dd6}
\end{eqnarray}
This central force can be equated with the centripetal force, \textit{i.e.}, $|\mathrm{F}_c|=\frac{\mu\,v^2}{r}$ in which $v$ is the orbital speed of the test particles at distances from the center of the BH and $\mu$ is the particles mass. After simplification, we find the orbital speed of the test particle given by
\begin{eqnarray}
    v=\sqrt{\frac{1}{\mu}\,\left|-\frac{M}{r}-\frac{b\,r}{2\,(r+r_s)^2}+\frac{\Lambda}{3}\,r^2\right|}.\label{dd7}
\end{eqnarray}

The expression above (\ref{dd7}) shows that the orbital speed of a timelike particle orbiting around the BH at far distances compared to the horizon radius is influenced by several geometric parameters. These include the string parameter ($\alpha$), the DM halo density ($\rho_s$), the core radius ($r_s$), the BH mass ($M$) as well as and the cosmological constant ($\Lambda$).

We now focus on the gyroscopic precession frequency of the timelike particle and analyze how various factors affect it. For that, we determine the orbital angular velocity of the timelike particle at a finite distance. This orbital velocity is given by \cite{VC}
\begin{equation}
    \Omega(r)=\frac{\dot{\phi}}{\dot{t}}=\sqrt{\frac{f'(r)}{2\,r}}=\sqrt{\frac{M}{r^3}+\frac{b}{2\,r\,(r+r_s)^2}-\frac{\Lambda}{3}}.\label{dd8}
\end{equation}

The proper angular velocity of timelike particle is defined by $\mathrm{L}=r^2\,\dot{\phi}=r^2\,\omega$, and so we have that
\begin{equation}
    \omega=\Omega\,\sqrt{\frac{2}{2\,f-r\,f'}}.\label{dd9}
\end{equation}

To determine the geodesic precession frequency $\Theta_{GPF}$ for a gyroscope within the GP-B, we use Eq. (\ref{dd8}) and follow the methodology described in \cite{RS}. This yields 
\begin{equation}
    \Theta_{GPF}=\Omega-\Omega_{GPF},\quad\quad \Omega_{GPF}=\Omega\,\sqrt{f(r)-\Omega^2\,r^2}.\label{dd10}
\end{equation}

Using (\ref{dd8}) and substituting the metric function $f(r)$, we find
\begin{eqnarray}
    \Theta_{GPF}=\sqrt{\frac{f'(r)}{2\,r}}\,\left[1-\sqrt{\frac{2\,f(r)-r\,f'(r)}{2}}\right]=\sqrt{\frac{M}{r^3}+\frac{b}{2\,r\,(r+r_s)^2}-\frac{\Lambda}{3}}\,\left[1-\sqrt{1-\alpha-\frac{3\,M}{r}-\frac{b\,\left(\frac{3\,r}{2}+r_s\right)}{(r+r_s)^2}}\right].\label{dd11}
\end{eqnarray}

The expression (\ref{dd11}) shows that the geodesic precession frequency is influenced by several parameters. These include the cosmic string parameter$\alpha$, the density $\rho_s$ of Hernquist DM, the BH mass $M$, and the cosmological constant $\Lambda$.

In the limit $\rho_s=0$, corresponding to the absence of the HDM halo, the geodesics precession frequency from Eq. (\ref{dd11}) reduces as follows: 
\begin{eqnarray}
    \Theta_{GPF}=\sqrt{\frac{M}{r^3}-\frac{\Lambda}{3}}\,\left[1-\sqrt{1-\alpha-\frac{3\,M}{r}}\right].\label{dd12}
\end{eqnarray}
Equation (\ref{dd12}) is the geodesic precession frequency of the timelike particle in the Letelier-AdS BH solution, which further reduces to the Schwarzschild result provided $\alpha=0$.

By comparing Eqs. (\ref{dd11}) and (\ref{dd12}), it is evident that the geodesic precession frequency of timelike particles decreases due to the combined effects of the halo of DM and the cloud of strings. Consequently, the result is modified by these physical contributions in comparison to the standard Schwarzschild BH solution.

\section{Perturbations of SHDMHCS} \label{sec4}

In gravity theories, perturbation theory plays a crucial role, particularly in the study of gravitational wave spectroscopy \cite{LIGO} and the stability of physical systems under small disturbances. Perturbation equations can often be written as second-order differential equations, which can be solved using numerical methods. Imposing appropriate boundary conditions-ingoing waves at the horizon and outgoing waves at spatial infinity-leads to discrete complex eigenvalues. These represent the QNMs, which characterize the BH's response to perturbations. Notably, QNM frequencies depend solely on the BH's properties, not on the initial perturbation. 

The study of perturbations with integer spin-such as scalar, vector, and gravitational perturbations-in BH spacetimes is a long-standing and extensively investigated problem. A significant body of literature has been devoted to this topic. In particular, it has been shown that the corresponding equations of motion can be recast into a compact form analogous to Schrödinger-like differential equations (see, for instance, \cite{isz26}). In the present work, we focus on spin-0 and spin-1/2 field perturbations in the selected BH spacetime and analyze the resulting behavior. In future studies, our aim is to extend our analysis to higher-spin fields, including spin-1, spin-3/2, spin-2, and spin-5/2 perturbations within the same BH background. Additionally, we plan to employ alternative methods-such as the asymptotic iteration method  \cite{HC}-to compute QNMs more effectively \cite{HTC}. 

\subsection{Scalar Perturbations: Spin-0 field}

In the subsection, we investigate the dynamics of a massless scalar field in the background of a selected SHDMHCS solution, incorporating a quintessence field surrounded by a cloud of strings. We begin by explicitly deriving the massless Klein–Gordon equation, reducing it to a second-order differential equation that governs the evolution of the scalar field within the given spacetime geometry. Among all types of perturbation, scalar perturbations play a particularly crucial role in assessing BH stability. They have been extensively studied in various BH backgrounds, yielding significant insights into the nature of BH stability and the propagation of test fields (see, {\it e.g.}, Refs. \cite{NPB, AHEP2,CJPHY,EPJC,AHEP3,AHEP4,AHEP5,AHEP6, AHEP7, AHEP8}, and references therein).

The massless scalar field wave equation is described by the Klein-Gordon equation as follows Refs. \cite{NPB,AHEP1,AHEP2,CJPHY,EPJC, AHEP3, AHEP4, AHEP5, AHEP6}:
\begin{equation}
\frac{1}{\sqrt{-g}}\,\partial_{\mu}\left[\left(\sqrt{-g}\,g^{\mu\nu}\,\partial_{\nu}\right)\,\Psi\right]=0,\label{ff1}    
\end{equation}
where $\Psi$ is the wave function of the scalar field, $g_{\mu\nu}$ is the covariant metric tensor, $g=\det(g_{\mu\nu})$ is the determinant of the metric tensor, $g^{\mu\nu}$ is the contravariant form of the metric tensor, and $\partial_{\mu}$ is the partial derivative with respect to the coordinate systems.

Before, writing explicitly, performing the following coordinate change (called tortoise coordinate). 
\begin{eqnarray}
    dr_*=\frac{dr}{f(r)}\label{ff2}
\end{eqnarray}
into the line-element Eq. (\ref{bb1}) results
\begin{equation}
    ds^2=f(r_*)\,\left(-dt^2+dr^2_{*}\right)+H^2(r_*)\,\left(d\theta^2+\sin^2 \theta\,d\phi^2\right),\label{ff3}
\end{equation}
where $f(r_*)$ and $h(r_*)$ are functions of $r_*$. 

Let us consider the following scalar field wave function ansatz form
\begin{equation}
    \Psi(t, r_{*},\theta, \phi)=\exp(i\,\omega\,t)\,Y^{m}_{\ell} (\theta,\phi)\,\frac{\psi(r_*)}{r_{*}},\label{ff4}
\end{equation}
where $\omega$ is the temporal frequency (possibly complex), $\psi (r)$ is a propagating scalar field in the candidate spacetime, and $Y^{m}_{\ell} (\theta,\phi)$ are the spherical harmonics. 

\begin{figure}[ht!]
    \centering
    \subfloat[$r_s=0.5\,M,\,M^2\,\rho_s=1$]{\centering{}\includegraphics[width=0.4\linewidth]{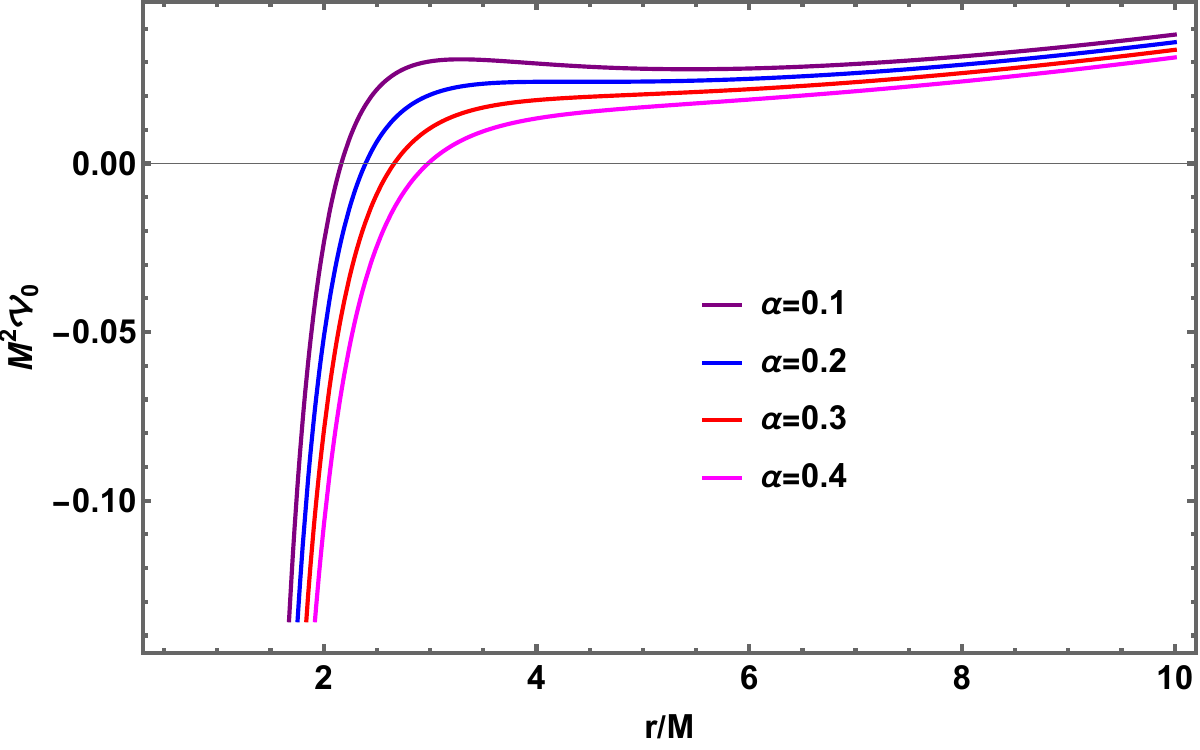}}\quad\quad
    \subfloat[$\alpha=0.1,\,M^2\,\rho_s=1$]{\centering{}\includegraphics[width=0.4\linewidth]{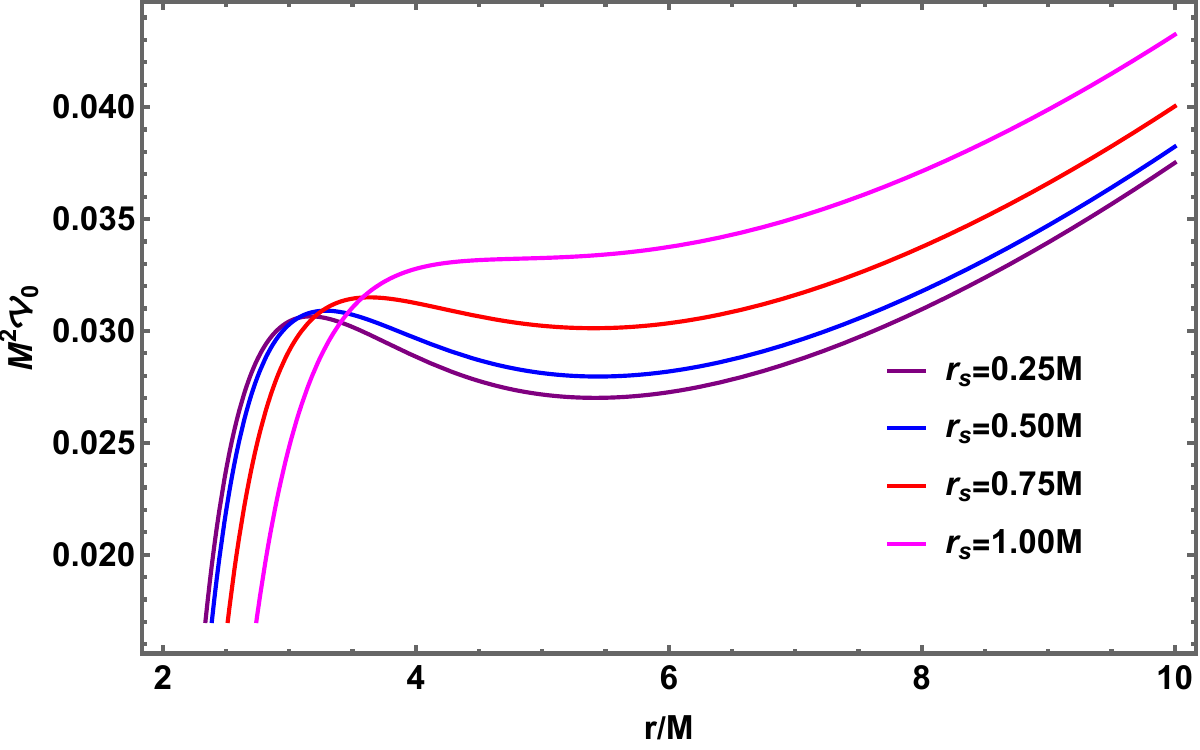}}\\
    \subfloat[$\alpha=0.1,\,r_s=1\,M$]{\centering{}\includegraphics[width=0.4\linewidth]{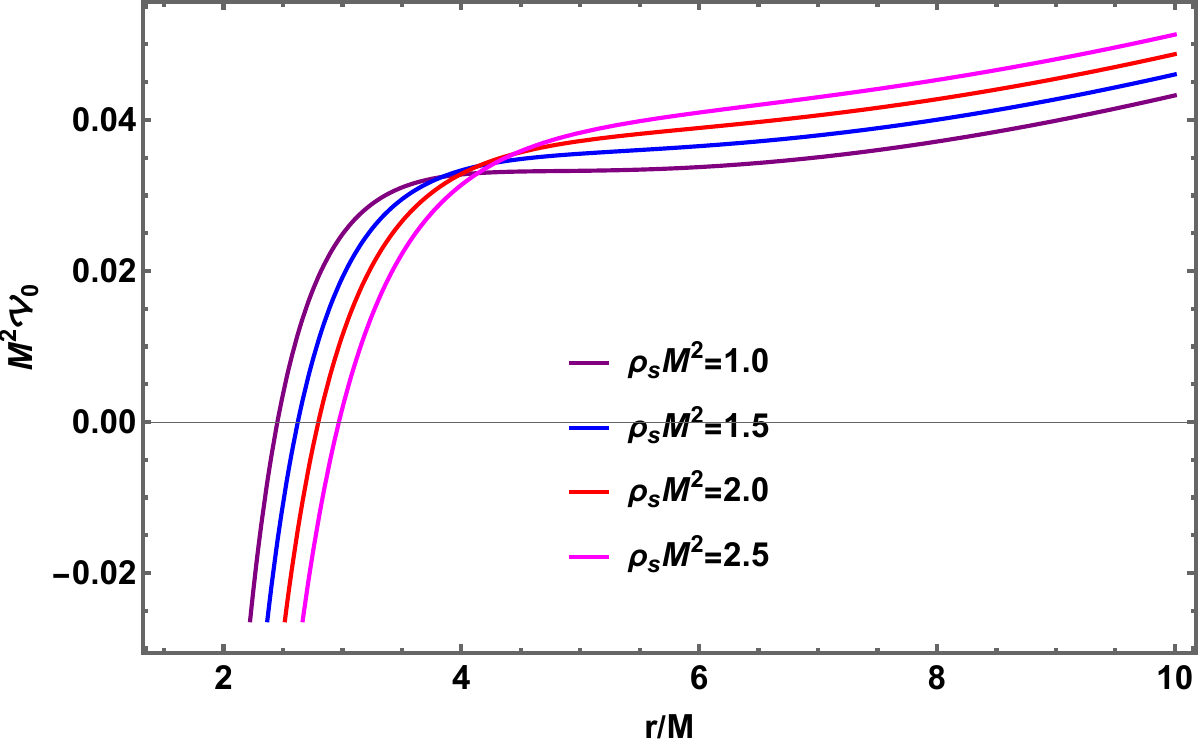}}\quad\quad
    \subfloat[$M^2\,\rho_s=1$]{\centering{}\includegraphics[width=0.4\linewidth]{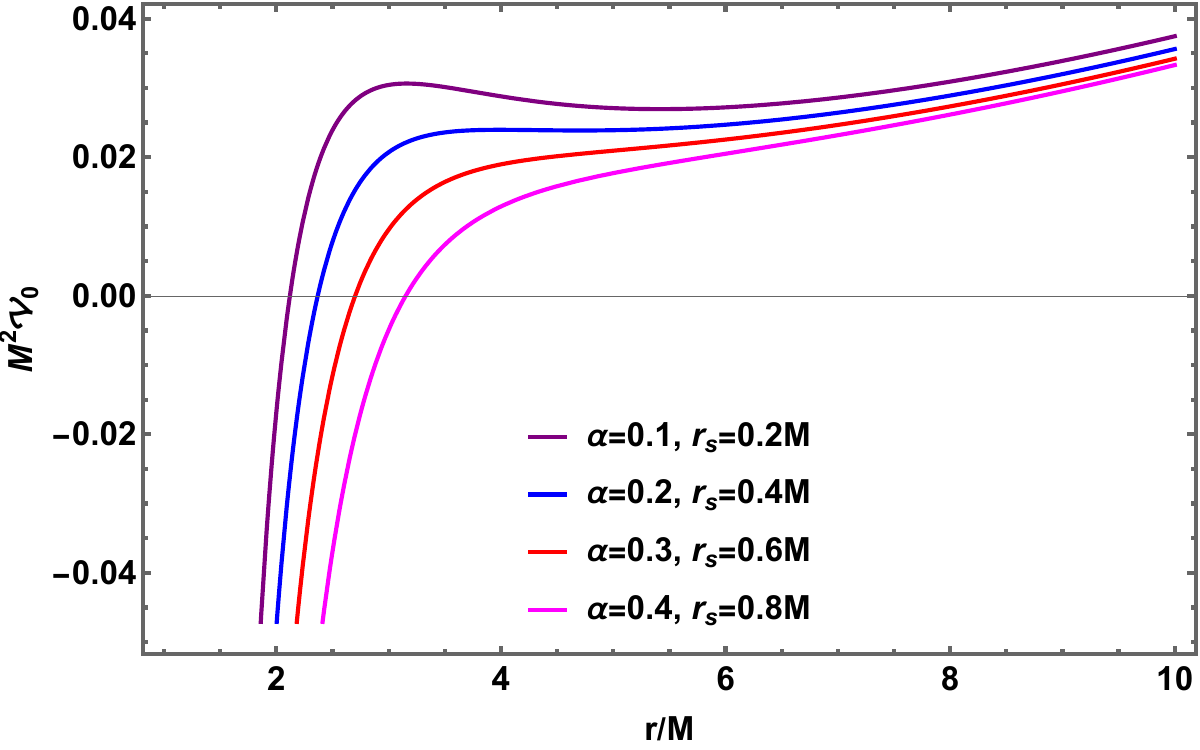}}\\
    \subfloat[$r_s=0.5\,M$]{\centering{}\includegraphics[width=0.4\linewidth]{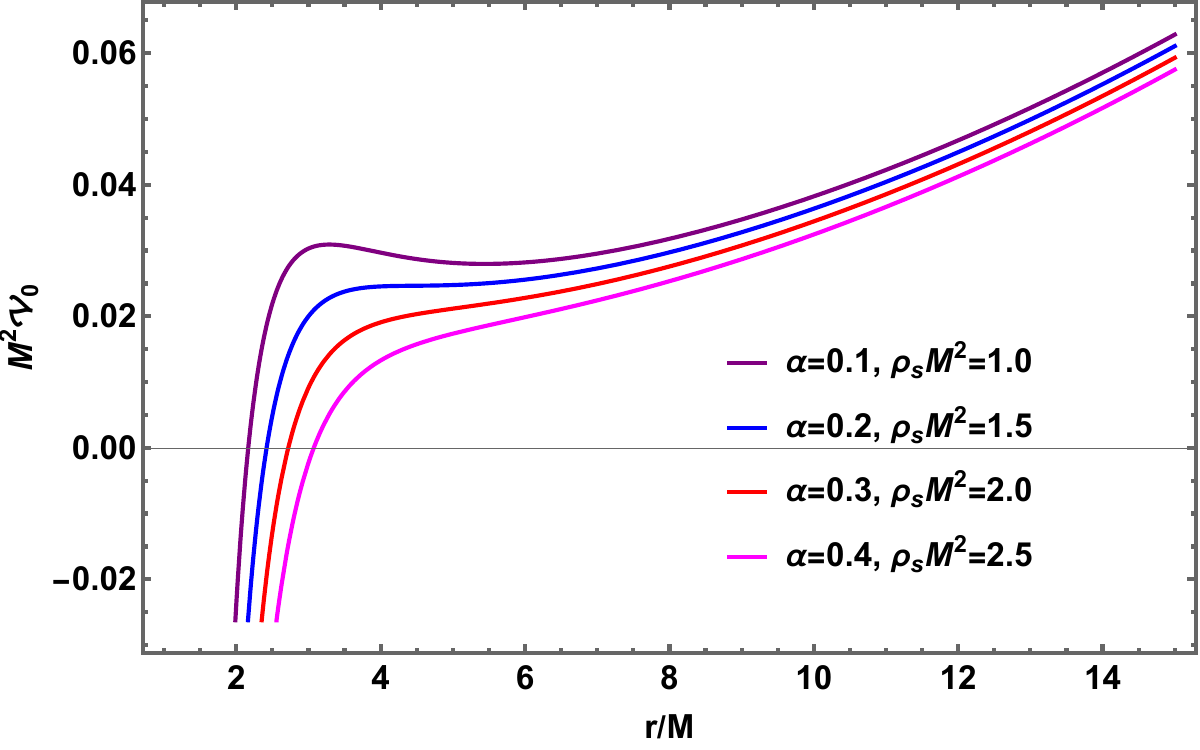}}\quad\quad
    \subfloat[]{\centering{}\includegraphics[width=0.4\linewidth]{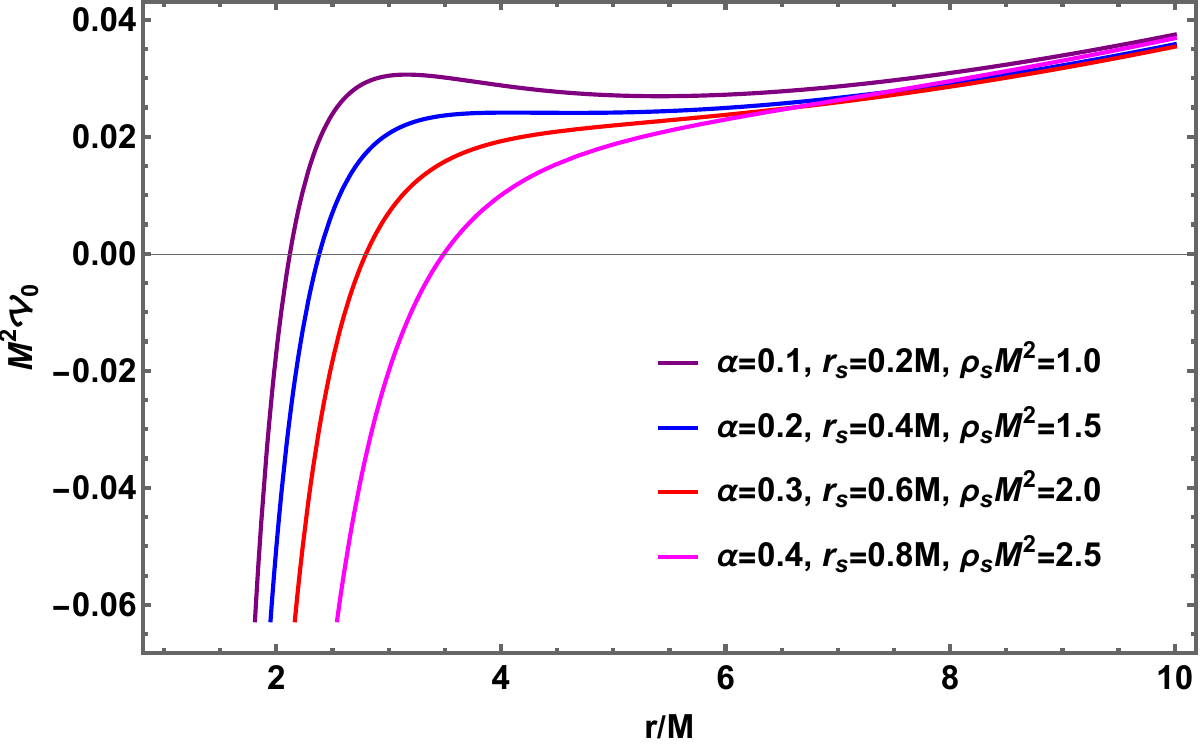}}
    \caption{\footnotesize Behavior of the scalar perturbative potential is shown for varying values of the string parameter \(\alpha\), the core radius \(r_s\), and the DM halo density \(\rho_s\). Sub-figures illustrate the individual effects of (a) \(\alpha\), (b) \(r_s\), and (c) \(\rho_s\,M^2\), as well as the combined effects of (d) \(\alpha\) and \(r_s\), (e) \(\alpha\) and \(\rho_s\,M^2\), and (f) \(\alpha\), \(r_s\), and \(\rho_s\,M^2\) together. Here, we set the multipole number $\ell=0$-state, corresponds to $s$-wave or dominant mode and the dimensionless parameter $k=M\,\sqrt{-\frac{\Lambda}{3}}=0.1$.}
    \label{fig:scalar-potential}
\end{figure}

\begin{figure}[ht!]
    \centering
    \subfloat[$r_s=0.5\,M$]{\centering{}\includegraphics[width=0.45\linewidth]{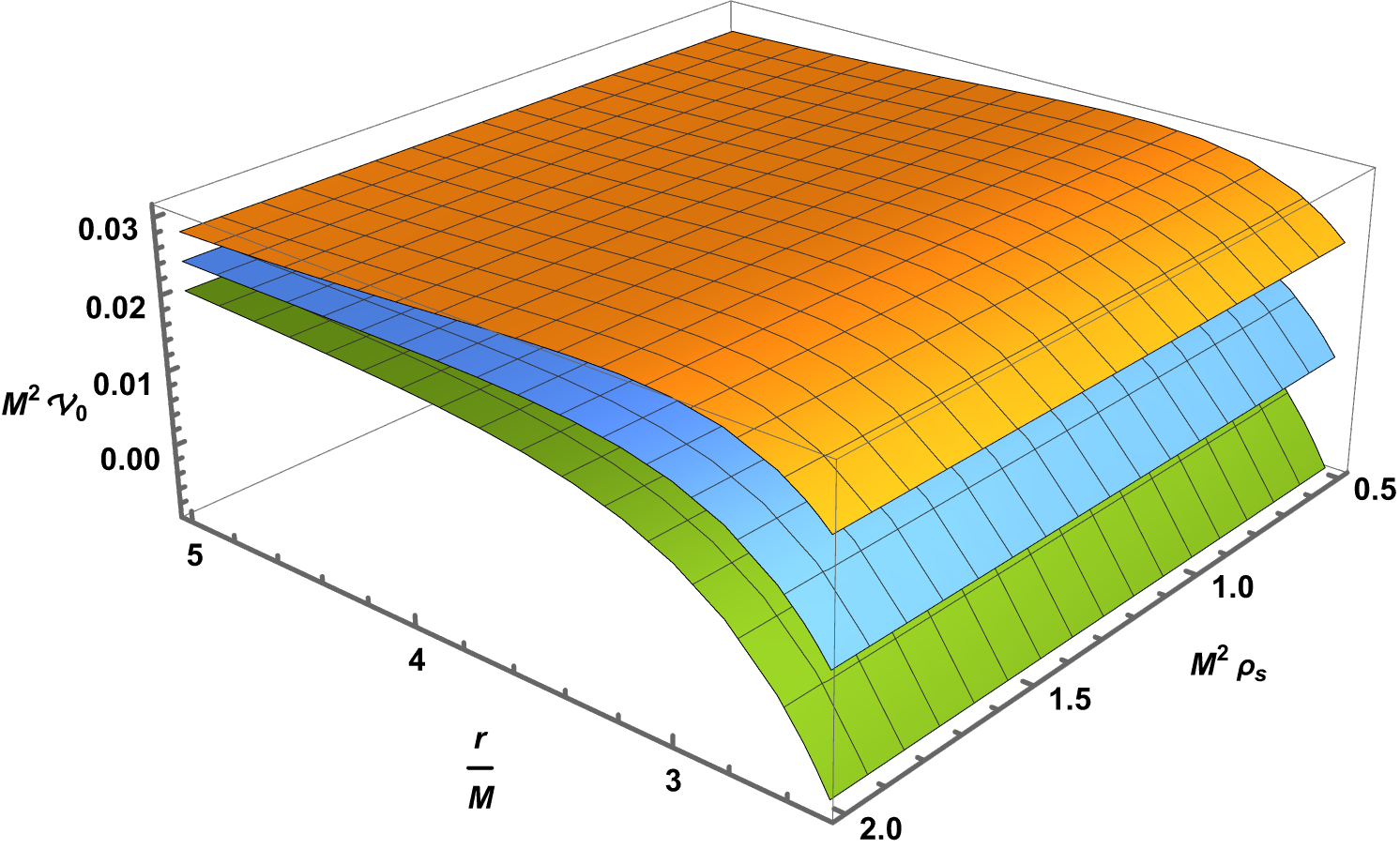}}\quad
    \subfloat[$M^2\,\rho_s=1$]{\centering{}\includegraphics[width=0.45\linewidth]{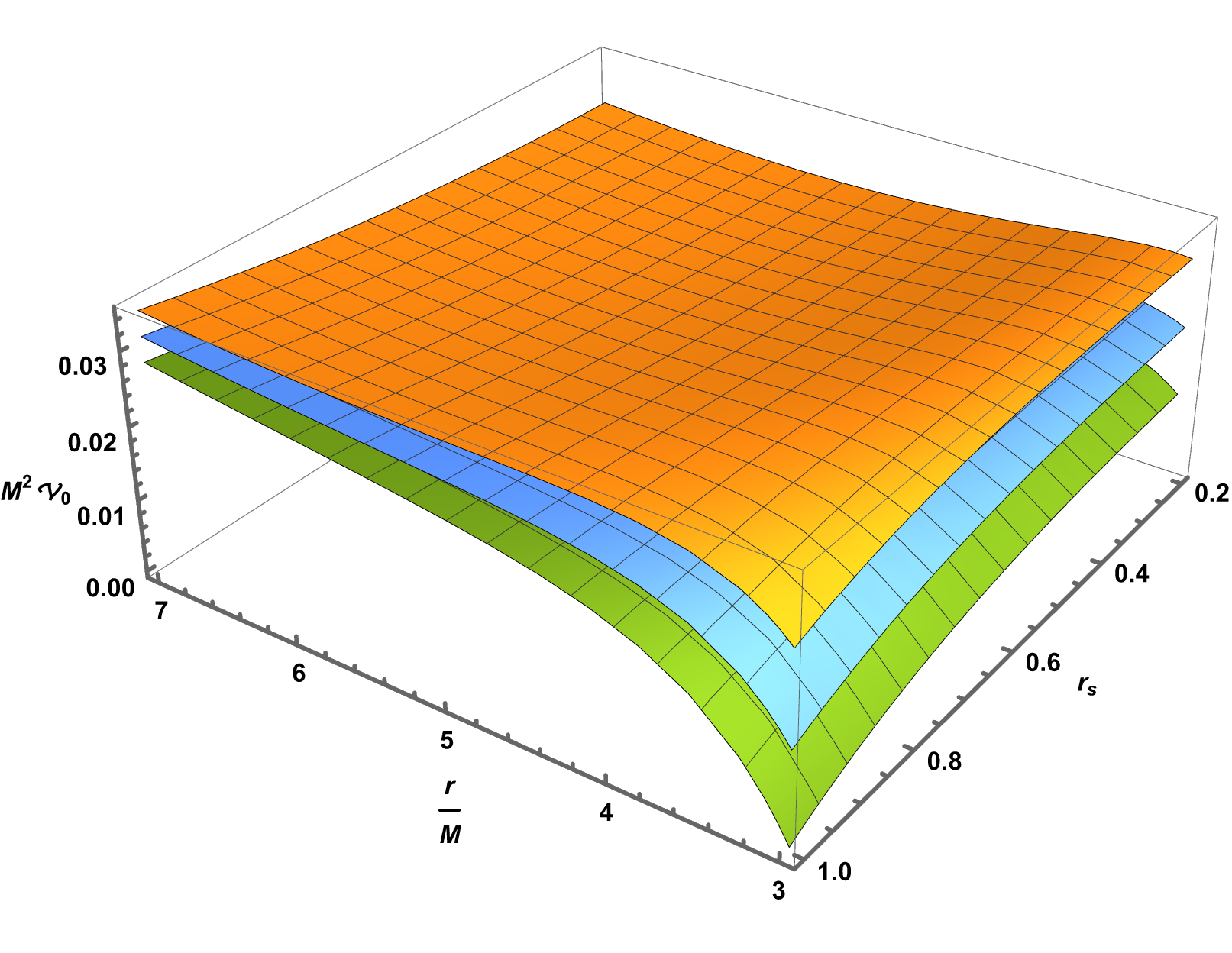}}
    \caption{\footnotesize Three-dimensional plot of $M^2\,\mathcal{V}_0$:  the qualitative features of spin-zero scalar perturbative potential for the dominant multipole number $\ell=0$ is shown by varying  values of the string parameter $\alpha$. Here, we set the dimensionless parameter $k=M\,\sqrt{-\frac{\Lambda}{3}}=0.1$. Orange curve: $\alpha=0.1$, blue: $\alpha=0.2$, green: $\alpha=0.3$.}
    \label{fig:scalar-potential-contour}
\end{figure}

With these, we can write the wave equation (\ref{ff1}) in the following Schrodinger-like wave equation:
\begin{equation}
    \frac{\partial^2 \psi(r_*)}{\partial r^2_{*}}+\left(\omega^2-\mathcal{V}_0\right)\,\psi(r_*)=0,\label{ff5}
\end{equation}
where the scalar perturbative potential $\mathcal{V}_0$ is given by 
\begin{eqnarray} 
\mathcal{V}_0&=&\left(\frac{\ell\,(\ell+1)}{r^2}+\frac{f'(r)}{r}\right)\,f(r)\nonumber\\
&=&\left(\frac{\ell\,(\ell+1)}{r^2}+\frac{2\,M}{r^3}+\frac{b}{r\,(r+r_s)^2}-\frac{2\,\Lambda}{3}\right)\,\left(1-\alpha-\frac{2\,M}{r}-\frac{b}{r+r_s}-\frac{\Lambda}{3}\,r^2\right).\label{ff6}
\end{eqnarray}

From the above expression (\ref{ff6}), it is evident that the scalar perturbative potential is influenced by several key parameters. These include the cosmic string parameter $\alpha$, the core radius of halo $r_s$, the density of HDM $\rho_s$, as well as the BH mass $M$ and the cosmological constant $\Lambda$. Moreover, this perturbative potential changes with the multipole number $\ell \geq 0$.

Now, we examine a special case corresponding to $\ell=0$, representing an $s$-wave. Moreover, we define dimensionless parameters $r/M=x$, $M^2\,\rho_s=y$,  $r_s/M=z$, and $k=M\,\sqrt{-\frac{\Lambda}{3}}$. Therefore, the scalar perturbative potential is from Eq. (\ref{ff6}) reduces as (setting $8\,\pi=1$)
\begin{eqnarray} 
M^2\,\mathcal{V}_0\Big{|}_{\ell=0}=\left(\frac{2}{x^3}+\frac{1}{2}\,\frac{y\,z^3}{x\,(x+z)^2}+2\,k\right)\,\left(1-\alpha-\frac{2}{x}-\frac{y\,z^3}{x+z}+k^2\,x^2\right).\label{ff7}
\end{eqnarray}

In Figure~\ref{fig:scalar-potential}, we present a series of plots that illustrate the behavior of the scalar perturbative potential as a function of \( r/M \), considering variations in the cosmic string parameter \( \alpha \), the radius of the core \( r_s \) and the density of the core \( M^2\,\rho_s \) of the HDM halo. In all panels except panels (b)-(c), it is evident that increasing the values of \( \alpha \), \( r_s/M \), \(M^2\, \rho_s \), or their combinations leads to a decrease in the perturbative potential.

\subsection{Electromagnetic perturbation}

In this subsection, we present a preliminary discussion on electromagnetic perturbation in the background of a selected SHDMHCS solution, incorporating a quintessence field surrounded by a cloud of strings.

\begin{figure}[ht!]
    \centering
    \subfloat[$r_s=0.5\,M,\,M^2\,\rho_s=1$]{\centering{}\includegraphics[width=0.4\linewidth]{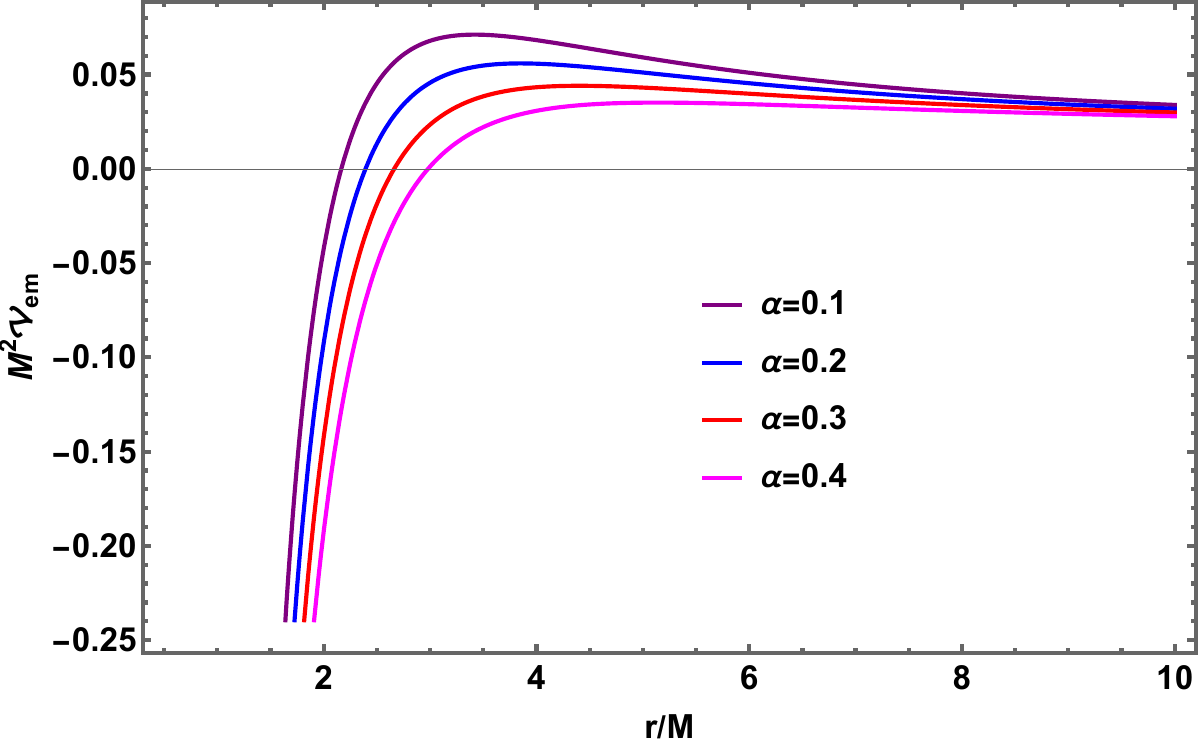}}\quad\quad
    \subfloat[$\alpha=0.1,\,M^2\,\rho_s=1$]{\centering{}\includegraphics[width=0.4\linewidth]{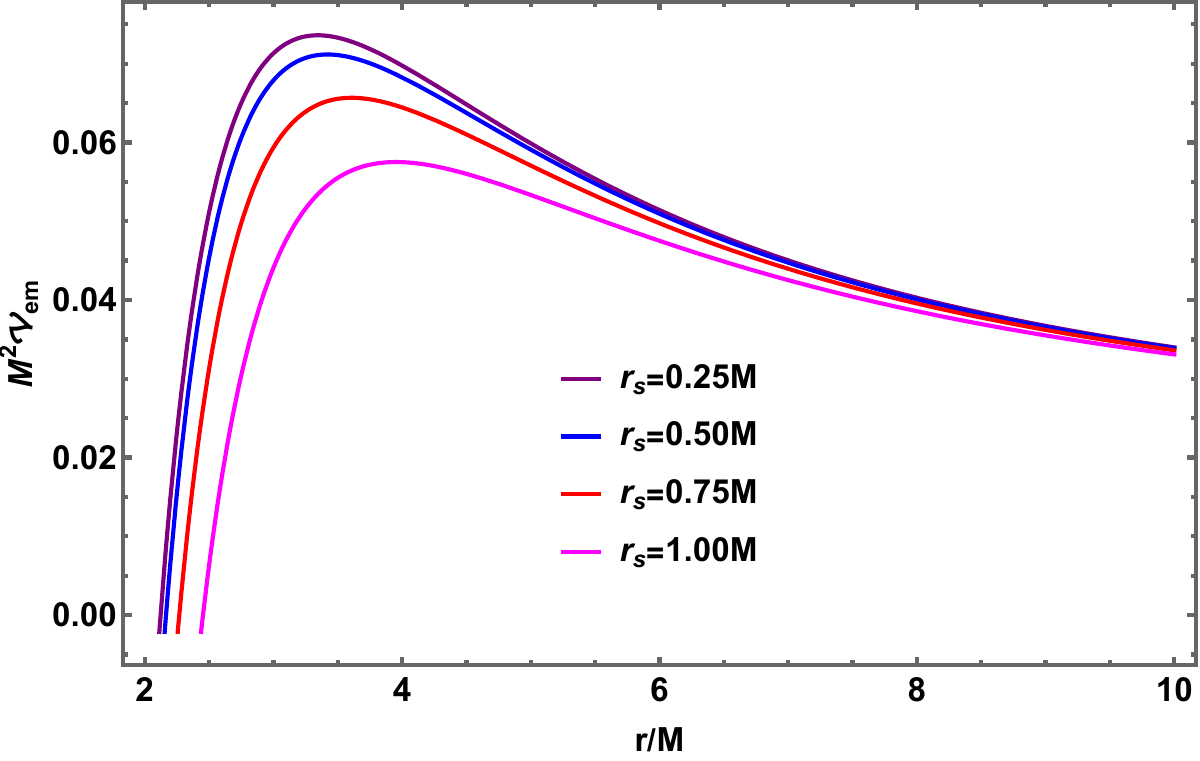}}\\
    \subfloat[$\alpha=0.1,\,r_s=1\,M$]{\centering{}\includegraphics[width=0.4\linewidth]{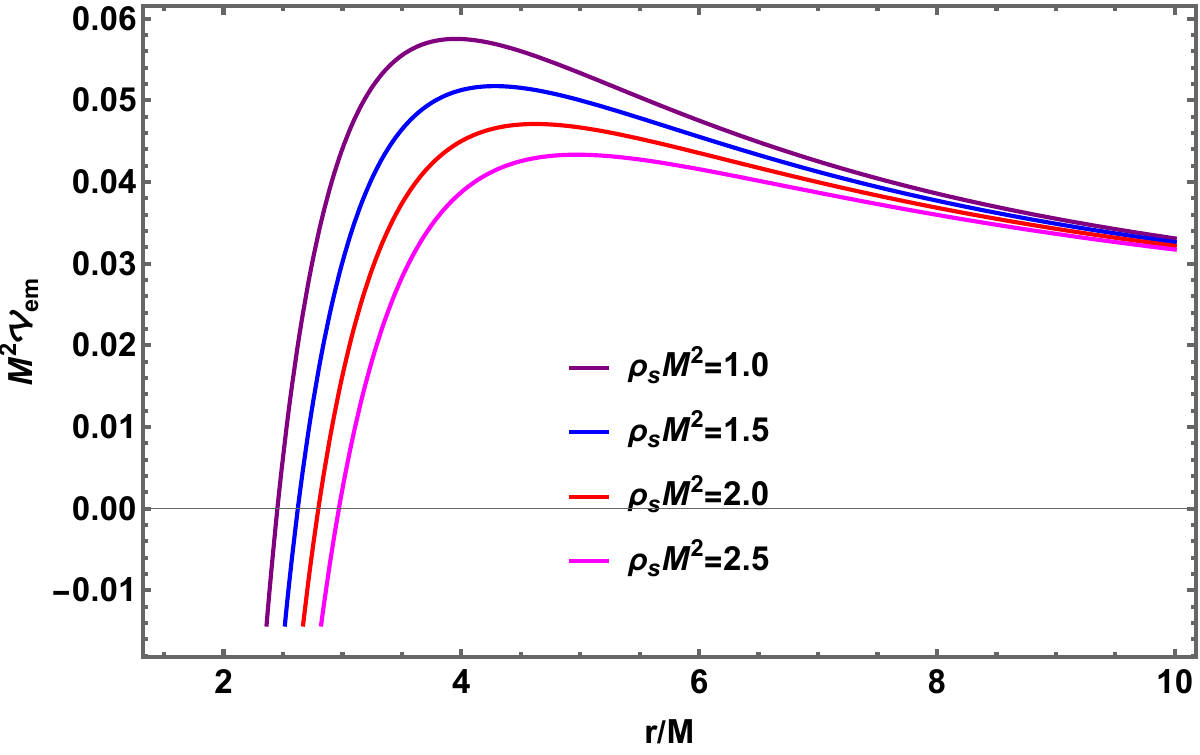}}\quad\quad
    \subfloat[$M^2\,\rho_s=1$]{\centering{}\includegraphics[width=0.4\linewidth]{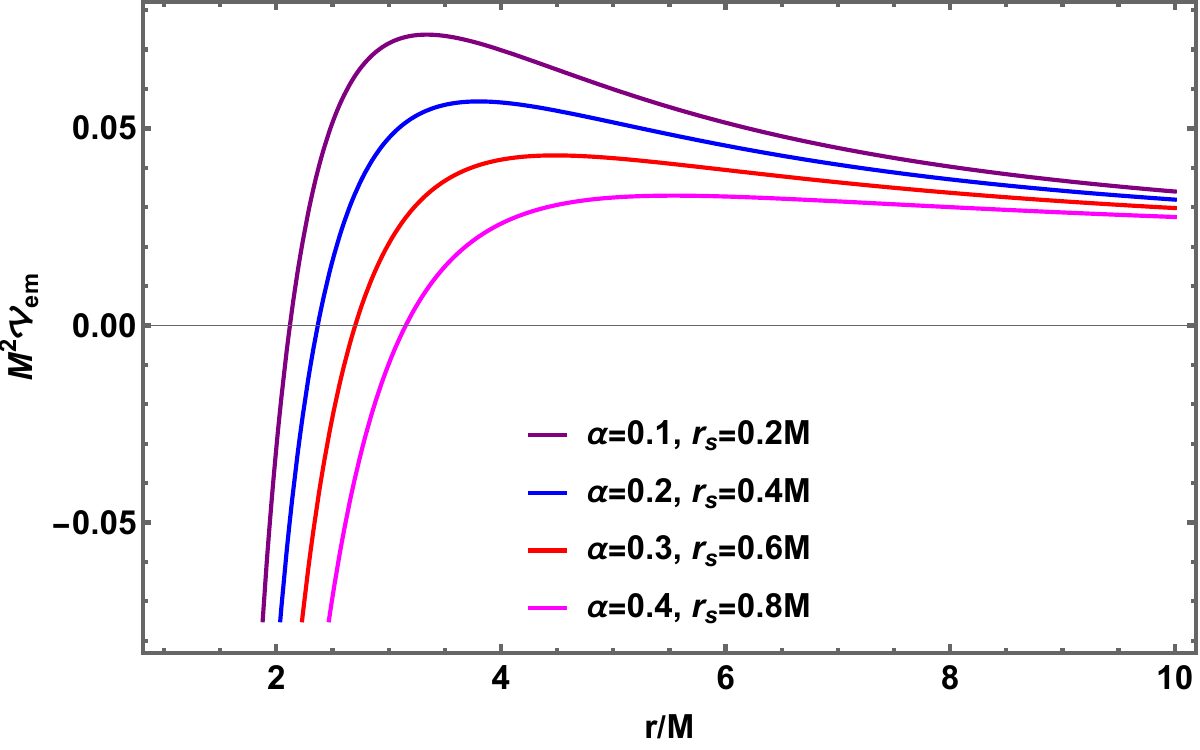}}\\
    \subfloat[$r_s=0.5\,M$]{\centering{}\includegraphics[width=0.4\linewidth]{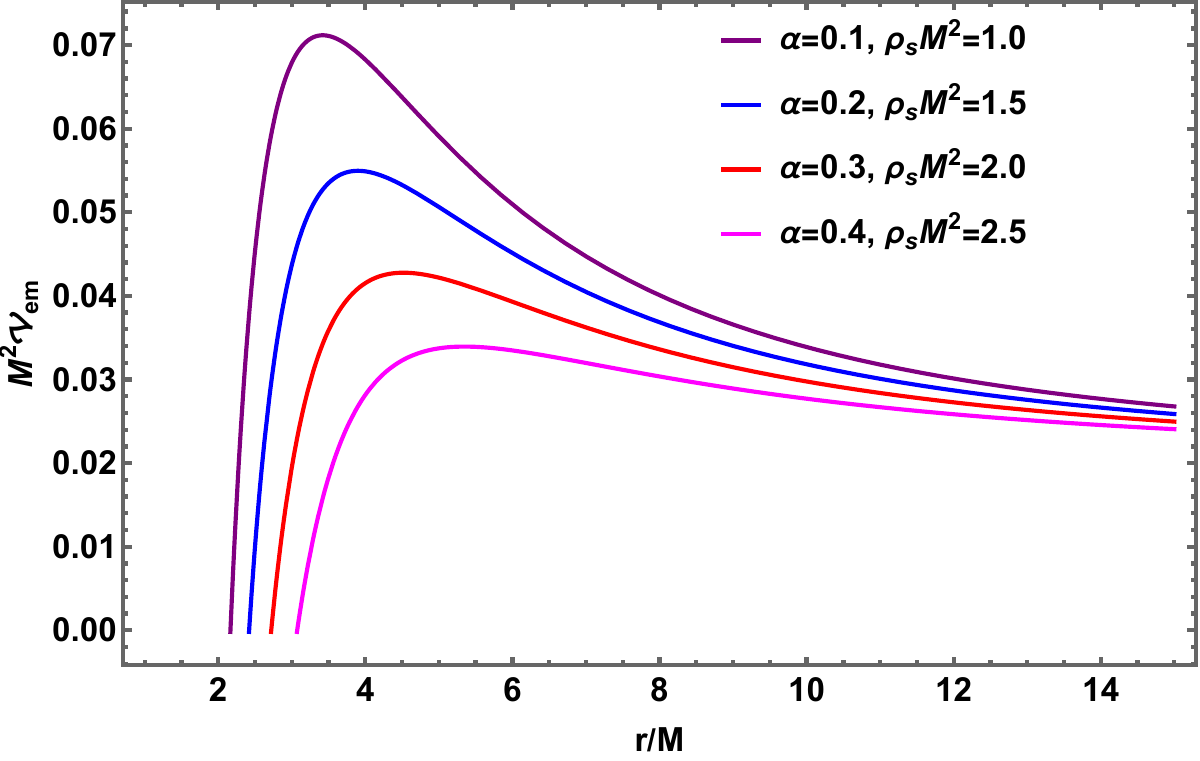}}\quad\quad
    \subfloat[]{\centering{}\includegraphics[width=0.4\linewidth]{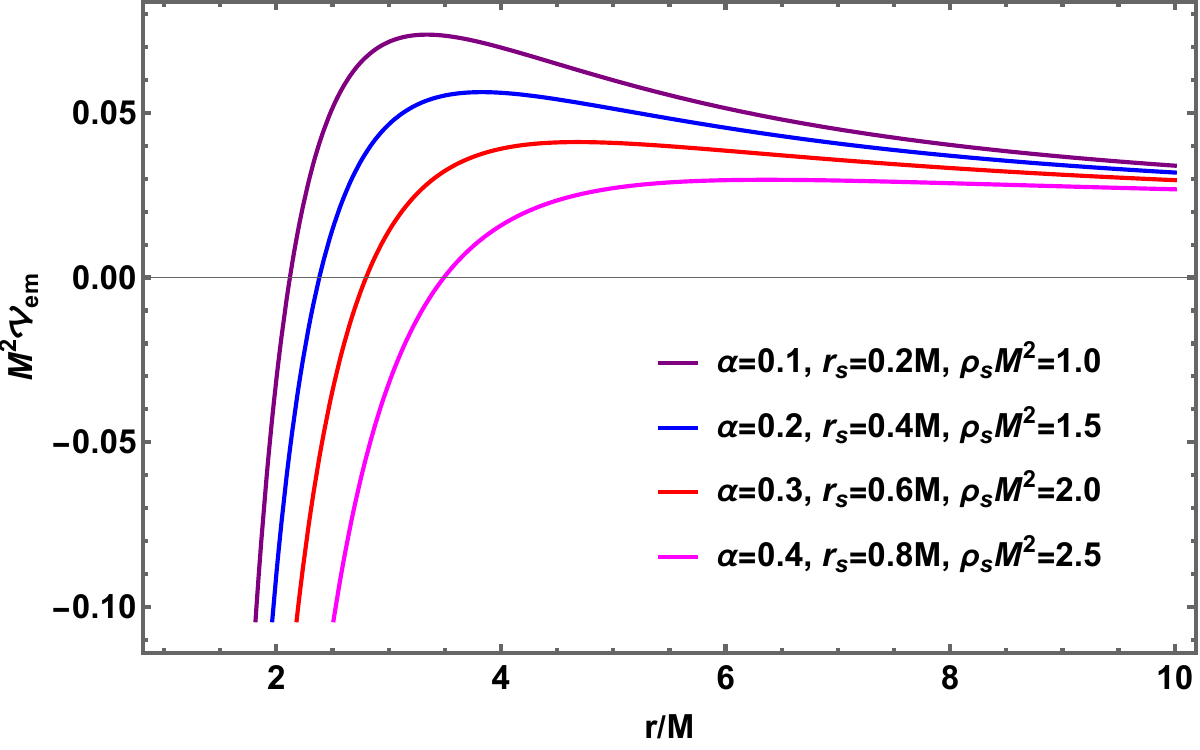}}
    \caption{\footnotesize Behavior of the electromagnetic perturbative potential $M^2\,\mathcal{V}_\text{em}$ is shown for varying values of the string parameter \(\alpha\), the core radius \(r_s\), and the DM halo density \(\rho_s\). Sub-figures illustrate the individual effects of (a) \(\alpha\), (b) \(r_s\), and (c) \(\rho_s\,M^2\), as well as the combined effects of (d) \(\alpha\) and \(r_s\), (e) \(\alpha\) and \(\rho_s\,M^2\), and (f) \(\alpha\), \(r_s\), and \(\rho_s\,M^2\) together. Here, we set the multipole number $\ell=1$ and the dimensionless parameter $k=M\,\sqrt{-\frac{\Lambda}{3}}=0.1$}
    \label{fig:em-potential}
\end{figure}

The electromagnetic field in curved spacetime follows the equation \cite{XZ}
\begin{equation}
    \frac{1}{\sqrt{-g}}\,\partial_{\mu}\left(\sqrt{-g}\,g^{\mu\sigma}\,g^{\nu\tau}\,F_{\sigma\tau}\,\partial_{\nu}\right)=0,\label{em1}
\end{equation}
where $F_{\sigma\tau}=\partial_{\sigma}\,A_{\tau}-\partial_{\tau}\,A_{\sigma}$ is the electromagnetic field tensor, with $A_{\mu}$ being the electromagnetic four-vector potential.  In the Regge-Wheeler-Zerilli formalism, one may decompose $A_{\mu}$ in terms of the scalar and vector spherical harmonics as follows:
\begin{equation}
    A_{\mu}=\sum_{\ell,m}\,e^{-i\,\omega\,t}\,\left(\left[\begin{array}{c}
         0\\
         0\\
         \psi_{em}(r)\,{\bf S}_{\ell,m}
    \end{array}\right]+\left[\begin{array}{c}
         j^{\ell,m}(r)\,Y^{m}_{\ell}\\
         h^{\ell,m}(r)\,Y^{m}_{\ell}\\
         k^{\ell,m}(r)\,{\bf Y}^{m}_{\ell}\\
    \end{array}\right] 
    \right),\label{em2}
\end{equation}
where $Y^{m}_{\ell}$ is the scalar spherical harmonics and (${\bf S}_{\ell,m}, {\bf Y}_{\ell,m}$) are the vector harmonics given by
\begin{equation}
    {\bf S}_{\ell,m}=\left(\begin{array}{c}
         \frac{1}{\sin \theta}\,\partial_{\phi}\,Y^{m}_{\ell}\\
         -\sin \theta\,\partial_{\theta}\,Y^{m}_{\ell} 
    \end{array}\right),\quad\quad {\bf Y}_{\ell,m}=\left(\begin{array}{c}
         \partial_{\theta}\,Y^{m}_{\ell}\\
         \partial_{\phi}\,Y^{m}_{\ell} 
    \end{array}\right),\label{em3}
\end{equation}
where $\omega$ is the frequency. Here, $\ell$ and $m$ denote the orbital quantum and azimuthal numbers, respectively. The first term on the right-hand side of Eq. (\ref{em2}) represents the axial mode, while the second term represents the polar mode. It is important to note that the axial and polar modes have parity $(-1)^{\ell+1}$ and $(-1)^{\ell}$, respectively.
Moreover, the polar and axial parts have equal contributions to the final result \cite{JAW,ARF}. Thus, we focus only on the
axial part. 

\begin{figure}[ht!]
    \centering
    \subfloat[$\ell=0,r_s=0.5\,M$]{\centering{}\includegraphics[width=0.4\linewidth]{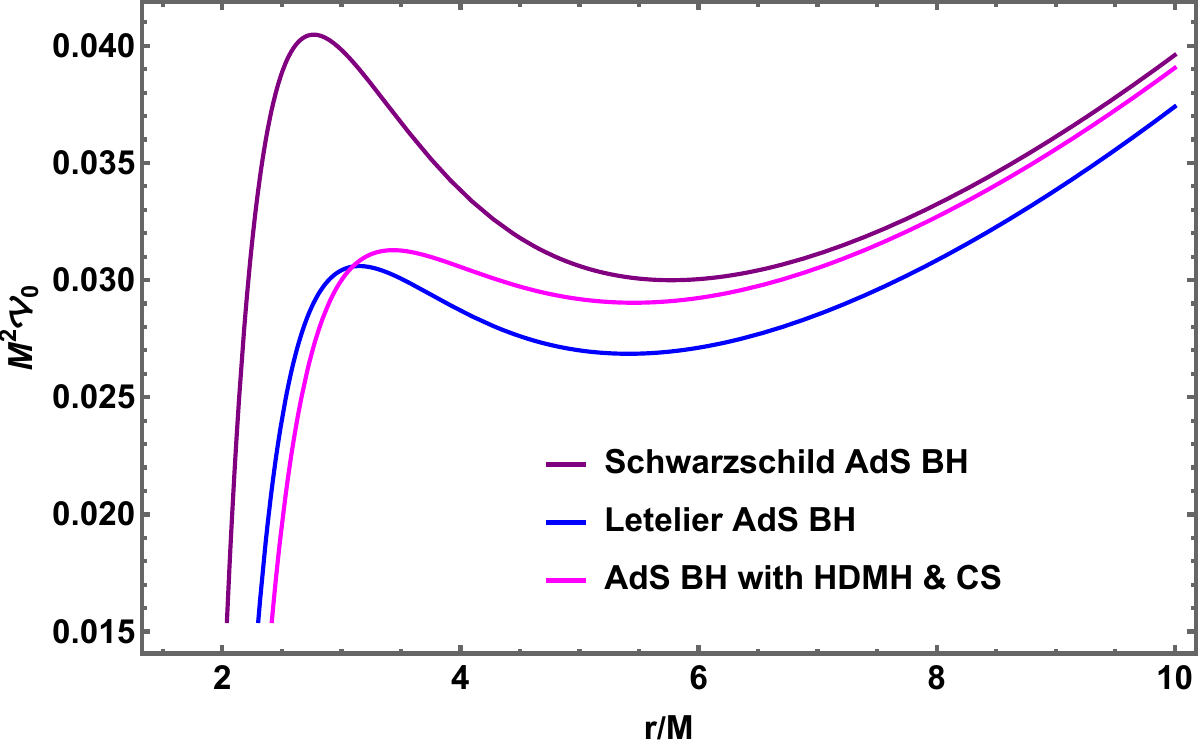}}\quad\quad
    \subfloat[$\ell=1,r_s=0.8\,M$]{\centering{}\includegraphics[width=0.4\linewidth]{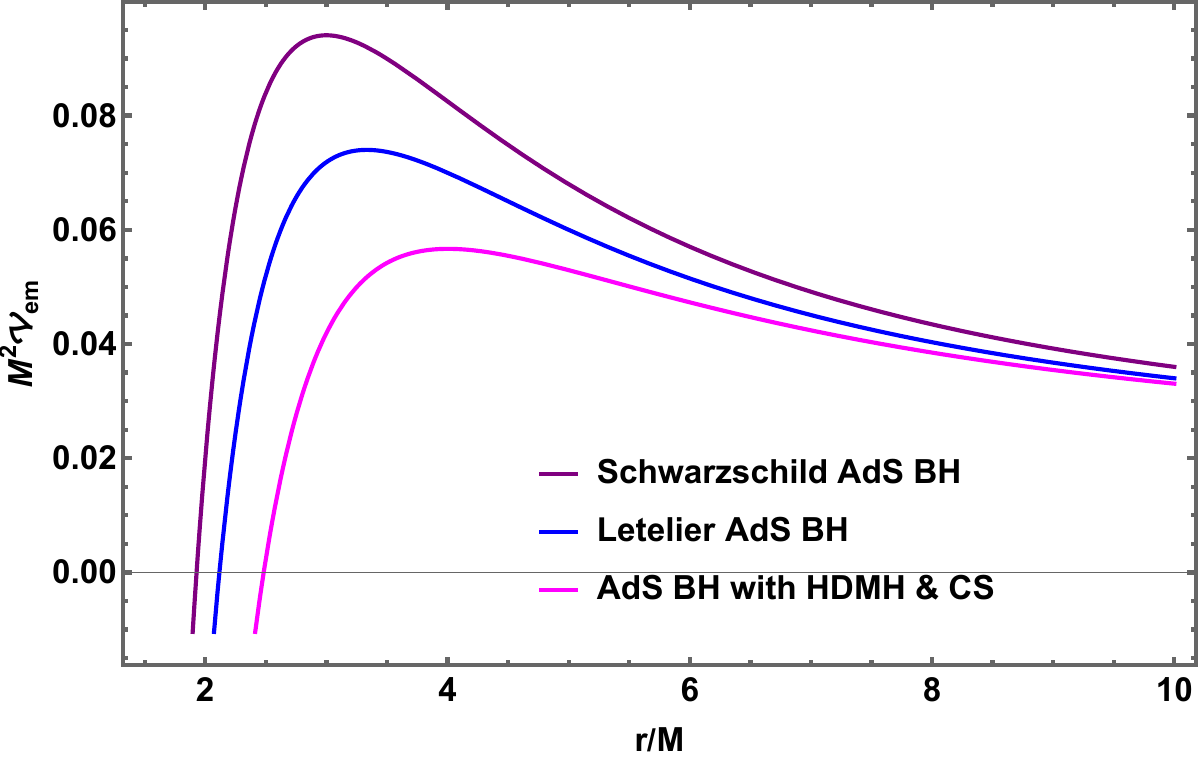}}
    \caption{\footnotesize Comparison of the scalar and electromagnetic potential under different BHs scenario. Here, we set the string parameter $\alpha=0.1$, the dimensionless parameters $M^2\,\rho_s=2$ and $M\,\sqrt{-\frac{\Lambda}{3}}=0.1$.}
    \label{fig:comparison-potential}
\end{figure}

Now, substituting Eq. (\ref{em2}) in Eq. (\ref{em1}) and applying the tortoise coordinate $r_{*}$, the radial part of Eq. (\ref{fermi1}) can be written in the form of Schrodinger-like wave equation as,
\begin{equation}
    \frac{\partial^2 \psi_{em}(r_*)}{\partial r^2_{*}}+\left(\omega^2-\mathcal{V}_{em}\right)\,\psi_{em}(r_*)=0,\label{em4}
\end{equation}
where the effective potential of the electromagnetic field is given by
\begin{equation}
    \mathcal{V}_{em}=\frac{\ell\,(\ell+1)}{r^2}\,f(r)=\frac{\ell\,(\ell+1)}{r^2}\,\left(1-\alpha-\frac{2\,M}{r}-\frac{b}{r+r_s}-\frac{\Lambda}{3}\,r^2\right).\label{em5}
\end{equation}

From the above expression (\ref{em5}), it is evident that the scalar perturbative potential is influenced by several key parameters. These include the cosmic string parameter $\alpha$, the core radius of halo $r_s$, the density of HDM $\rho_s$, as well as the BH mass $M$ and the cosmological constant $\Lambda$. Moreover, this perturbative potential changes with the multipole number $\ell \geq 1$.

In Figure~\ref{fig:em-potential}, we present a series of plots that illustrate the behavior of the electromagnetic perturbative potential as a function of \( r/M \), considering variations in the cosmic string parameter \( \alpha \), the radius of the core \( r_s \), and the density of the core \( M^2\,\rho_s \) of the HDM halo. In all panels, it is evident that increasing the values of \( \alpha \), \( r_s/M \), \(M^2\, \rho_s \), or their combinations, leads to a decrease in the perturbative potential.

In Figure~\ref{fig:comparison-potential}, we compare the perturbative scalar and electromagnetic potentials for different scenarios of BH: Schwarzschild-AdS BH (\( r_s = 0 = \rho_s,\, \alpha = 0 \)); Letelier-AdS BH (\( r_s = 0 = \rho_s,\, \alpha = 0.1 \)); and 
AdS BH with HDMH and CS (\( r_s = 0.5\,M,\, M^2 \rho_s = 2,\, \alpha=0.1 \)). Panel (a) shows the scalar perturbation potential, while panel (b) corresponds to the electromagnetic perturbation. We observe that the scalar potential in the selected BH solution with HDMH and CS lies between those of the Schwarzschild-AdS and Letelier-AdS BHs. In contrast, the electromagnetic potential is reduced compared to both of these cases.

\subsection{Fermionic perturbations: Spin-1/2 field }

In this section, we present a preliminary analysis of spin-1/2 field perturbations through the Dirac equation in the background of the selected SHDMHCS solution, which includes a cloud of strings and quintessence. Spin-1/2 field perturbations have been studied in various BH spacetimes, such as the Schwarzschild background, and are governed by the Dirac equation (for both massive and massless cases) \cite{DRB}, given by:
\begin{equation}
\left[\gamma^{a}\,e^{\mu}_{a}\,(\partial_{\mu}+\Gamma_{\mu})+\mu_0\right]\,\Psi=0,\label{fermi1}
\end{equation}
Where $\mu_0$ is the mass of the Dirac field, and $e^{\mu}_{a}$ is the inverse of the tetrad $e^{a}_{\mu}$ defined by the metric tensor $g_{\mu\nu}$, as $g_{\mu\nu}=\eta_{ab}\, e^{a}_{\mu}\, e^{b}_{\nu}$ with $\eta_{ab}$ being the Minkowski metric, $\gamma^{a}$ are the Dirac matrices, and $\Gamma_{\mu}$ is the spin connection given by
\begin{equation}
    \Gamma_{\mu}=\frac{1}{8}\,[\gamma^{a},\gamma^{b}]\,e^{\nu}_{a}\,e_{b\,\nu;\mu}.\label{fermi4}
\end{equation}

For the spherically symmetric BH background given by Eq. (\ref{aa1}), the equation of motion for massless spin-1/2 particles can be reduced to a one-dimensional Schr\"{o}dinger-like wave equation \cite{DRB}:
\begin{equation}
\frac{d^{2}\psi_{\pm 1/2}}{dr_{\ast }^{2}}+\left( k^{2}-\mathcal{V}_{\pm 1/2}\right)\, \psi_{\pm 1/2}=0,  \label{fermi2}
\end{equation}
where the tortoise coordinate $r_{\ast }$ is defined earlier and the effective potential for spin-half particle is given by:
\begin{equation}
\mathcal{V}_{\pm 1/2}=\frac{\left( \frac{1}{2}+\ell\right) }{r^{2}}\,\left[\ell+\frac{1}{2} \pm \frac{r\,f^{\prime }(r)}{2\sqrt{f(r)}}\mp \sqrt{f(r)}\right]\,f(r),\quad \ell \geq 1.\label{fermi3}
\end{equation}
where $\ell$ is the standard spherical harmonics indices. 

By substituting the metric function \( f(r) \) into the expression (\ref{fermi3}), one can analyze how the core radius \( r_s \), the halo density \( \rho_s \), and the string parameter \( \alpha \) influence the effective perturbative potential of the fermionic field.

\section{Shadow of SHDMHCS} \label{sec5}

To study the shadow cast by the SHDMHCS, we need to derive the equations of motion for photons (null geodesics) in this spacetime. The DM halo modifies the metric function, typically making the shadow larger or smaller depending on the DM density profile. The string cloud tends to increase the effective mass of the BH, which can enlarge the shadow. \\ To obtain the equation of the photon sphere we consider the  effective potential for null geodesics (\ref{cc1}) as well as the following condition: 
\begin{equation}
\left. V_{eff}\left( r\right) \right\vert _{r=r_{ph}}=0, \,\,\,
\left. \frac{d}{dr}V_{eff}\left( r\right) \right\vert _{r=r_{ph}}=0, \,\,\,
\text{and}  \,\, \left. \frac{d^{2}}{dr^{2}}V_{eff}\left( r\right) \right\vert_{r=r_{p}}<0, \label{con1}
\end{equation}
which indicates a maxima for the effective potential at $r=r_{ph}$. Thus, 
\begin{equation}
    3M(r_{ph}+r_s)^2+r_{ph}\left( (\alpha-1)(r_{ph}+r_s)^2-\frac{1}{2}b(3\,r_{ph}+2\,r_s)  \right)=0.
\end{equation}
Analytical solution for the photon sphere is 
\begin{equation}
    r_{ph}=\frac{A}{3B}-\frac{(   (1+i\sqrt{3})\left( 4A^2+12BC  \right) }{   6(2)^{3/2}B\left( D +\sqrt{D}+4A^2-12BC\right)^{1/3}}-\frac{   (1+i\sqrt{3}) \left( D +\sqrt{D}+4A^2-12BC\right)^{1/3} }{\left( 12(2)^{3/2}B  \right)},\label{phton3}
\end{equation}
where \begin{equation}
    A=6\,r_s^3\,\pi\,\rho_s-2\,r_s\,B-3,\,\,\,B=\alpha-1,\,\,\,C=4\,r_s^4\,\pi\,\rho_s-\,r_s\,B-6r_s,
\end{equation}
and \begin{equation}
    D=\left(16\left( r_s\,B-3   \right)^3    +9\,r_s^3\,\pi\,\rho_s(18+r_s^2\,B^2)-108r_s^2\,\pi^2\,\rho_s^2(3+r_s\,B)+216r_s^9\,\pi^3\,\rho_s^3 \right).
\end{equation}
In the limit when there are no DM and CS ($\alpha=0=r_s=\rho_s$), then Eq. (\ref{phton3}) reduces to $r_{ph}=3M$. 

Our shadow radius $R_s$ can be calculated directly from the photon sphere radii using  \begin{equation}
    R_s=\frac{r_{ph}}{\sqrt{f(r_{ph})} }.\label{shadeq1}
\end{equation}

Figure \ref{figph1} shows three-dimensional visualizations of the photon sphere radius (left panel) and shadow radius (right panel) influenced by core density, core radius, and CS parameter.  Our analysis shows that raising $\alpha$ expands both radii $(r_{ph},R_s)$. In contrast, core density and radius have an opposite effect: they reduce both.

\begin{figure}[ht!]
    \centering
    \includegraphics[width=0.45\linewidth]{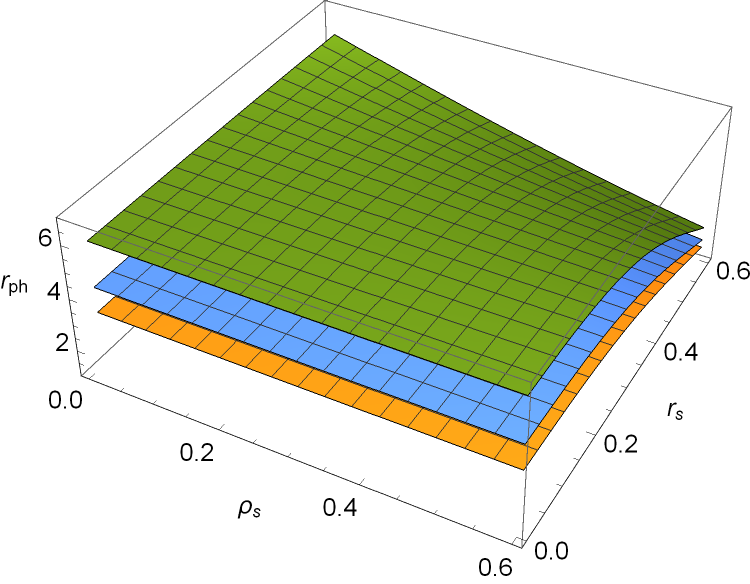}
    \includegraphics[width=0.45\linewidth]{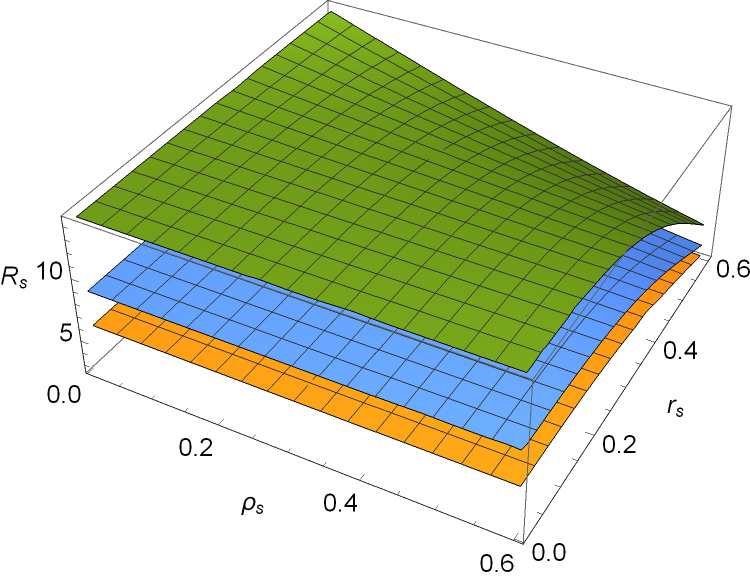}
    \caption{\footnotesize Plot of  the photon sphere $ r_{ph}$ and shadow radius $R_s$ versus  the  parameters ($\rho_s, r_s$) for different values of CS parameter $\alpha=0.5$ (top curve), $\alpha=0.3$ (middle curve) and $\alpha=0.1$ (bottom curve).}
    \label{figph1}
\end{figure}

Table \ref{taba13} depicts the qualitative change in the photon sphere and shadow radius caused by a change in the DM and CS parameters.

\begin{center}
\begin{tabular}{|c|c c|c c|c c|}
 \hline 
 \multicolumn{7}{|c|}{ $\alpha=0.1$}
\\  \hline
 & $r_{ph}$ & $R_{s}$ & $r_{ph}$ & $R_{s}$ & $r_{ph}$ & $R_{s}$ \\ \hline
$r_{s} $ &  \multicolumn{2}{|c|}{$\rho_{s}=0.2$}   & \multicolumn{2}{|c|}{$\rho_{s}=0.4$}  & \multicolumn{2}{|c|}{$\rho_{s}=0.6$}    \\ \hline
$0.2$ & $3.30234$ & $6.0172$ & $3.27139$ & $5.95988$ & $3.24048$ & $5.90263$
\\ 
$0.4$ & $3.10488$ & $5.64328$ & $2.88165$ & $5.21944$ & $2.66449$ & $4.8045$
\\ 
$0.6$ & $2.64152$ & $4.73644$ & $2.04082$ & $3.54428$ & $1.55607$ & $2.55709$
\\ 
 \hline 
 \multicolumn{7}{|c|}{ $\alpha=0.3$}
\\  \hline 
$0.2$ & $4.24519$ & $8.75336$ & $4.2047$ & $8.66922$ & $4.16427$ & $8.58514$
\\ 
$0.4$ & $3.98203$ & $8.19626$ & $3.68417$ & $7.56356$ & $3.39323$ & $6.94152$
\\ 
$0.6$ & $3.34913$ & $6.8163$ & $2.52598$ & $4.99053$ & $1.86111$ & $3.47127$
\\ 
 \hline 
 \multicolumn{7}{|c|}{ $\alpha=0.5$}
\\  \hline 
$0.2$ & $5.94228$ & $14.4023$ & $5.88461$ & $14.2638$ & $5.82698$ & $14.1253$
\\ 
$0.4$ & $5.55991$ & $13.4704$ & $5.12636$ & $12.4078$ & $4.70075$ & $11.356$
\\ 
$0.6$ & $4.61409$ & $11.1001$ & $3.37378$ & $7.91793$ & $0.617369$ & $\mathrm{x}$
\\ 
 \hline
\end{tabular}
\captionof{table}{Numerical values for the photon radius and shadow radius with various BH parameters.} \label{taba13}
\end{center} 

Next, we introduce celestial coordinates $X$ and $Y$, to describe the actual shadow of the BH seen in an observer's frame \cite{Vazquez}  
\begin{eqnarray}
X&=&\lim_{r_{o}\rightarrow \infty }\left( -r_{o}^{2}\sin \theta _{o}\frac{d\varphi }{dr}\right) ,  \label{XX} \\
Y&=&\lim_{r_{o}\rightarrow \infty }\left( r_{o}^{2}\frac{d\theta }{dr}\right). \label{YY}
\end{eqnarray}%
Here, $r_{o}$ is the location of the observer, and $\theta _{o}$ is the angle of inclination. When we consider the equatorial plane with $\theta _{0}=\pi /2$, the celestial coordinates simplify to 
\begin{equation}
X^{2}+Y^{2}=R_{s}^{2}.
\end{equation} 
We now show how the core density and radius affect the BH shadows. First, in Fig. \ref{figsh14}, we demonstrate the impact of the core density. The shadows have decreasing radii as $\rho_s$ increases (left panel). A similar trend is shown for the parameter $r_s$ (right panel).
\begin{figure}[ht!]
    \centering
    \includegraphics[width=0.45\linewidth]{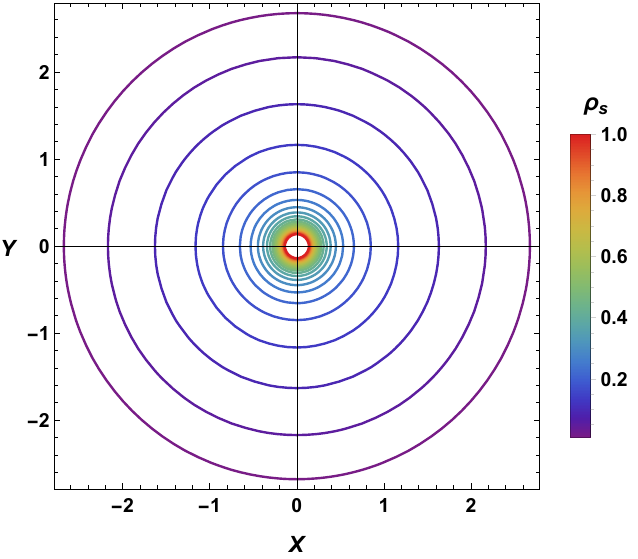}\quad\quad
    \includegraphics[width=0.45\linewidth]{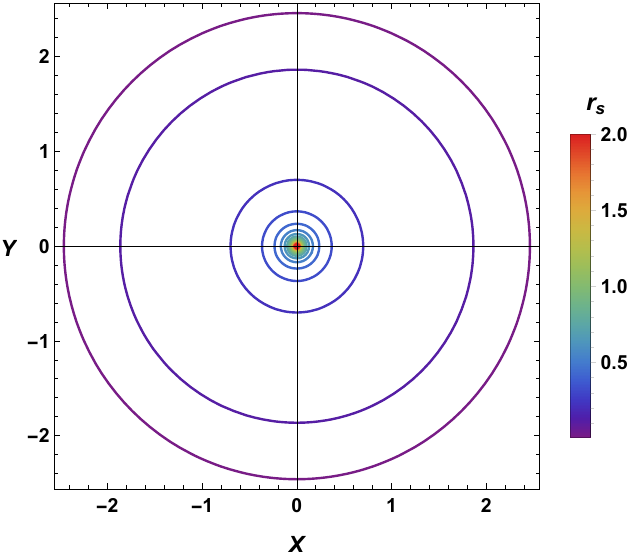}
    \caption{The geometrical shape of the shadow radius in terms of celestial coordinates for   several values of $\rho_s$  (left), and $r_s$ (right). Here, $\alpha=0.2$.}
    \label{figsh14}
\end{figure}

\section{Conclusions} \label{sec6}

{\color{black}In this comprehensive investigation, we presented a detailed theoretical analysis of a novel BH solution that describes an SBH embedded within an HDMH and surrounded by a cloud of CS, which we termed the SHDMHCS configuration. Our study systematically explored the observable signatures of this composite gravitational system, extending from fundamental geodesic motion to the formation of BH shadows, thereby providing a complete theoretical framework for understanding the rich phenomenology that emerges when multiple exotic matter components coexist in the vicinity of a compact object.

We began our analysis in Section \ref{sec2} by deriving the complete spacetime geometry of the SHDMHCS system, starting from the Hernquist density profile given in Eq. (\ref{aa2}) and progressing through the Einstein field equations to obtain the final metric function presented in Eq. (\ref{function}). The embedding diagrams shown in Figures \ref{figizzet} and \ref{figizzet2} provided compelling visual evidence of how the combined presence of HDMH and CS dramatically alters the spacetime curvature compared to the classical Schwarzschild geometry. We observed that increasing the core density $\rho_s$ leads to progressively more pronounced spacetime deformation, with the system exhibiting exotic "inner BH geometry" when $\rho_s$ exceeds critical threshold values. The scalar curvature analysis, encapsulated in Eqs. (\ref{scalar1})-(\ref{scalar3}), confirmed that our solution represents a genuine singularity at $r = 0$ while maintaining asymptotically AdS behavior, as demonstrated in Eq. (\ref{scalar4}).

Our comprehensive geodesic analysis in Section \ref{sec3} revealed the intricate dynamics of both massless and massive test particles in the SHDMHCS background. For null geodesics, we derived the effective potential in Eq. (\ref{cc1}) and demonstrated through Figure \ref{fig:null-potential} that increasing values of the CS parameter $\alpha$, core radius $r_s$, and core density $\rho_s$ collectively lead to a systematic decrease in the effective potential. The force analysis, presented in Eq. (\ref{cc2}) and visualized in Figure \ref{fig:force}, unveiled the remarkable result that the combined influence of CS and HDMH induces a confining-like effect, where the effective force on photons becomes increasingly attractive at larger distances, contrary to the behavior expected in pure Schwarzschild spacetime. We derived the photon trajectory equation (\ref{cc5}), which explicitly demonstrated how both the CS parameter and HDMH density modify the orbital dynamics of light rays. The stability analysis of circular null geodesics provided crucial insights into the response characteristics of the system. We computed the Lyapunov exponent in Eq. (\ref{cc11}), showing that the presence of HDMH decreases the stability of photon orbits compared to the pure Letelier-AdS case given in Eq. (\ref{cc12}). Similarly, the geodesic angular velocity analysis, presented in Eqs. (\ref{cc14}) and (\ref{cc15}), revealed that HDMH presence leads to slower orbital motion of photons around the BH, with all system parameters playing coordinated roles in determining the orbital characteristics. For timelike geodesics, our investigation of massive particle dynamics revealed equally rich phenomenology. The effective potential analysis in Eq. (\ref{dd1}) and the corresponding visualizations in Figures \ref{fig:timelike-potential}-\ref{fig:energy} demonstrated systematic trends in how the various parameters influence particle motion. We derived the expressions for angular momentum and energy of circular orbits in Eqs. (\ref{dd2}) and (\ref{dd3}), respectively, showing that the maximum particle energy approaches $\sqrt{1-\alpha}$ at spatial infinity, indicating a fundamental modification of the energy landscape due to the CS parameter. The orbital velocity analysis in Eq. (\ref{dd7}) provided insights into the motion of distant astronomical objects, while the geodesic precession frequency calculation in Eq. (\ref{dd11}) demonstrated how HDMH presence decreases precession effects compared to the pure Letelier-AdS case shown in Eq. (\ref{dd12}).

Our perturbation analysis in Section \ref{sec4} investigated the stability and response characteristics of the SHDMHCS system under small field perturbations. For scalar perturbations, we derived the effective potential in Eq. (\ref{ff6}) and demonstrated through Figure \ref{fig:scalar-potential} how the various system parameters influence the propagation of spin-0 fields. The three-dimensional visualization in Figure \ref{fig:scalar-potential-contour} provided additional insights into the parameter space dependence of the scalar potential. For electromagnetic perturbations, the effective potential given in Eq. (\ref{em5}) and analyzed in Figure \ref{fig:em-potential} showed similar trends, with all parameters contributing to potential modifications. The comparative analysis in Figure \ref{fig:comparison-potential} revealed that the SHDMHCS configuration produces intermediate behavior between pure Schwarzschild-AdS and Letelier-AdS cases for scalar perturbations, while electromagnetic potentials show more pronounced deviations.

The BH shadow analysis in Section \ref{sec5} provided the most direct connection to observational astronomy. We derived the photon sphere radius through the complex analytical expression in Eq. (\ref{phton3}) and connected it to the shadow radius via Eq. (\ref{shadeq1}). The three-dimensional visualizations in Figure \ref{figph1} demonstrated that increasing the CS parameter $\alpha$ expands both the photon sphere and shadow radii, while increasing core density and radius have the opposite effect. The numerical results presented in Table \ref{taba13} quantified these effects across different parameter ranges, showing substantial variations in shadow properties that could potentially be observable with current and future astronomical facilities. The celestial coordinate analysis visualized in Figure \ref{figsh14} provided direct representations of how the shadow appears to distant observers, with clear dependencies on both $\rho_s$ and $r_s$ parameters.

Throughout our investigation, we consistently observed that the SHDMHCS system exhibits behavior that cannot be captured by simpler models containing only single exotic matter components. The synergistic effects of HDMH and CS create unique signatures in geodesic motion, perturbation propagation, and shadow formation that distinguish this configuration from both pure vacuum solutions and single-component modifications. These results have significant implications for observational astronomy, as they suggest that careful analysis of BH shadow properties, orbital dynamics of test particles, and gravitational wave signatures could potentially reveal the presence of composite exotic matter environments around astrophysical BHs. Our findings also contribute to the theoretical understanding of modified gravity and exotic matter effects in strong-field regimes. The analytical expressions we derived offer practical tools for comparing theoretical predictions with astronomical observations, particularly in the era of high-precision BH imaging and gravitational wave astronomy.

Looking toward future research directions, several promising avenues emerge from this work. First, we plan to extend our perturbation analysis to include gravitational perturbations and higher-spin fields, which will provide insights into gravitational wave emission and quasinormal mode spectra of SHDMHCS systems. Second, we intend to investigate the thermodynamic properties of these composite BH solutions, including Hawking radiation, entropy, and phase transition behavior. Third, we aim to develop numerical simulations to study the nonlinear dynamics of matter accretion onto SHDMHCS BHs and the resulting electromagnetic signatures. Fourth, we plan to conduct detailed comparisons with observational data from the EHT and future space-based gravitational wave detectors to constrain the parameters of realistic astrophysical systems. Finally, we intend to explore generalizations of our work to rotating BH solutions and investigate how angular momentum affects the interplay between HDMH and CS components.}

{\footnotesize

\section*{Acknowledgments}

F.A. acknowledges the Inter University Centre for Astronomy and Astrophysics (IUCAA), Pune, India, for granting a visiting associateship. \.{I}.~S. expresses gratitude to T\"{U}B\.{I}TAK, ANKOS, and SCOAP3 for helpful support. He also acknowledges COST Actions CA22113, CA21106, and CA23130 for their contributions to networking.

\section*{Data Availability Statement}

This manuscript has no associated data.

\section*{Code Availability Statement}

This manuscript has no associated code/software.

\section*{Conflict of interest}

Author(s) declare no such conflict of interest.
}

\end{document}